\documentclass[a4paper,fleqn,usenatbib]{mnras}

\usepackage{amsmath}	
\usepackage{amssymb}	
\usepackage{txfonts}

\usepackage[T1]{fontenc}
\usepackage{ae,aecompl}

\usepackage{graphicx}	
\usepackage{grffile}  
\usepackage{natbib}
\usepackage{url}

\usepackage{multirow}

\bibpunct[; ]{(}{)}{;}{a}{}{,}

\newcommand{\eqnref}[1]{equation~(\ref{#1})}

\newcommand{\Herschel}{\textit{Herschel}}
\newcommand{\tphot}{\textsc{t-phot}}

\newcommand{\Lir}{\ensuremath{L_\text{IR}}}
\newcommand{\Luv}{\ensuremath{L_\text{UV}}}
\newcommand{\Lsun}{\ensuremath{\textrm{L}_{\sun}}}
\newcommand{\sfr}{\ensuremath{\textrm{M}_{\sun} \textrm{yr}^{-1}}}
\newcommand{\MHtwo}{\ensuremath{M_{\mathrm{H}_2}}}
\newcommand{\Mstar}{\ensuremath{M_{\ast}}}
\newcommand{\Mdust}{\ensuremath{M_{d}}}
\newcommand{\Msun}{\ensuremath{\textrm{M}_{\sun}}}

\newcommand{\LUV}{\ensuremath{L_\text{UV}}}
\newcommand{\ci}{\mbox{[C{\sc i}]}}

\newcommand{\kms}{\mbox{km\,s$^{-1}$}}

\newcommand{\cofour}{\mbox{CO(4--3)}}
\newcommand{\coone}{\mbox{CO(1--0)}}
\newcommand{\cothree}{\mbox{CO(3--2)}}

\newcommand{\Htwo}{\ensuremath{\textrm{H}_2}}
\newcommand{\Xci}{\ensuremath{X_\mathrm{CI}}}

\title[Dust and {[CI]} at z=1]{The relationship between dust and \ci\ at $z = 1$ and beyond}
\author[N. Bourne et al.]
 {N. Bourne,$^{1}$\thanks{nbourne22@gmail.com}
J.\,S. Dunlop,$^{1}$
J.\,M. Simpson,$^{2}$
K.\,E. Rowlands,$^{3}$
J.\,E. Geach,$^{4}$\newauthor
D.\,J. McLeod$^{1}$
   \\
   $^1$Institute for Astronomy, University of Edinburgh, Royal Observatory, Edinburgh EH9 3HJ, UK\\
   $^2$Academia Sinica Institute of Astronomy and Astrophysics, No. 1, Sec. 4, Roosevelt Road, Taipei 10617, Taiwan\\
   $^3$Department of Physics \& Astronomy, Johns Hopkins University, Bloomberg Center, 3400 N. Charles St., Baltimore, MD 21218,USA\\
   $^{4}$Centre for Astrophysics Research, Science and Technology Research Institute, University of Hertfordshire, Hatfield AL10 9AB, UK
 }

\date{Accepted XXX. Received YYY; in original form ZZZ}

\pubyear{2018}

\begin{document}
\label{firstpage}
\pagerange{\pageref{firstpage}--\pageref{lastpage}}
\maketitle

\begin{abstract}
Measuring molecular gas mass is vital for understanding the evolution of galaxies at high redshifts ($z\gtrsim1$).
Most measurements rely on CO as a tracer, but dependences on metallicity, dynamics and surface density lead to systematic uncertainties in high-$z$ galaxies, where these physical properties are difficult to observe, and where the physical environments can differ systematically from those at $z=0$.
Dust continuum emission provides a potential alternative assuming a known dust/gas ratio, but this must be calibrated on a direct gas tracer at $z\gtrsim1$.
In this paper we consider the \ci\ 492-GHz emission line, which has been shown to trace molecular gas closely throughout Galactic clouds and has the advantages of being optically thin in typical conditions (unlike CO), and being observable at accessible frequencies at high redshifts (in contrast to the low-excitation lines of CO). 
We use the Atacama Large Millimetre/submillimetre Array (ALMA) to measure \ci, \cofour\ and dust emission in a representative sample of star-forming galaxies at $z=1$, and combine these data with multi-wavelength spectral energy distributions to study relationships between dust and gas components of galaxies.
We uncover a strong \ci--dust correlation, suggesting that both trace similar phases of the gas. By incorporating other samples from the literature, we show that this correlation persists over a wide range of luminosities and redshifts up to $z\sim4$.
Finally we explore the implications of our results as an independent test of literature calibrations for dust as a tracer of gas mass, and for predicting the C\textsc{i} abundance.
\end{abstract}

\begin{keywords}
Galaxies: evolution -- 
galaxies: ISM --
submillimetre: galaxies
\end{keywords}



\section{Introduction}
\label{sec:intro}

Research in galaxy evolution at high redshifts ($z\geq 1$) is concerned with understanding how the stellar mass present in galaxies today was built up over the history of the Universe. 
Thanks to a wealth of photometric data in the rest-frame UV, optical, near-infrared (NIR) and far-infrared (FIR), we have built up an impressive level of understanding of galaxy demographics, stellar masses and star-formation rates (SFRs) throughout a large fraction of cosmic history \citep[][]{Madau2014}.
We know, for example, that galaxies at high-$z$ had much higher SFRs than today (by a factor $\gtrsim10$), and as a result, most of the stellar mass was assembled at $z>1$, residing for the most part in massive galaxies ($\log\Mstar/\Msun\gtrsim10$). 
In constrast, the SFR density of the Universe has declined rapidly in the last 8--10 Gyr, and today's star-forming galaxies are typically less active, less massive, and less obscured by dust. 
Yet these results tell us nothing about the supply of cold gas necessary to fuel star formation, for which we still lack sufficiently strict observational constraints to validate theoretical models \citep[e.g.][]{Genel2014, Lagos2015}

Recent studies are beginning to reveal a consensus picture in which the high-redshift Universe was much richer in cold gas than the present day \citep[e.g.][]{Decarli2016}, and in which star-formation efficiency (SFE~=~SFR$/\MHtwo$, i.e. the inverse of gas depletion time) increases with lookback time and with specific SFR (SSFR=SFR/\Mstar; \citealp{Combes2013, Genzel2015, Tacconi2018}). 
Not only did galaxies have much more fuel for star formation in the past, but it appears that they may also have been able to process that fuel more efficiently into stellar mass.
However, interpretation of these results is significantly hampered by systematic uncertainties in the calibration of molecular gas mass ($\MHtwo$).
The most widely used tracers of molecular gas are the CO rotational transitions. 
The low-$J$ transitions have low excitation temperatures and critical densities, hence are easily excited throughout the molecular phase in galaxies \citep{Carilli2013}. 
However, systematic variations are introduced by the dependence of CO abundance on gas metallicity \citep{Leroy2011a, Narayanan2012a}.
Another problem is that the CO emission lines are optically thick, so the line luminosity is not directly related to the mass of emitting material, but is chiefly dependent on the line width. 
A constant conversion factor $\alpha_\mathrm{CO} = \MHtwo / L^{\prime}_{\mathrm{CO(1-0)}}\approx 4$~\Msun\,(K\,\kms\,pc$^2$)$^{-1}$ can be assumed in kinematically well-ordered disk galaxies like the Milky Way, on the basis that individual giant molecular clouds (GMCs) are offset from each other in velocity and have approximately uniform surface density, so that the total line width is proportional to the number of clouds and hence to their total mass \citep{Bolatto2013}. This value for disk galaxies still has a factor $\sim2$ scatter, and the assumption breaks down in systems with a higher gas filling factor such as starburst nuclei in the local Universe. In such systems, line luminosity depends on total dynamical mass and on gas pressure, and a lower conversion factor $\alpha_\mathrm{CO} \sim0.8$~\Msun\,(K\,\kms\,pc$^2$)$^{-1}$ has been determined \citep{Downes1998}.

These dependences on the metallicity, dynamics and surface density of gas become even more problematic at high redshifts where such properties are likely to differ from galaxies in the local Universe, and require detailed spectroscopic observations to measure directly.  
An additional complication is that, while a library of several hundred CO measurements has now been assembled \citep{Carilli2013, Sargent2014, Tacconi2018}, most of those at $z\geq1$ are from \cothree, which has a relatively high excitation temperature and critical density, introducing further variability in the conversion to total gas mass. 
Such systematic uncertainties obfuscate the true relationship between molecular gas and star formation as seen in observations, especially across the wide range of physical conditions spanning star-forming disk galaxies and luminous starbursts \citep{Narayanan2011c}.

An idea that is becomingly increasingly popular is the utility of dust continuum emission in the sub-millimetre (submm) as a tracer of total gas mass \citep{Hughes1997,Eales2012, Scoville2016}. This has the considerable advantage that measurements exist for large samples across a very wide range of redshifts, thanks to the negative $k$-correction \citep{Blain2002} combined with the availability of sensitive, wide-area submm sky mapping from the \Herschel/SPIRE \citep{Griffin2010} and JCMT/SCUBA-2 \citep{Holland2013} bolometer array cameras. Notwithstanding the difficulties presented by low-resolution submm imaging with these single-dish facilities, dust emission represents a promising avenue for gas calibrations in large samples, since the uncertainties surrounding the variation in the dust/gas ratio (chiefly as a function of metallicity) are smaller than the uncertainties surrounding the CO/H$_2$ conversion \citep{Eales2012}. 
However, this method depends upon accurate calibration of the dust/gas ratio and its dependencies in representative samples of star-forming galaxies, using a direct tracer that is not prone to large systematic uncertainties.
In particular, while it has been shown that the dust continuum is strongly correlated with the CO luminosity with remarkably small scatter over a wide range of luminosities and SSFRs \citep{Scoville2016}, we also know that the CO/H$_2$ ratio is elevated in galaxies with high SSFR \citep[e.g.][]{Genzel2015}. It is likely, therefore, that variations in the CO/H$_2$ ratio are mirrored by variations in the dust/gas ratio, both being dependent on metallicity, so that an unbroken CO/dust correlation may exist in spite of non-linearities in both tracers.

It is imperative, therefore, to develop and validate independent tracers of cold gas to validate the current observational framework based on CO and dust.
An alternative tracer that is less well studied is atomic carbon.
The 492-GHz \mbox{\ci($^3P_1$--$^3P_0$)} transition has similar excitation conditions to \coone, hence traces similar physical environments, 
but {several important factors give it distinct advantages as a tracer of \Htwo}.  
Firstly, this transition is typically optically thin, and has weakly varying excitation \citep{Gerin2000, Papadopoulos2004, Popping2014, Jiao2017}. As a result, its luminosity is more directly correlated with the mass of emitting material than the CO lines are, without recourse to strong assumptions about dynamics. 
Secondly, theoretical predictions from photo-dissociation region (PDR) modelling suggest that the \ci/H$_2$ ratio is {robust over a wider range of metallicities and gas densities} than \coone\ \citep{Papadopoulos2004, Geach2012}.
\footnote{The abundances of both C and CO depend on metallicity, but modelling shows that in low-metallicity environments CO is dissociated by far-UV radiation and by cosmic rays \citep[e.g.][]{Papadopoulos2004,Bisbas2017}; and also that CO is a very poor tracer of young molecular clouds before the onset of star formation \citep{Glover2016}. These theoretical studies conclude that \ci\ is the preferred tracer in low-metallicity environments, which is supported by observations \citep[and references therein]{Hunt2017}.}

Historically, \ci\ has been largely overlooked in comparison to CO. This is partly due to observational limitations (the opacity of the atmosphere at the rest-frame frequency, and the lack of sensitive receivers) preventing investigation of \ci\ in Galactic environments. 
The lack of observational evidence was compounded by theoretical indications that atomic carbon exists only in a narrow transition layer in the photo-dissociation region (PDR) surrounding a GMC. Within the GMC, carbon is locked up in CO, and is shielded from the interstellar radiation field (ISRF), but in the narrow transition layer the ISRF is strong enough to dissociate CO but not strong enough to ionise carbon.
This picture was challenged by observations of the Orion clouds and Galactic centre by \citet{Ikeda2002} and \citet{Ojha2001}, which revealed \ci\ emission throughout the molecular clouds. \citet{Papadopoulos2004} showed that by including turbulence on short dynamical timescales in PDR models, the clouds are not generally in equilibrium and the C\textsc{i} layer penetrates deep within, such that \ci\ emission traces the full H$_2$ mass of the cloud (see also \citealp{Tomassetti2014, Glover2015}).

The frequency of \ci\ makes it more accessible at high redshifts than the low-excitation lines of CO, yet observations at $z>1$ have so far been limited to small samples of luminous submm galaxies (SMGs) and quasars. 
For example, \citet{Walter2011b} and \citet{Alaghband-Zadeh2013} found \ci-derived H$_2$ masses consistent with CO results, although \citet{Bothwell2017} reported some tensions. 
Similarly, at low redshifts, there is no strong consensus as to whether \ci\ is a better H$_2$ tracer than CO in local ultra-luminous IR galaxies (ULIRGs; \citealp{Israel2015, Jiao2017, Lu2017}).
More importantly, however, neither local ULIRGs nor the luminous SMGs studied at high redshifts are representative of the general galaxy population, and they tell us nothing about the cold-gas content of typical galaxies or of the Universe as a whole. 
We need to build up a library of \ci\ measurements in the typical galaxies that dominate the star-forming budget of the Universe, that is massive ($\log\Mstar/\Msun>10$) galaxies at $z>1$ undergoing sustained high-SFR activity. The conditions of star formation in these galaxies are distinct from disk galaxies at $z=0$ (which have relatively low gas densities and low SSFR and SFE) and from merger-driven starbursts (which have very high gas densities and high SSFR in a compact, dense, gas-filled nucleus).
Most high-$z$ star-forming galaxies are characterised by a common relationship between stellar mass and SFR (the star-forming ``main sequence''; \citealp{Speagle2014,Schreiber2015}) indicating that high SFRs are regulated over long timescales, driven by secular processes within galaxies.

The aim of the current work is to conduct a pilot study of the relationship between \ci\ and dust continuum emission in a small sample of star-forming galaxies at $z=1$.
We have selected a sample that is representative of typical galaxies on and around the main sequence at a redshift where \ci\ is readily observable, and in an epoch when many massive galaxies were still actively forming stars.
We have obtained observations of \ci\ and dust continuum from the Atacama Large Millimetre/Submm Array (ALMA). In this paper we study the \ci/dust correlation to determine whether the two tracers are sensitive to the same gas phase, and we look for variations to identify limitations of either tracer that may result from variability in the dust/gas ratio or the properties of \ci\ emission. 
We describe our sample and existing data in Section~\ref{sec:sample}, present the ALMA data in Section~\ref{sec:data}, and discuss correlations in the results in Section~\ref{sec:discussion}. In Section~\ref{sec:discussion2} we compare alternative calibrations of the molecular gas mass, presenting our final conclusions in Section~\ref{sec:conclusions}.
Throughout this paper we adopt a flat $\Lambda$CDM cosmology with $\Omega_M=0.3$ and $h=H_0/100$~km\,s$^{-1}$\,Mpc$^{-1}=0.7$. All magnitudes are in the AB system \citep{Oke1974,Oke1983} and we assume the \citet{Kroupa2003} initial mass function (IMF) throughout, unless otherwise stated.

\section{The sample}
\label{sec:sample}

\begin{figure}
\begin{center}
\includegraphics[width=0.5\textwidth,clip,trim=2mm 0 0 2mm ]{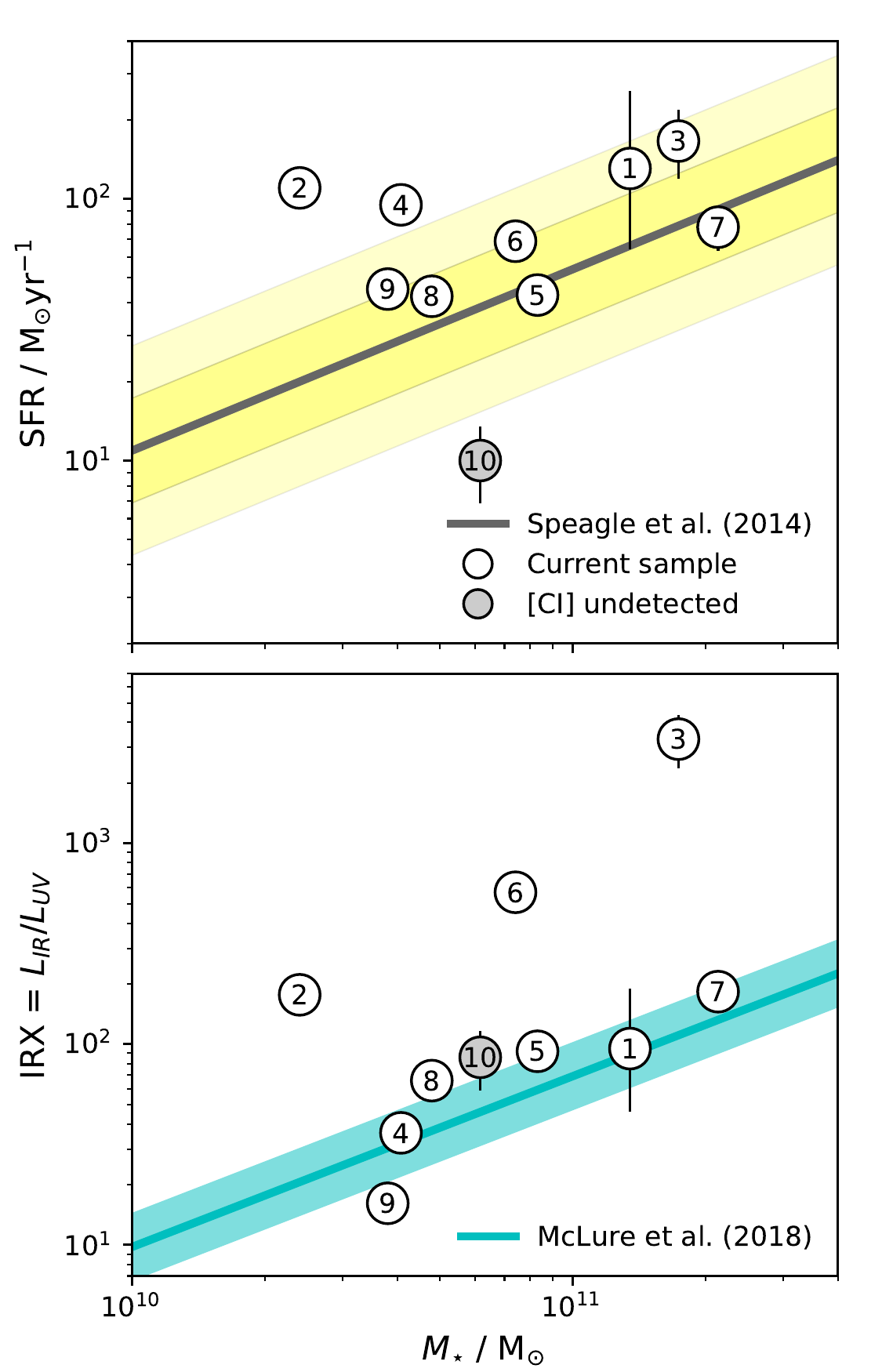}
\caption{Global properties of galaxies in the sample: the relationship between stellar mass and SFR (top panel); and infrared excess, IRX (bottom panel). Stellar masses and UV luminosities are from 3DHST \citep{Skelton2014}. SFRs are derived from UV plus IR luminosities as described in Section~\ref{sec:seds}.
Galaxies are numbered according to the IDs in Table~\ref{tab:physparams}.
In the top panel, the thick grey line indicates the star-forming main sequence at the mean redshift of our sample ($\bar{z}=1.02$) from \citet{Speagle2014}, with the shaded bands representing the scatter of $\pm 1\sigma$ and $\pm 2\sigma$ ($\sigma=0.2$~dex).
In the bottom panel, the cyan band indicates the mean $\pm 1\sigma$ scatter in the IRX--$\Mstar$ relation of $z\sim2.5$ mass-selected star-forming galaxies derived by \citet{McLure2018}. Although our sample is at $z=1$, we do not expect significant evolution in IRX at fixed mass between these epochs \citep[e.g.][]{Bourne2017}.
}
\label{fig:msfr}
\end{center}
\end{figure}

\subsection{Selection}
In order to identify star-forming galaxies at $z=1$ with good measurements of the dust spectral energy distribution (SED), we selected sources detected by SCUBA-2 at 450\,\micron.  While traditional surveys of SMGs selected at wavelengths $\lambda\geq850\,\micron$ tend to favour starbursts with exceptionally high SSFRs (which dominate the very bright end of the luminosity function and source counts), the much smaller beam size of JCMT/SCUBA-2 at 450\,\micron\ (7.5 arcsec FWHM compared with 14 arcsec at 850\,\micron) means that the confusion limit is substantially lower and galaxies of much lower luminosities can be individually detected across a wide range of redshifts. In particular, typical star-forming galaxies of stellar mass $\Mstar\gtrsim10^{10}\Msun$ are detected at 450\,\micron\ at redshifts up to $z\approx 4$ \citep{Roseboom2013, Bourne2017}. 
We selected our sample from the catalogue of prior-based submm photometry for stellar-mass selected galaxies that we used for the analysis in \citet{Bourne2017}.
This data set benefits from deblended flux measurements with uncertainties that take into account full covariances between blended sources, using the \tphot\ deconfusion code \citep{Merlin2015, Merlin2016}. It takes advantage of some of the deepest fields surveyed at 450\,\micron\ from the SCUBA-2 Cosmology Legacy Survey (S2CLS; \citealp{Geach2016}), as well as deep \textit{Hubble Space Telescope} (\textit{HST}) imaging from CANDELS \citep{Koekemoer2011}, extensive multi-wavelength photometry \citep{Skelton2014} and grism redshifts for a large subset of galaxies from 3DHST \citep{Momcheva2015}.

We selected galaxies detected at 450\,\micron\ (deblended $S_{450}>4.2$~mJy, S/N~$>3$) from the UDS-CANDELS and COSMOS-CANDELS fields of S2CLS. We selected 10 galaxies with grism redshifts in the range $0.84<z<1.29$, so that \ci\ falls within ALMA Band 6 (see Table~\ref{tab:physparams}).
The sample encompasses a representative variety of star-forming galaxies from across the stellar mass/SFR plane, as shown in Fig.~\ref{fig:msfr} (top panel). 
The population of star-forming galaxies at all redshifts is dominated by the main sequence, a roughly linear relation between stellar mass and SFR with relatively small dispersion of about 0.2 dex \citep[][]{Speagle2014}. Galaxies in our sample are distributed on and above the main sequence, hence can be considered representative of the types of galaxies responsible for most of the stellar mass growth throughout the history of the Universe. 

It might be expected that a 450-\micron-selected sample should be somewhat biased in terms of dust obscuration. However, as we show in Fig.~\ref{fig:msfr} (bottom panel), all but three of the galaxies have an IR/UV excess (IRX) consistent with the average star-forming galaxies of the same stellar mass from \citet{McLure2018}. The reference IRX--\Mstar\ relation from \citet{McLure2018} was determined by stacking ALMA continuum emission from stellar-mass-selected galaxies at $2<z<3$, and although the current sample is at $z=1$, observational studies agree that the relation does not evolve significantly between these epochs \citep{Heinis2014,Pannella2015,Bourne2017,Fudamoto2017}.

It is also relevant to consider the gas-phase metallicities of galaxies in our sample, since the carbon abundance is expected to scale with metallicity \citep{Papadopoulos2004, Glover2016}. 
The metallicities in our sample are unknown, but we can adopt the widely-used \citet{Mannucci2010} relation to predict metallicity as a function of \Mstar\ and SFR, taking Fig.~\ref{fig:msfr} as evidence that we do not expect most of these galaxies to be exceptionally dusty (hence metal-rich) for their stellar mass and SFR. Using the data in Table~\ref{tab:physparams}, we predict moderate values of oxygen abundance $8.6<12+\log{O/H}<9.0$, i.e. within a factor 2 of solar abundance, indicating that we should expect \ci\ to trace \Htwo\ in a similar way in this sample as in the Milky Way.

\begin{figure*}
\begin{center}
\vspace{-6pt}
\begingroup
\setlength{\tabcolsep}{-2pt} 
\begin{tabular}{llll}
\includegraphics[scale=0.345,clip,trim=2mm 3mm 0mm 3mm]{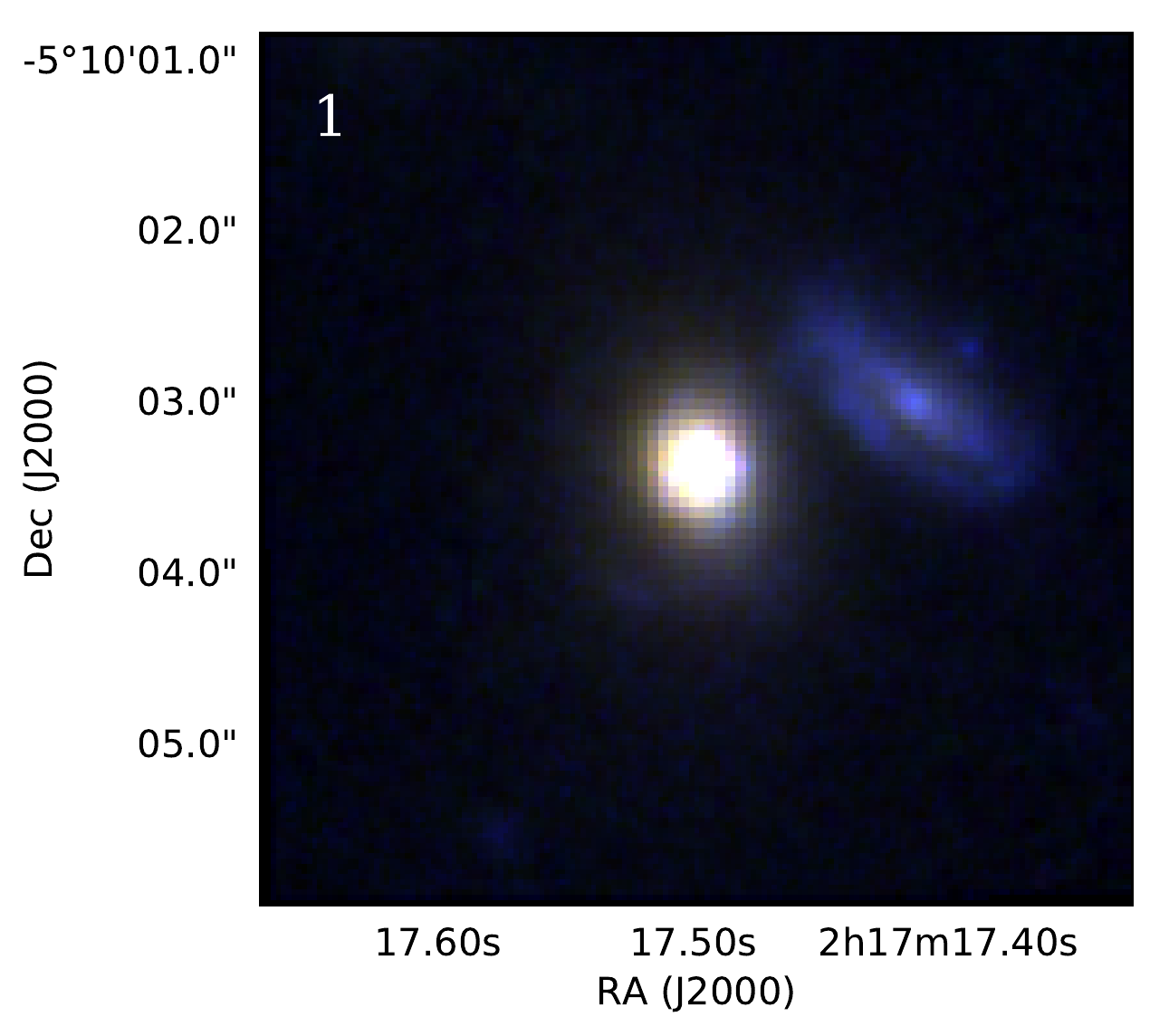} &
\includegraphics[scale=0.345,clip,trim=2mm 3mm 3.5mm 3mm]{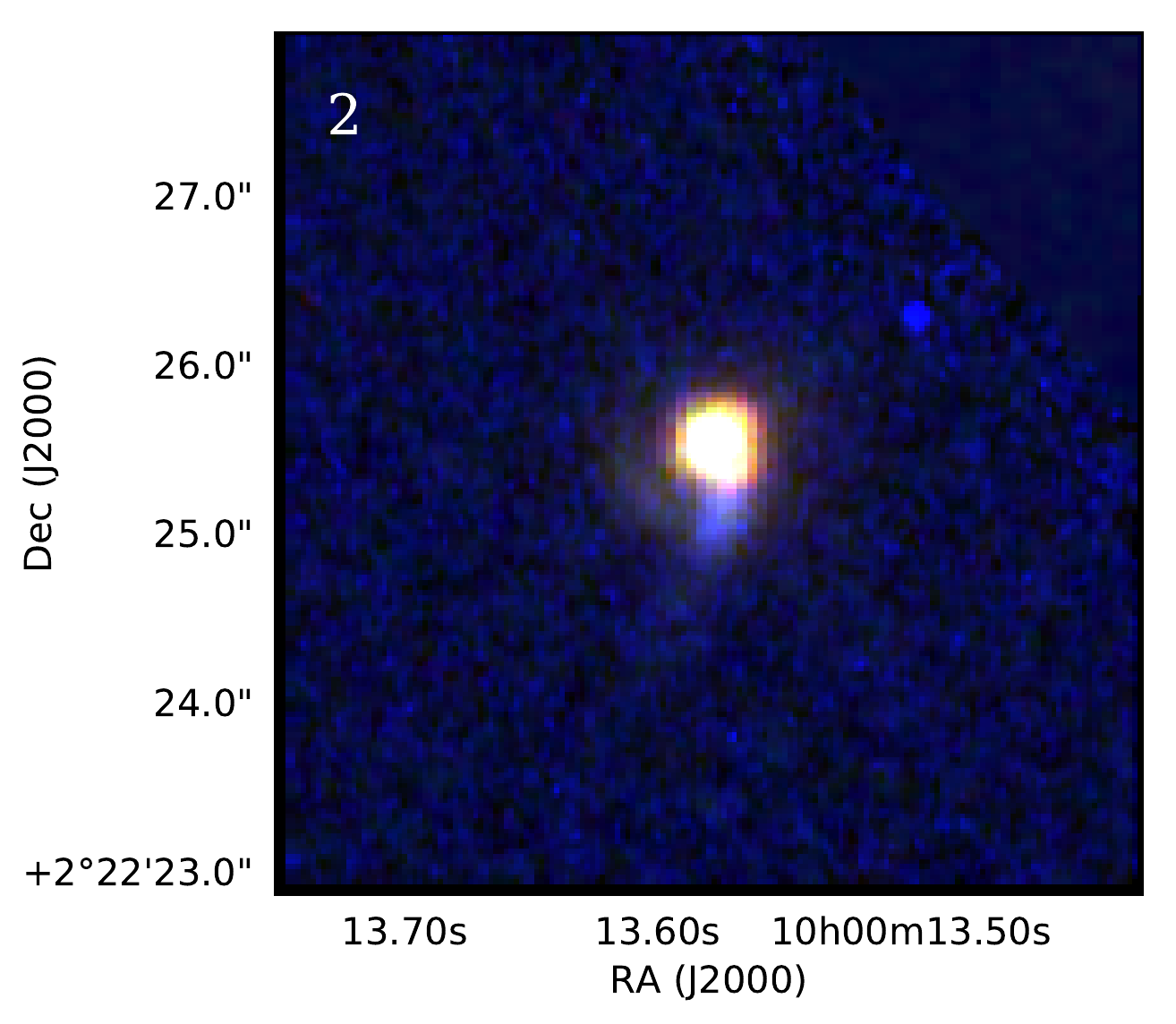} &
\includegraphics[scale=0.345,clip,trim=2mm 3mm 3.5mm 3mm]{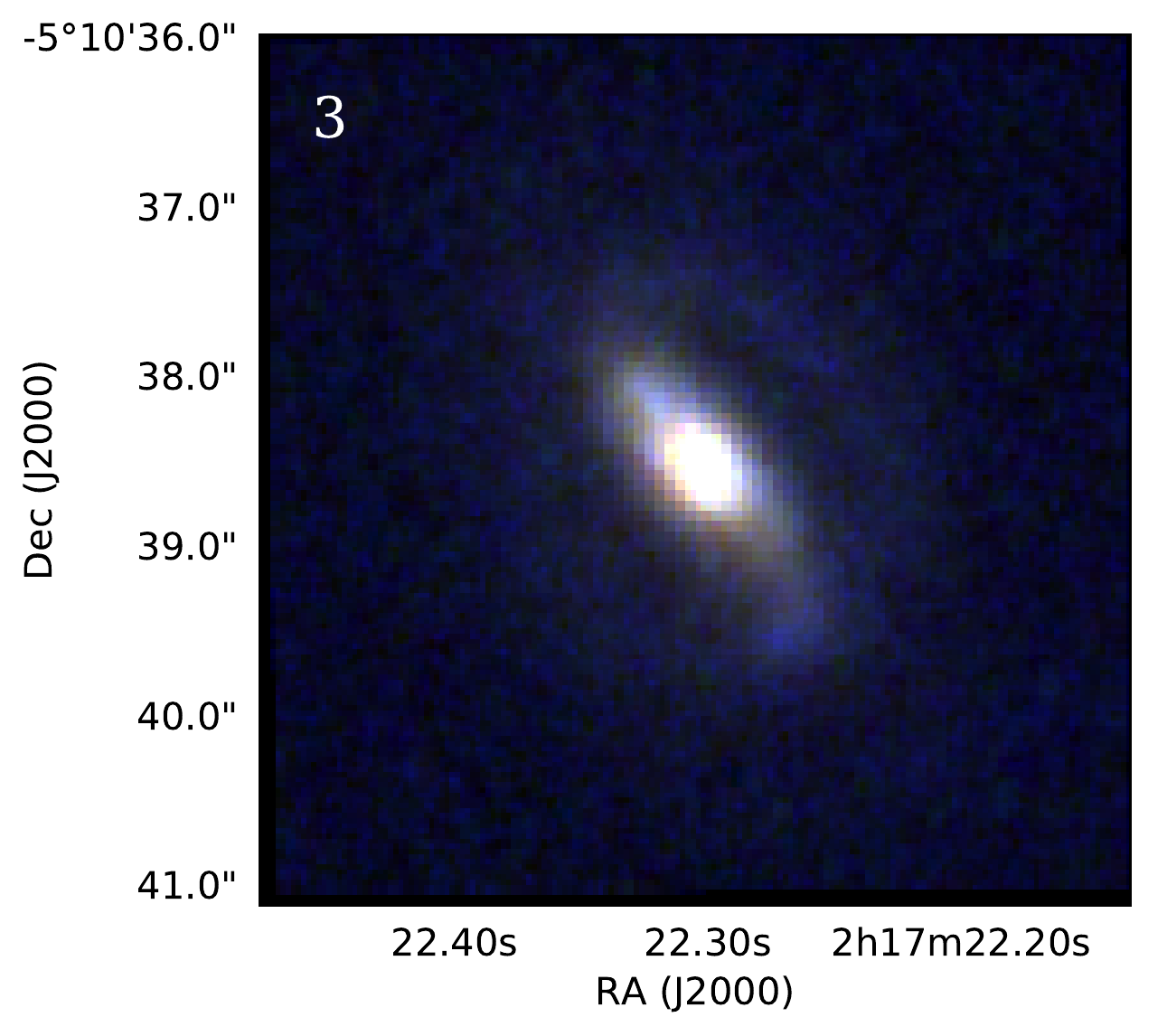} &
\includegraphics[scale=0.345,clip,trim=2mm 3mm 0mm 3mm]{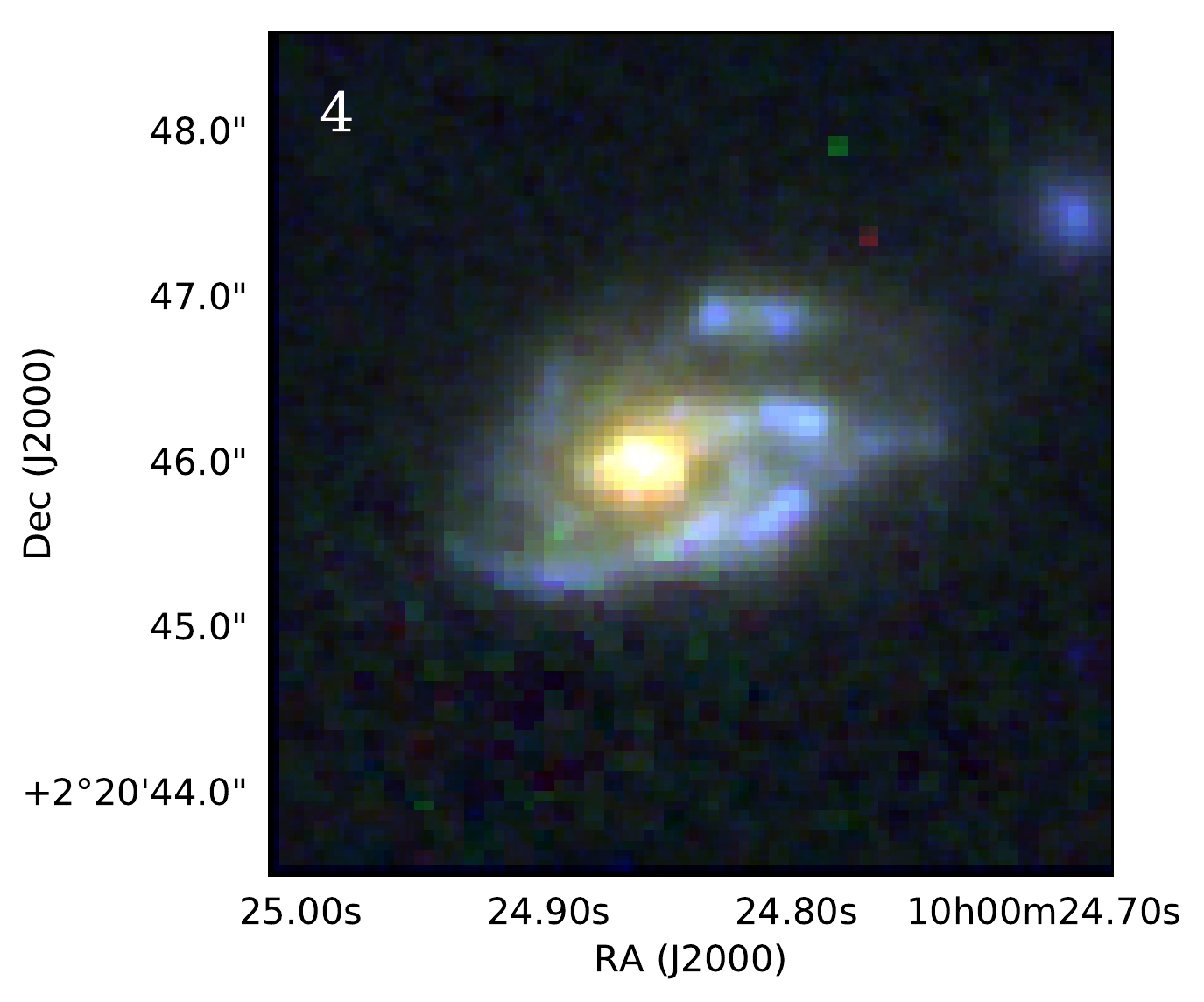} \\ 
\includegraphics[scale=0.345,clip,trim=2mm 3mm 0mm 3mm]{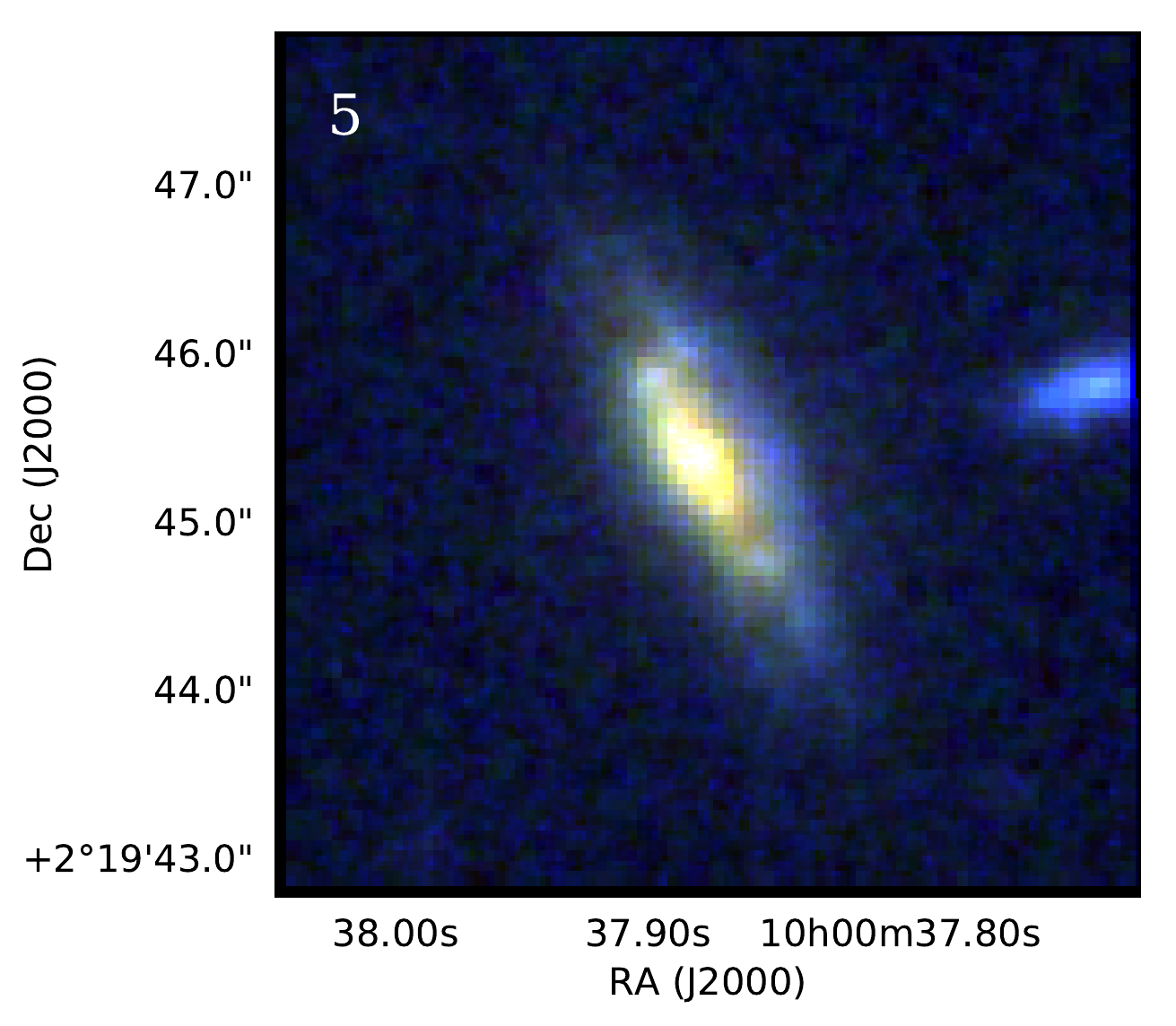} &
\includegraphics[scale=0.345,clip,trim=2mm 3mm 3.5mm 3mm]{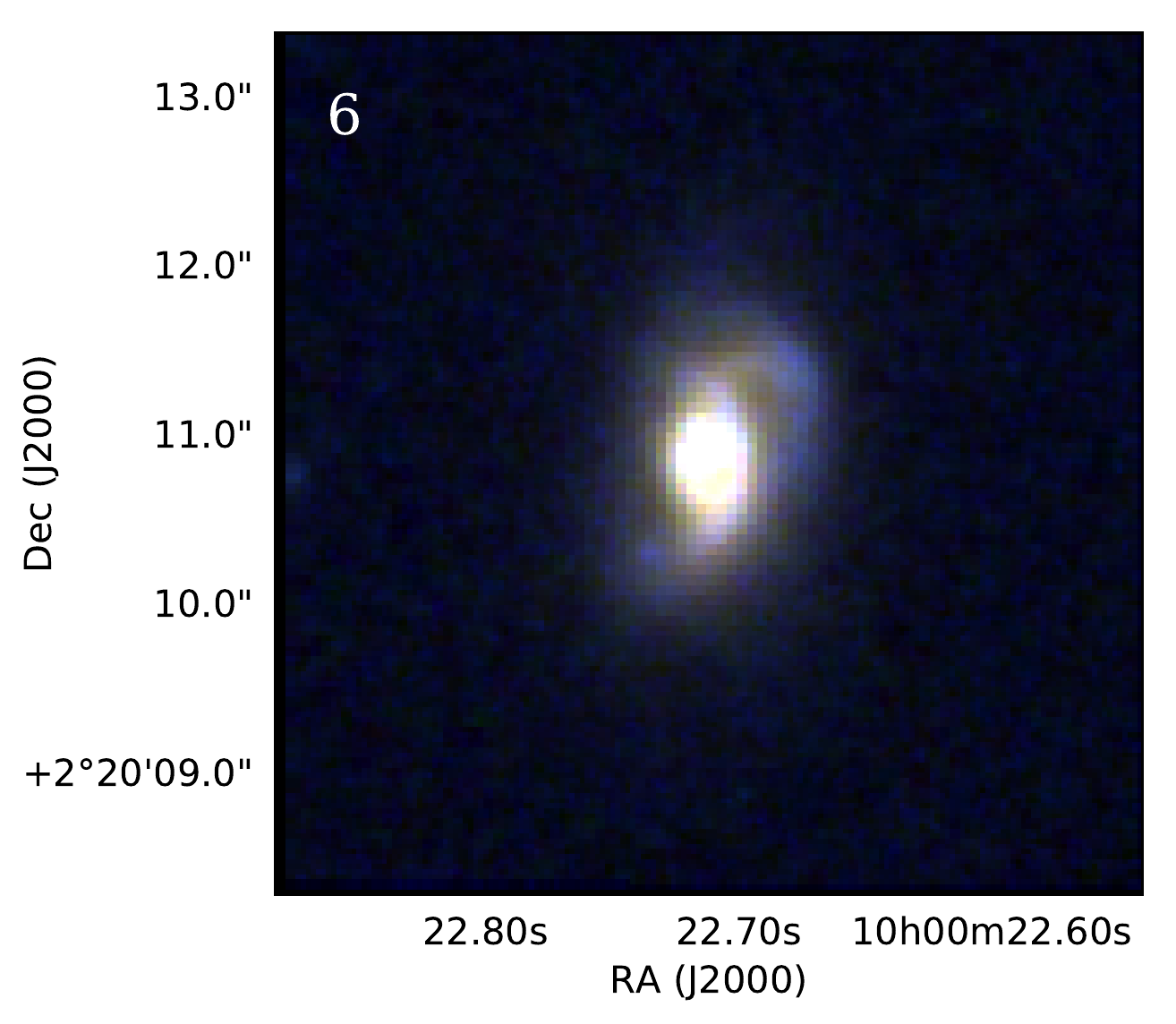} &
\includegraphics[scale=0.345,clip,trim=2mm 3mm 3.5mm 3mm]{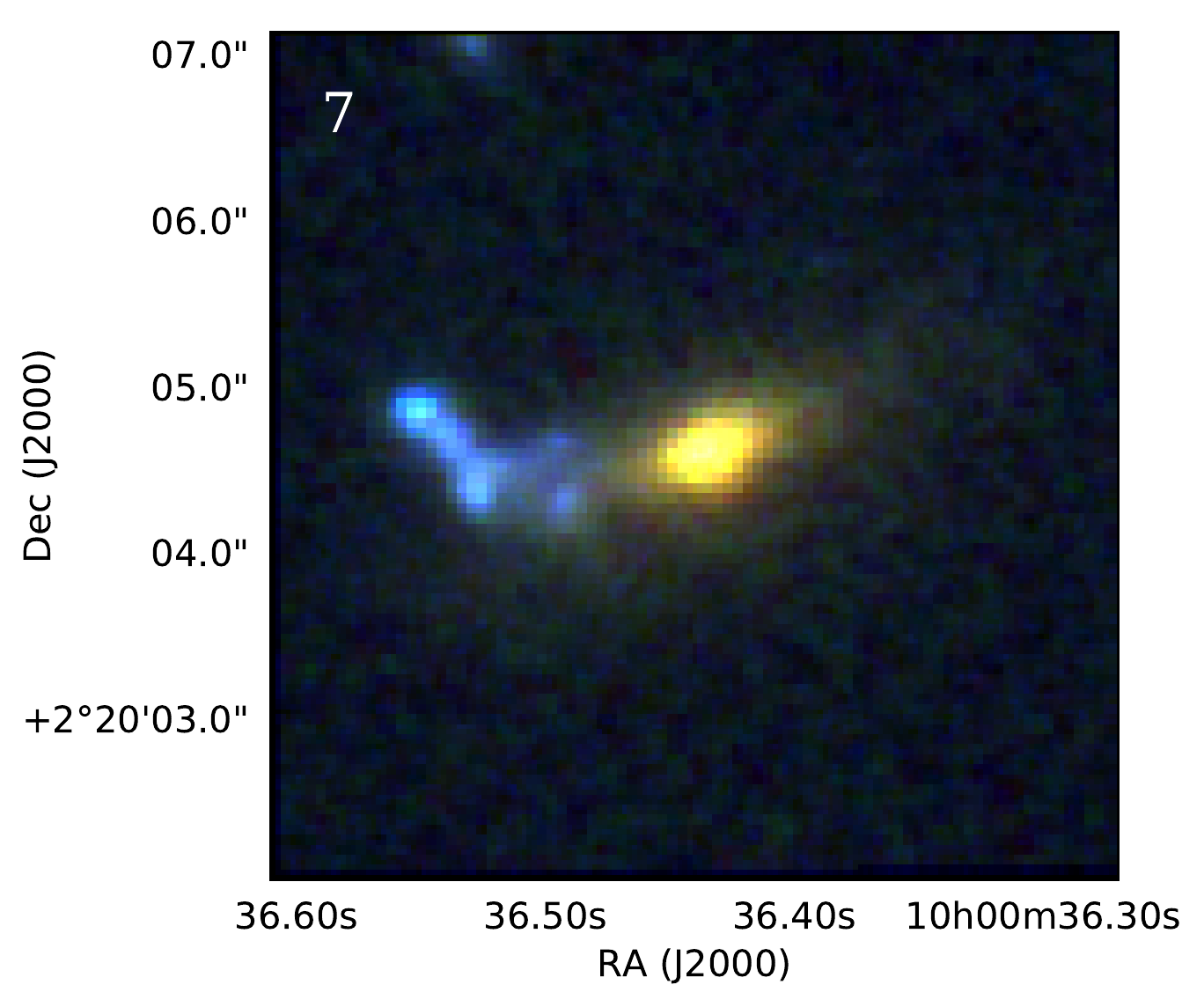} & 
\includegraphics[scale=0.345,clip,trim=2mm 3mm 0mm 3mm]{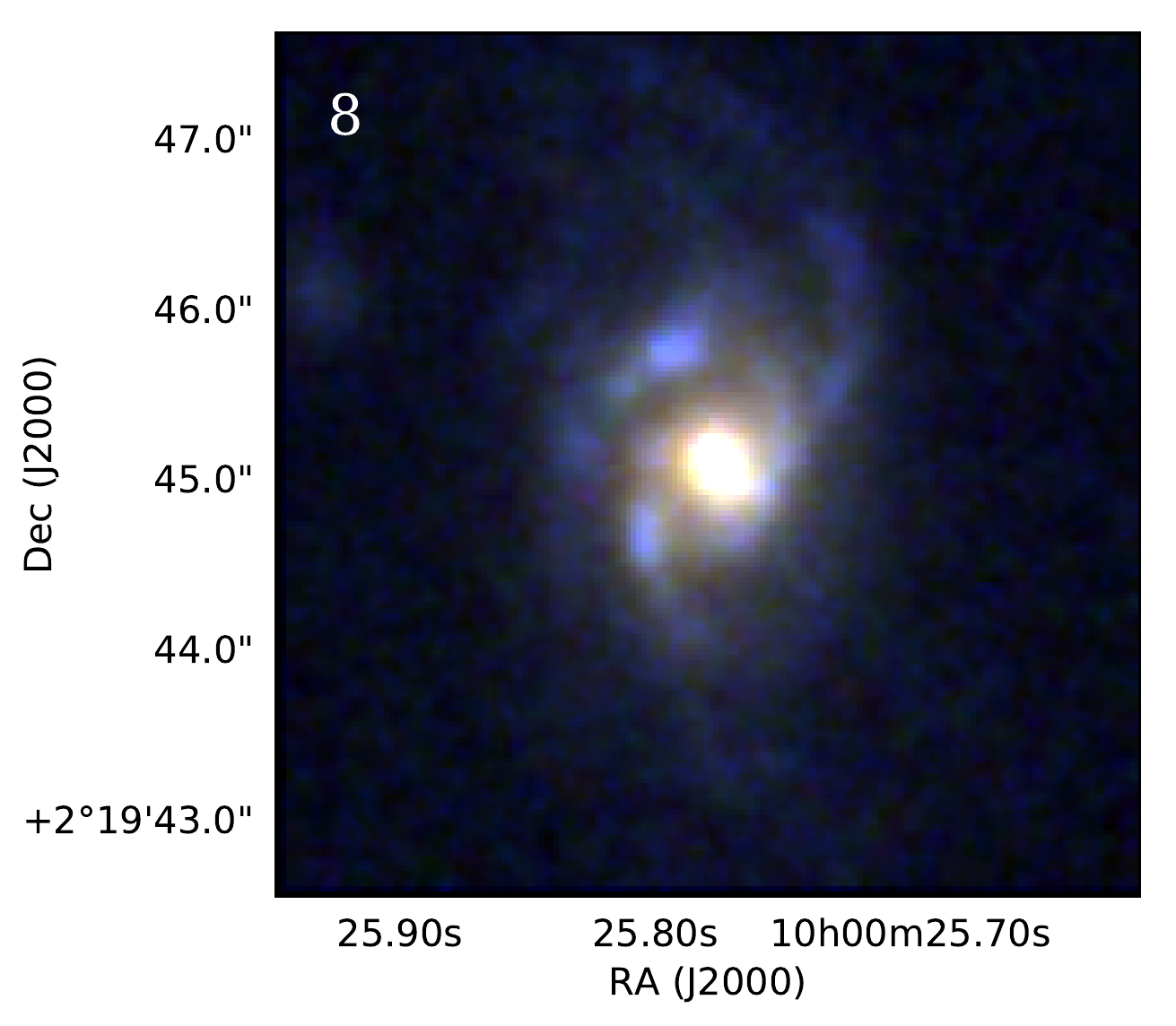} \\ 
& 
\includegraphics[scale=0.345,clip,trim=2mm 3mm 3.5mm 3mm]{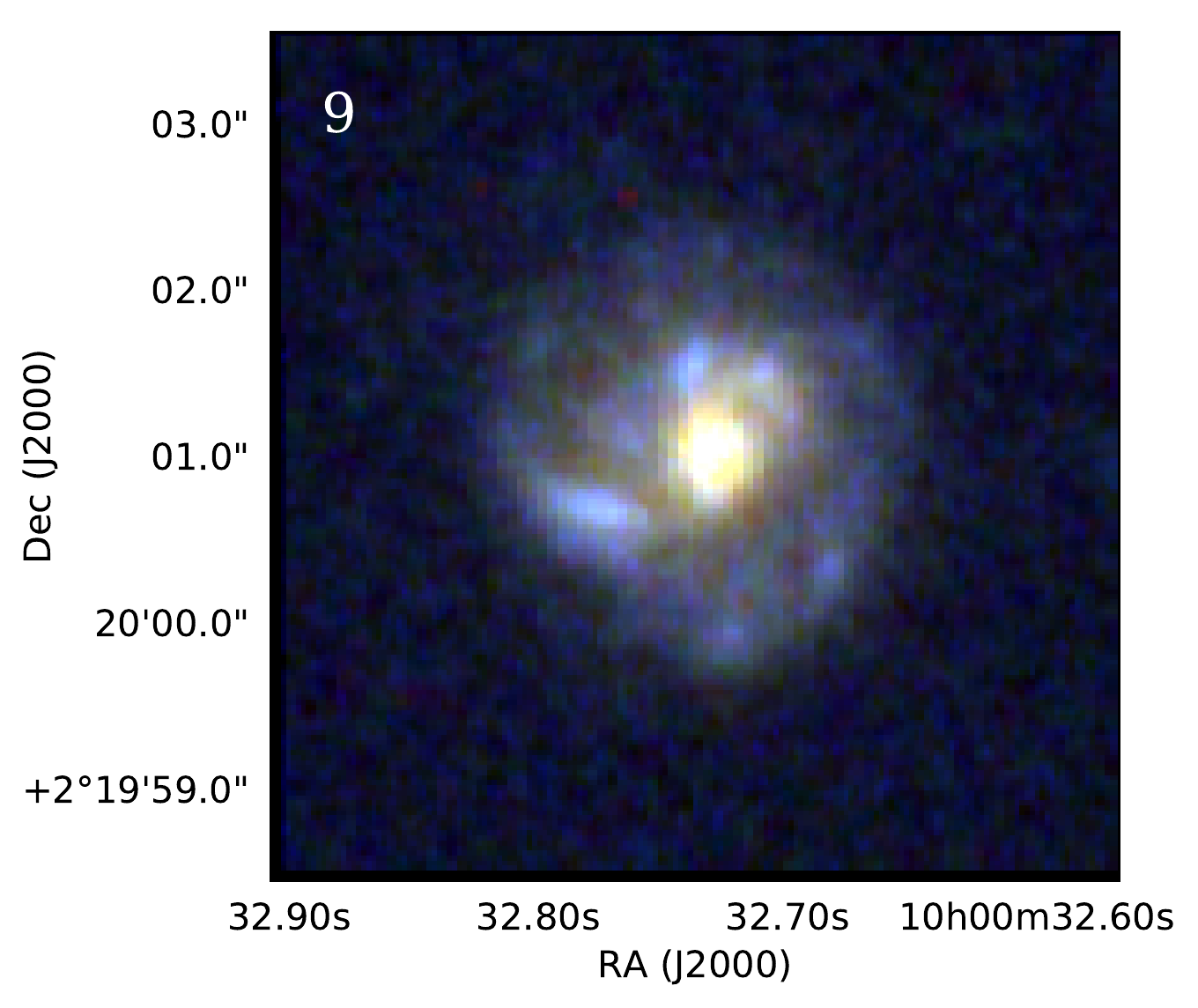} &
\includegraphics[scale=0.345,clip,trim=2mm 3mm 3.5mm 3mm]{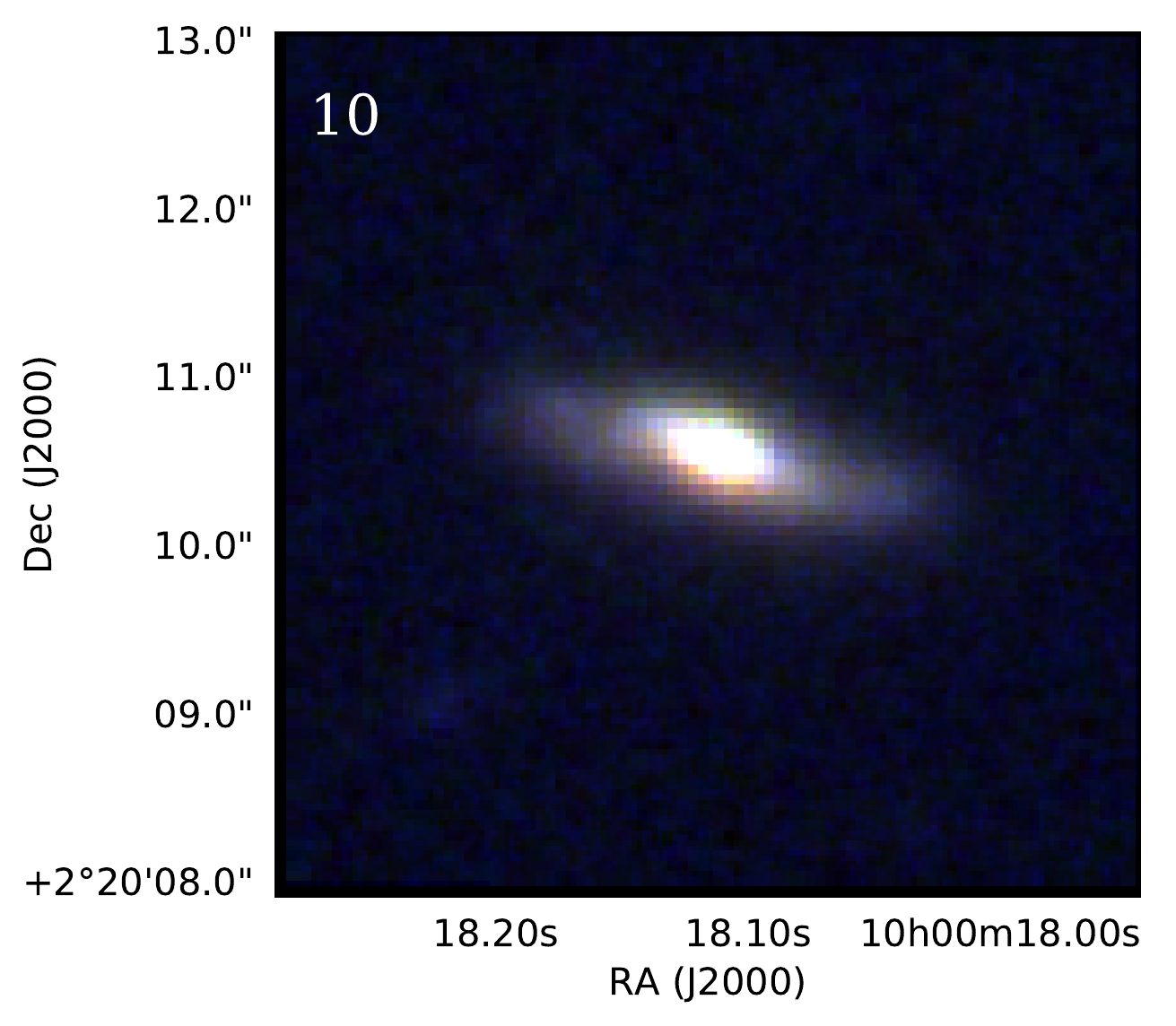} &\\
\end{tabular}
\endgroup
\caption{RGB images (5 arcsec on a side) of the target galaxies from CANDELS \emph{HST} imaging in F814W, F125W, F160W filters (approximately rest-frame $B$, $R$, $I$).}
\label{fig:rgb}
\end{center}
\end{figure*}

Optical \textit{HST} images of the galaxies in our sample are shown in Fig.~\ref{fig:rgb}. Most have disk-like morphology, and some of these have large clumps (IDs 4, 8, 9). The sample also includes one very compact object with high SSFR $>3\sigma$ above the main sequence (ID 2), and two that are potentially interacting with neighbouring galaxies at similar redshifts (IDs 1, 7), although neither of these has significantly elevated SSFR in relation to the main sequence. 
We discuss the nature of the individual galaxies in more detail in Section~\ref{sec:data}.
Before describing the ALMA observations in Section~\ref{sec:data}, we will provide details of the archival observations and derived physical properties of our sample in the following subsections.

\begin{table*}
\caption{Identifiers and physical properties of galaxies in the sample.
Stellar masses (\Mstar) and UV luminosities $L_\mathrm{UV}\equiv\nu L_{\nu}(\lambda=1600\AA)$ are from 3DHST (see Section~\ref{sec:seds}). 
The remaining columns list the parameters derived from our IR SED fitting (Section~\ref{sec:sedfits}): integrated \Lir\ (8--1000\,\micron); SFR$=$SFR$_\mathrm{UV}$+SFR$_\mathrm{IR}$($\approx$SFR$_\mathrm{IR}$); dust mass \Mdust; temperature $T_{d}$; optically-thick limit $\lambda_0$; and mid-IR power-law index $\alpha$. 
Values (and 1-$\sigma$ errors) for the SED parameters $T_d$, $\lambda_0$, $\alpha$ are directly estimated from the median (and 16th \&\ 84th percentiles) of the posterior distribution, and are marginalised over all parameters. 
The integrated \Lir\ (and 1-$\sigma$ errors) are estimated from the median (and 16th \&\ 84th percentiles) of all SEDs sampled by the posterior distribution of parameters. 
The dust masses are similarly estimated from the values of these SEDs at 850\,\micron, converted to dust mass using $\kappa_{850}=0.077$~m$^2$\,kg$^{-1}$.}
\renewcommand{\arraystretch}{1.3}
\begin{tabular}{rccccrrrrrrrr}
\hline
ID & CID$^{(1)}$ & R.A. & Dec. & z$^{(2)}$ & \Mstar &    \Luv & \Lir & SFR & $M_d$ & $T_d$ & $\lambda_0$ & $\alpha$ \\
 &  & J2000 & J2000 & & $10^{10}$\Msun  & $10^{10}\Lsun$ & $10^{10}\Lsun$ & \Msun\,yr$^{-1}$ & $10^7\Msun$ & K & \micron & \\
\hline
 1 & U.32979 & 02:17:17.50 & $-$5:10:03.39 & 1.032 & 13.5 &    0.91 & $ 86^{+ 85}_{- 44}$ & $131^{+127}_{- 66}$ & $ 66^{+ 48}_{- 29}$ & $27.2^{+8.9}_{-8.7}$ & $ 35^{+246}_{- 29}$ & $1.7^{+0.6}_{-0.5}$ \\
 2 & C.19048 & 10:00:13.58 & $+$2:22:25.47 & 1.078 & 2.4  &   0.42 & $ 73^{+  6}_{-  5}$ & $110^{+  8}_{-  8}$ & $232^{+ 22}_{- 27}$ & $18.1^{+1.3}_{-0.7}$ & $ 37^{+ 54}_{- 30}$ & $1.4^{+0.1}_{-0.0}$ \\
 3 & U.30192 & 02:17:22.30 & $-$5:10:38.53 & 1.092 & 17.4 &   0.03 & $111^{+ 35}_{- 32}$ & $166^{+ 53}_{- 47}$ & $ 34^{+  7}_{-  7}$ & $28.7^{+3.1}_{-2.9}$ & $ 13^{+ 32}_{-  9}$ & $2.0^{+0.2}_{-0.3}$ \\
 4 & C.16172 & 10:00:24.84 & $+$2:20:46.09 & 1.033 & 4.1  &  1.70 & $ 61^{+  6}_{-  6}$ & $ 95^{+  9}_{-  9}$ & $ 20^{+  3}_{-  3}$ & $27.8^{+1.7}_{-1.5}$ & $ 26^{+ 33}_{- 20}$ & $1.7^{+0.1}_{-0.1}$ \\
 5 & C.14475 & 10:00:37.87 & $+$2:19:45.37 & 0.907 & 8.3  &  0.31 & $ 28^{+  4}_{-  4}$ & $ 43^{+  6}_{-  5}$ & $ 13^{+  3}_{-  3}$ & $26.8^{+3.9}_{-2.0}$ & $ 35^{+ 86}_{- 30}$ & $1.6^{+0.1}_{-0.1}$ \\
 6 & C.15161 & 10:00:22.71 & $+$2:20:10.86 & 0.931 & 7.4  &  0.08 & $ 46^{+  5}_{-  5}$ & $ 69^{+  7}_{-  7}$ & $ 20^{+  4}_{-  3}$ & $26.7^{+2.3}_{-1.4}$ & $ 16^{+ 56}_{- 12}$ & $1.7^{+0.1}_{-0.1}$ \\
 7 & C.15066 & 10:00:36.44 & $+$2:20:04.60 & 1.220 & 21.4 &  0.28 & $ 52^{+ 10}_{- 10}$ & $ 78^{+ 15}_{- 15}$ & $ 15^{+  3}_{-  3}$ & $29.2^{+3.5}_{-2.2}$ & $ 28^{+ 65}_{- 24}$ & $1.9^{+0.1}_{-0.2}$ \\
 8 & C.14371 & 10:00:25.78 & $+$2:19:45.12 & 0.928 & 4.8  &   0.42 & $ 28^{+  3}_{-  3}$ & $ 42^{+  5}_{-  5}$ & $ 18^{+  4}_{-  4}$ & $25.9^{+4.3}_{-3.0}$ & $ 77^{+ 83}_{- 65}$ & $1.5^{+0.1}_{-0.1}$ \\
 9 & C.14796 & 10:00:32.74 & $+$2:20:01.03 & 1.125 & 3.8  &   1.75 & $ 28^{+  3}_{-  3}$ & $ 45^{+  5}_{-  5}$ & $ 16^{+  2}_{-  2}$ & $26.0^{+2.9}_{-1.6}$ & $ 32^{+ 73}_{- 27}$ & $1.8^{+0.1}_{-0.1}$ \\
10 &C.15236 & 10:00:18.11 & $+$2:20:10.53 & 0.841 & 6.2  &   0.08 & $7^{+  2}_{-  2}$ & $ 10^{+  3}_{-  3}$ & $ 42^{+ 31}_{- 25}$ & $20.0^{+10.2}_{-5.6}$ & $198^{+406}_{-179}$ & $1.4^{+0.2}_{-0.2}$ \\
\hline
\end{tabular}
Notes: (1) Catalogue ID in 3DHST \citep{Skelton2014}, prefixed with ``C'' for the COSMOS field and ``U'' for the UDS field;
(2) maximum-likelihood grism redshift $z_\textrm{max,grism}$
\label{tab:physparams}
\end{table*}

\subsection{Multi-wavelength data and derived properties}
\label{sec:seds}

The sample is based upon the 3DHST galaxy catalogue described by \citet{Skelton2014}. 
This means that all galaxies benefit from extensive photometry in the $u$--8-\micron\ bands from surveys including CFHTLS, Subaru, CANDELS, NMBS, WIRDS, UKIDSS-UDS, UltraVISTA, SEDS, S-COSMOS and SpUDS.
The grism redshifts are based on observations with the WFC3 G141 grism as described by \citet{Momcheva2015}. Redshifts were determined by fitting the 2D spectra and multi-band photometry simultaneously, with templates modified from \textsc{eazy} \citep{Brammer2008} and additional emission-line templates from \citet{Dobos2012}. 
We also use the stellar mass estimates provided in the 3DHST catalogue, as described by \citet{Skelton2014}. These were derived by SED fitting with \textsc{fast} \citep{Kriek2009}, assuming \citet{Bruzual2003} SPS models with a \citet{Chabrier2003} IMF; 
solar metallicity; exponentially-declining star-formation histories with a minimum e-folding time of $10^7$~yr; minimum age of 40~Myr; $0<A_V<4$ and a \citet{Calzetti2000} dust attenuation law. We have made no correction for the difference between \citet{Chabrier2003} and \citet{Kroupa2003} IMFs, since they are essentially identical for our purposes \citep{Chomiuk2011}.

We determine SFRs as in \citet{Bourne2017}, from the sum of the SFR derived from the raw UV luminosity (without extinction correction) and that derived from the total IR luminosity, representing unobscured and obscured star formation respectively.
We use the 3DHST measurements of $\LUV\equiv\nu L_{\nu}(\lambda=1600\AA)$ (i.e. FUV-band), which were estimated from the rest-frame SED fits from \textsc{eazy} \citep{Skelton2014}. 
The UV SFR is calibrated following \citet{Kennicutt2012}: 
\begin{equation}
\text{SFR}_\text{UV}/\sfr=1.70\times10^{-10}\,\Luv/\Lsun 
\end{equation}
\citep{Hao2011,Murphy2011}.
The IR SFR is calibrated from the integrated 8--1000-\micron\ luminosity (\Lir) obtained from SED fitting (described in Section~\ref{sec:sedfits}). Following \citet{Kennicutt2012} again, 
\begin{align}
\text{SFR}_\text{IR}/\sfr= 1.49\times10^{-10}\,\Lir/\Lsun
\end{align}
\citep{Murphy2011}.
The SFRs for these galaxies are almost completely dominated by the obscured $\text{SFR}_\text{IR}$, since they all have $\Lir/\Luv>10$.

\newpage

\begin{figure*}
\begin{center}
\raisebox{3cm}{1}
\includegraphics[width=0.98\textwidth]{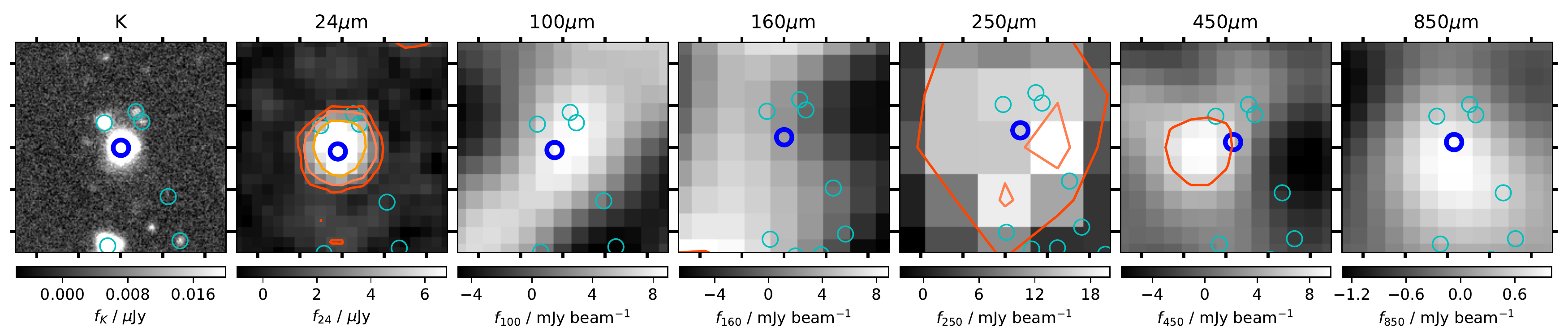}\\
\raisebox{3cm}{2}
\includegraphics[width=0.98\textwidth]{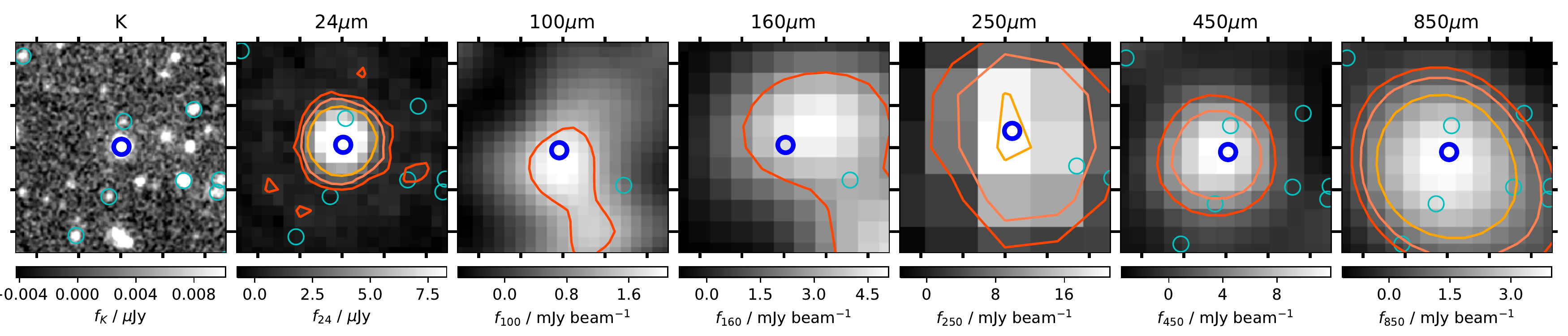}\\
\raisebox{3cm}{3}
\includegraphics[width=0.98\textwidth]{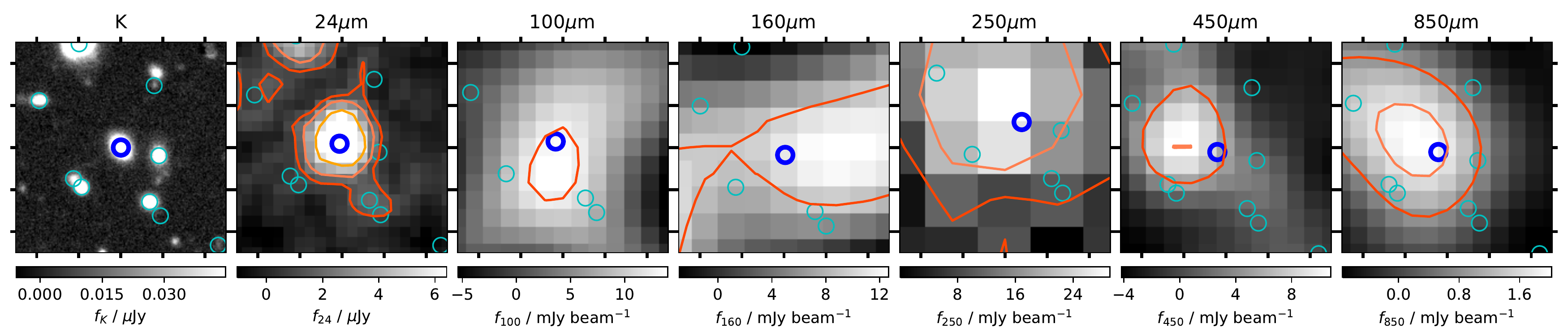}\\
\raisebox{3cm}{4}
\includegraphics[width=0.98\textwidth]{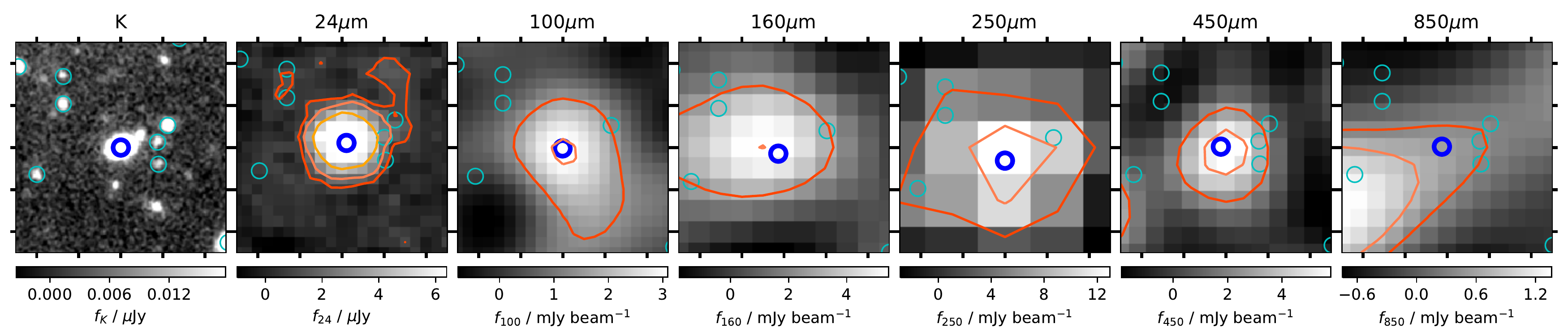}\\
\raisebox{3cm}{5}
\includegraphics[width=0.98\textwidth]{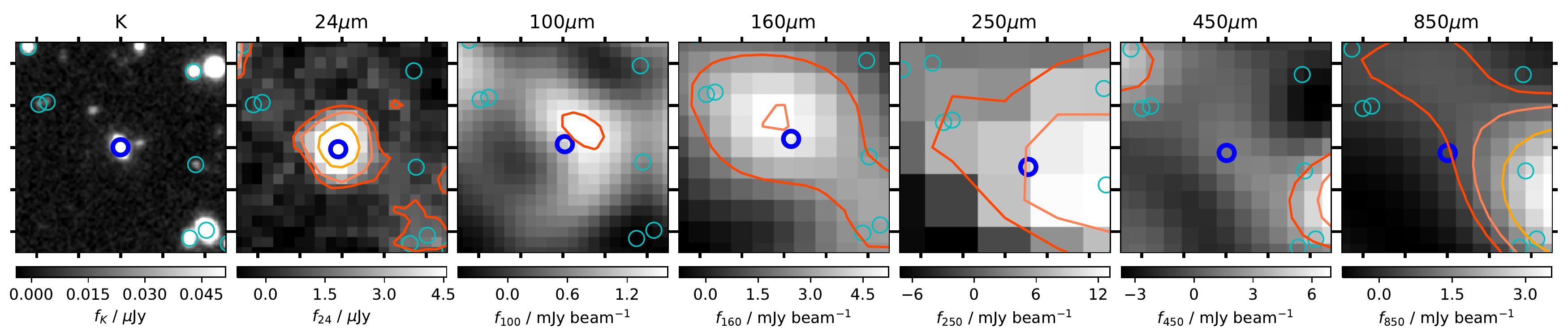}\\
\raisebox{3cm}{6}
\includegraphics[width=0.98\textwidth]{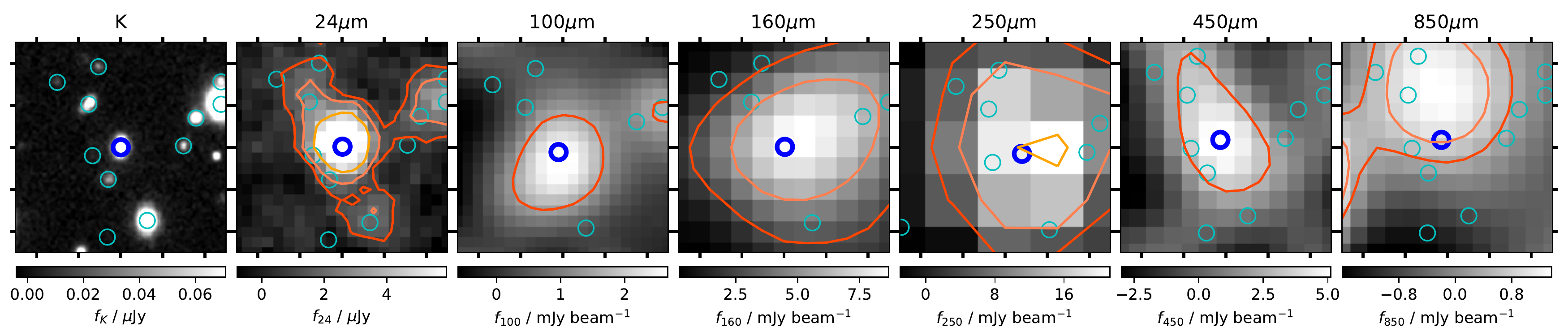}\\
\caption{Cutouts centred on the targets in various IR bands. Blue circles mark the targets, with cyan circles indicating other priors used in the \tphot\ fitting of each image. The 24--850-\micron\ images shown are match-filtered with contours drawn at 3, 6 \&\ 12-$\sigma$ levels. Each cutout is 25 arcsec on a side, with tick marks drawn at 5-arcsec intervals.}
\label{fig:fir}
\end{center}

\end{figure*}
\renewcommand{\thefigure}{\arabic{figure} (continued)}
\addtocounter{figure}{-1}
\begin{figure*}
\begin{center}
\raisebox{3cm}{7}
\includegraphics[width=0.98\textwidth]{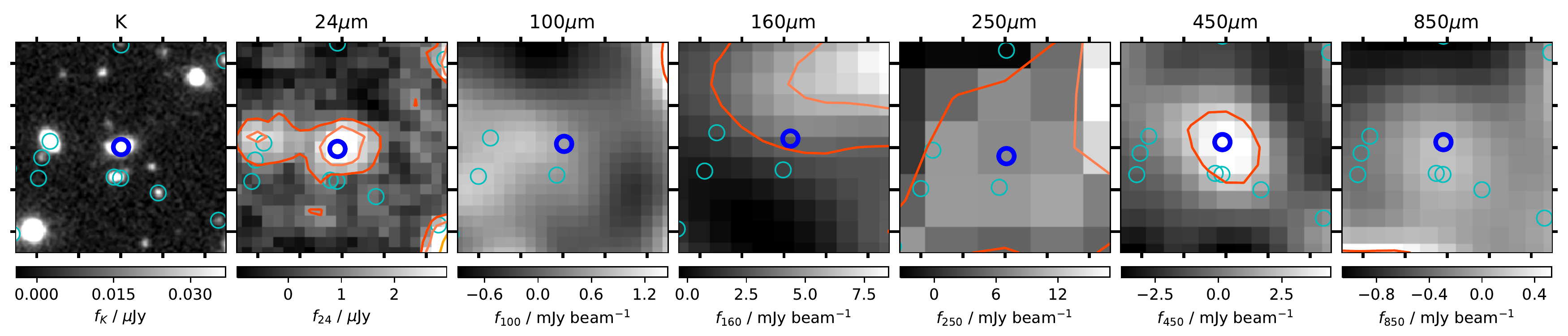}\\
\raisebox{3cm}{8}
\includegraphics[width=0.98\textwidth]{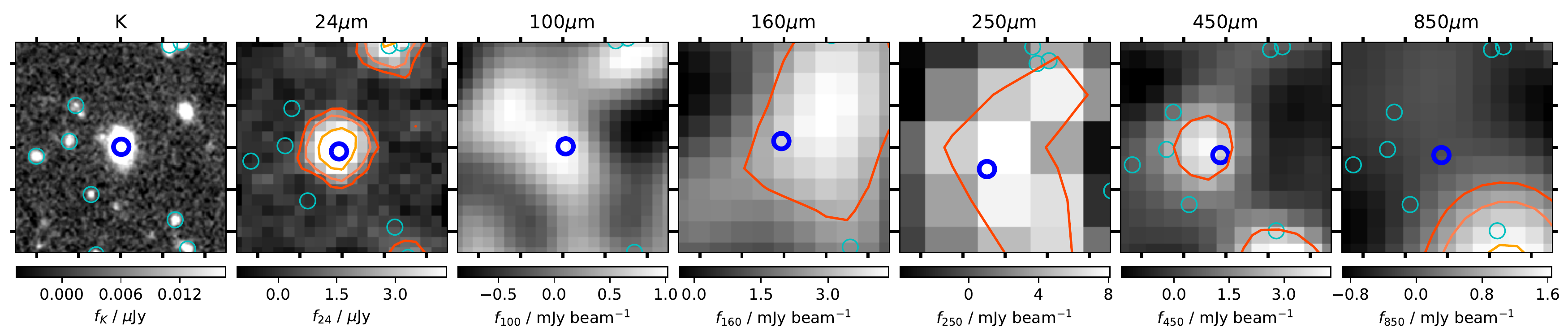}\\
\raisebox{3cm}{9}
\includegraphics[width=0.98\textwidth]{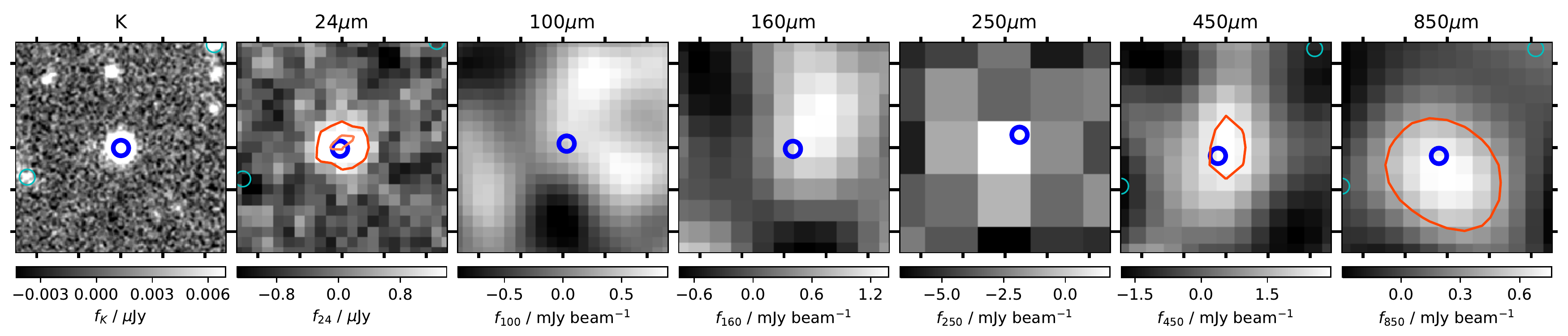}\\
\caption{}
\end{center}
\end{figure*}
\renewcommand{\thefigure}{\arabic{figure}}

\subsection{Far-infrared spectral energy distributions}
\label{sec:sedfits}

We measured the FIR SEDs using the following photometry: \textit{Spitzer}/MIPS 24\,\micron\ from S-COSMOS \citep{Sanders2007} or SpUDS (PI: Dunlop); \textit{Herschel}/PACS 100 \&\ 160\,\micron\ from PEP \citep{Lutz2011}; \textit{Herschel}/SPIRE 250\,\micron\ from HerMES \citep{Oliver2012} and JCMT/SCUBA-2 450 \&\ 850\,\micron\ from S2CLS.
The 24-\micron\ flux measurements were taken from the 3DHST catalogue and were measured using prior-based deblending with an \textit{HST}/WFC3 detection image as described by \citet{Whitaker2014}.
Due to the coarse angular resolution of \Herschel\ and SCUBA-2, the FIR/submm images consist of a highly confused distribution of point sources, and so we use the methods developed in \citet{Bourne2017} to measure deblended photometry with \tphot\ \citep{Merlin2015, Merlin2016}.

The methodology used here is very similar to that described in \citet{Bourne2017}, and we refer the reader to that publication for full details. 
The only modification we made was to refine the list of prior positions around each target in this sample, to ensure that the \tphot\ modelling provides unbiased and precise photometry in each waveband. 
Fig.~\ref{fig:fir} shows image cutouts in the $K$ band and the six FIR/submm bands, highlighting the target and nearby prior positions used as inputs to \tphot. 
We found that the optimal approach was to use the following source lists as positional priors: 
\begin{enumerate}
\item All galaxies with the photometry flag \textit{USE}, and with either UDS $K$, UltraVISTA $K_s$ or \textit{Spitzer}/IRAC [3.6] magnitude $<24$, and $\log(\Mstar/\Msun)>9$ in the 3DHST catalogue
\item For 100 \&\ 160\,\micron, we additionally required a prior-based measurement of $S_{24}$>10\,$\mu$Jy
\item For 250\,\micron, we additionally required either $S_{24}$>10\,$\mu$Jy or $\log(\Mstar/\Msun)>10$
\end{enumerate}
These criteria ensure that the prior lists are sufficient to describe all significant sources of emission in the FIR/submm images (so that all sources of blending are accounted for), without being so crowded that the deblended flux measurements have very large covariances and therefore low signal-to-noise. 
In particular, note that a 24-\micron\ flux of 10\,$\mu$Jy is only just above the $1\sigma$ level in the prior-based catalogue, so this is a very weak condition, but it allows us to exclude from the priors any sources which are certain not to be bright at 24\,\micron, and are therefore certain not to be bright in the (much shallower) PACS images. 
At 250\,\micron, we allow for the possibility that high-$z$ massive galaxies might be bright at 250\,\micron\ but not at 24\,\micron, since at $z\gtrsim3$ the 24-\micron\ band does not trace emission from dust. In the SCUBA-2 bands, the original prior selection (i) is sufficient to achieve deblended measurements with good signal-to-noise.

We made further adjustments to the prior lists for two sources, IDs 1 and 5, as follows. 
In the field of ID 5, we excluded the nearby source with catalogue ID (CID) 14710, which is 4~arcsec to the East and has $\log(\Mstar/\Msun)=9.9$, but is undetected at 24\,\micron\ and in our ALMA continuum image. This source does not appear to be significant in any of the FIR images, but if we include it in the priors then the \tphot\ estimates for 250 and 850-\micron\ fluxes for ID 5 are biased towards unreasonably high values (this is because 14710 happens to fall on a negative noise peak in these images, so is fitted with a negative flux, and ID 5 is biased towards a high flux to compensate for blending). Occasionally, in individual cases like these, the assumption made in \tphot\ of a flat prior on relative fluxes is inadequate to obtain optimal flux-sharing between blended sources.

In the field of ID 1, we removed two nearby objects from the prior lists. First, CID 32422 (which is 3~arcsec to the East, has $\log(\Mstar/\Msun)=9.7$ but is undetected at 24\,\micron\ and in our ALMA continuum image) was excluded for similar reasons to CID 14710 described above. Secondly, we excluded CID 32863 which is very close to ID 1 (1~arcsec separation), at a similar redshift, and indeed may be interacting with it.  These two galaxies are blended even in the $K$ band, but our ALMA continuum image suggests that there is not significant dust emission from 32863, and since it is impossible to deblend two such close priors, we exclude 32863 in order to measure the total flux of the system with just one prior at the position of the target ID 1. We discuss the dust distribution in this system further in Section~\ref{sec:s1}.

\begin{figure*}
\begin{center}
\includegraphics[width=0.99\textwidth,clip,trim=10mm 12mm 10mm 12mm]{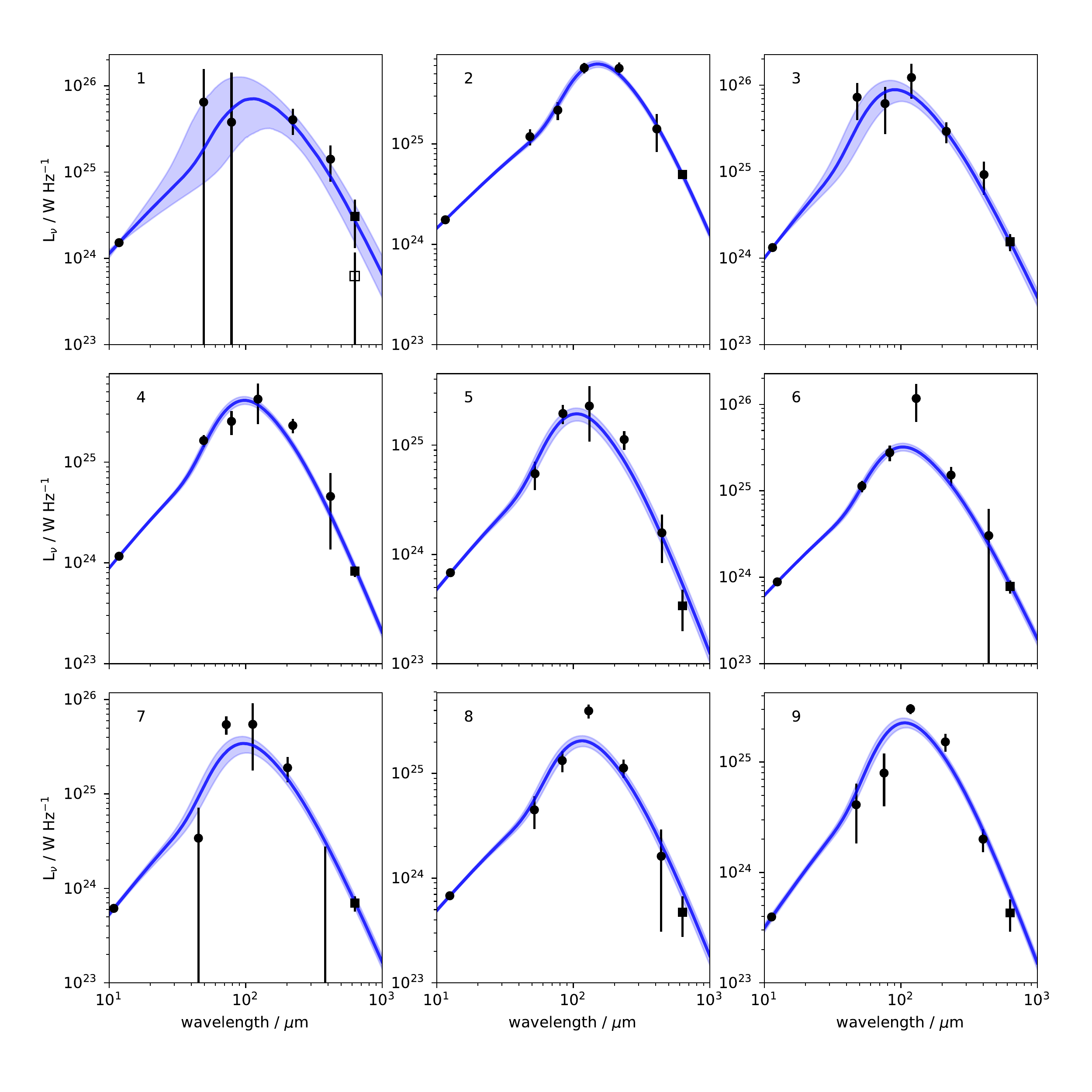}
\caption{Results of fitting the SEDs of galaxies with ALMA detections. Photometry at 100, 160, 250, 450 and 850\,\micron\ are measured using \tphot\ deblending as shown in Fig.~\ref{fig:fir} and described in Section~\ref{sec:sedfits}. We also include 24-\micron\ photometry from 3DHST \citep{Whitaker2014} and the ALMA continuum measurements described in Section~\ref{sec:extraction}. We used an MCMC method to fit models that combine an optically-thick modified blackbody (fixed emissivity coefficient $\beta=1.8$) with a mid-IR power-law, following \citet{Casey2012}, as described in Section~\ref{sec:sedfits}. Blue lines and shaded bands indicate the median and 16th--84th percentile range of the SEDs given by the posterior distribution of parameters from the MCMC samples.}
\label{fig:seds}
\end{center}
\end{figure*}

The SEDs of the ALMA-detected sources are shown in Fig.~\ref{fig:seds}. For brevity, we do not show the ALMA-undetected galaxy (ID 10) since it is excluded from the analysis in this paper, although we have performed a similar SED fit to the FIR/submm photometry to obtain its \Lir. For the remaining nine galaxies, we fit SED models to the photometry from 24 to 850\,\micron\ as well as our measured ALMA continuum fluxes, which are indicated by square symbols in Fig.~\ref{fig:seds}. 
These are at observed wavelengths between 1.2--1.4\,\micron, depending on redshift, and the data are described in Section~\ref{sec:data}).
For ID 1, two ALMA fluxes are plotted: the open square indicates the flux in a small aperture matched to the optical extent of the target galaxy, while the filled square includes extended flux as described in Section~\ref{sec:s1}. Note that the extended flux agrees with the SED extrapolated from the SCUBA-2 bands.

We modeled the SEDs with a modified blackbody function combined with a power law, as in \citet{Casey2012}. We fixed the emissivity coefficient to $\beta=1.8$ (the submm data are not sufficiently precise to constrain $\beta$ as a free parameter, although the data are more closely described by this than by $\beta=1.5$). An optically-thin blackbody gives a similarly good fit to an optically-thick blackbody, but we show the optically-thick fits here so that the uncertainties on temperature are marginalised over the uncertainties on the optically-thick wavelength limit, $\lambda_0$ (as well as the other parameters).  We used the \textsc{python} package \textsc{emcee} \citep{Foreman-Mackey2013} to conduct Monte Carlo Markov Chain (MCMC) sampling with 100 walkers over 500 steps, with 100-step burnin. The posterior median $\pm1\sigma$ estimates of the temperature ($T_d$), optically-thick limit ($\lambda_0$) and power-law index ($\alpha$) are given in Table~\ref{tab:physparams}.
Note that $\lambda_0$ was parametrised in log space (as was the normalisation, not shown) although we show the linearised parameter here for clarity. 
The total IR luminosity ($\Lir$) was calculated by integrating from 8--1000\,\micron\ the median SED from the MCMC samples, and the $\pm1\sigma$ uncertainty is given by the 16th and 84th percentile SEDs.  
As one would expect, the constraints on $\Lir$ are almost identical if we substitute the model for an optically-thin SED or allow $\beta$ to vary. Either of these modifications would result in a small systematic reduction in the median temperatures but do not affect our overall conclusions.
The dust mass and $\pm1\sigma$ uncertainty was calculated from the median, 16th and 84th percentiles of the 850-\micron\ flux density converted to dust mass using
\begin{equation}
M_{d} = \dfrac{S_{\nu,850}\,d_L^2}{\kappa_{850}B_{\nu,850}(T_d)}
\end{equation}
where we assumed the widely-adopted value of $\kappa_{850}=0.077$~m$^2$\,kg$^{-1}$ \citep{James2002}.

\section{ALMA data}
\label{sec:data}
\subsection{Observations and data reduction}
\label{sec:imaging}

The ALMA data for this project originate from Cycle 4 proposal number \mbox{2016.1.01184.S} {and Cycle 5 proposal number \mbox{2017.A.00013.S}}.
Observations were conducted between 2016 November 28 and December 14 {in Cycle 4, and 2017 December 24 in Cycle 5}. We used configurations C40-3 (15-460m baselines) and C40-4 (15-704m) in Cycle 4, and C43-6 (15-2400m) in Cycle 5. The correlator was set to time division mode (TDM), which provides the maximal bandwidth of 2~GHz per baseband appropriate for both line and continuum observations of extragalactic sources.
The spectral setup was designed with one spectral window centred on the \ci\ line at rest-frame 492\,GHz, and a second spectral window was centred on the \cofour\ line at 461\,GHz. 
The integration times (listed in Table \ref{tab:almaresults}) were calculated to reach a sensitivity level sufficient to detect the \ci\ line at SNR~$>5$ in 40\kms\ channels across an assumed line width of 250\kms. The \ci\ fluxes were predicted by assuming a \ci/\coone\ line ratio $R_{CI}=L^\prime_{CI}/L^\prime_{CO(1-0)}=0.2$, which is the typical value measured in local galaxies \citep{Gerin2000}, the Milky Way \citep{Ikeda2002} and high-redshift SMGs and quasars \citep{Walter2011b, Alaghband-Zadeh2013}. 
The \coone\ flux was predicted from the \Lir-$L_\mathrm{CO}$ relation from \citet{Sargent2014}, and we measured \Lir\ by SED fitting to the deblended SCUBA-2 and \Herschel\ flux measurements from \citet{Bourne2017}.

All data were reduced using the ALMA pipeline in \textsc{casa} version 4.7.2.
We first conducted bandpass, phase and flux calibration using the scripts that were packaged with the raw data from the ALMA Science Portal.
We then designed our own scripts to reduce the data and image the lines and continuum. 
We created separate data cubes for the \ci, \cofour\ and continuum imaging. For the continuum, we flagged any channels containing line emission (determined from the spectral line cubes) and then collapsed the full spectral range across both sidebands using the task \textsc{tclean}. During cleaning, we masked any emission associated with either the target or other galaxies in the field.
For each of the emission lines, we split off the appropriate spectral window and identified channels containing line emission, then used the task \textsc{uvcontsub} to fit and subtract the continuum using channels containing no line emission. We imaged the continuum-subtracted spectral lines using \textsc{tclean}, again masking any line emission for cleaning. We collapsed the channels containing line emission to form moment-0 maps, and inspected the spectrum within apertures of varying size to confirm the line widths (see Section~\ref{sec:extraction}). The process was iterative since it was necessary to revise the line masking, continuum subtraction and cleaning masks after inspecting the moment maps and aperture spectra.

Most of the data were imaged with natural weighting of the visibilities, which resulted in beam sizes of approximately $0.8\times0.6$ arcsec FWHM. 
We also tried imaging with a \emph{u-v} taper to down-weight the contributions from long baselines and increase sensitivity to extended flux. In cases where this led to higher flux measurements, we used the tapered images, which have beam sizes of $1.1\times1.1$ arcsec FWHM (IDs 1, 5, 8, 9).
Two of the sources were imaged with higher resolution in Cycle 5. The reason for this is that, in the original Cycle 4 data for these targets, the emission lines were found at the very edge of the spectral windows, due to errors in the grism redshifts that were assumed for the spectral setup. We obtained Director's Discretionary Time in 2017.A.00013.S to re-observe IDs 2 and 9 with a revised spectral setup to confirm the full line width. The natural-weighted images have beam sizes of $0.3\times0.2$ arcsec FWHM.  The data presented here for ID 2 are the Cycle 5 data imaged with natural weighting, but for ID 9, which is very extended with low surface brightness, we have combined both Cycle 4 and Cycle 5 data sets to maximise continuum signal-to-noise. Since the observations were conducted in different configurations, we imaged each set of visibilities for ID 9 separately, 
using a \emph{u-v} taper to image with a $1.1\times1.1$ arcsec beam, then combined the two continuum images with an inverse-variance weighted mean. 
To image the \ci\ and CO emission lines, we used only the Cycle 5 spectral windows, which are centred on the lines, and we again tapered the visibilities to produce an image with a $1.1\times1.1$ arcsec beam.

\subsection{Line and continuum measurements}

\label{sec:extraction}
\begin{figure*}
\begin{center}
\raisebox{4cm}{1}
\includegraphics[width=0.48\textwidth,clip,trim=8mm 0 3mm 0]{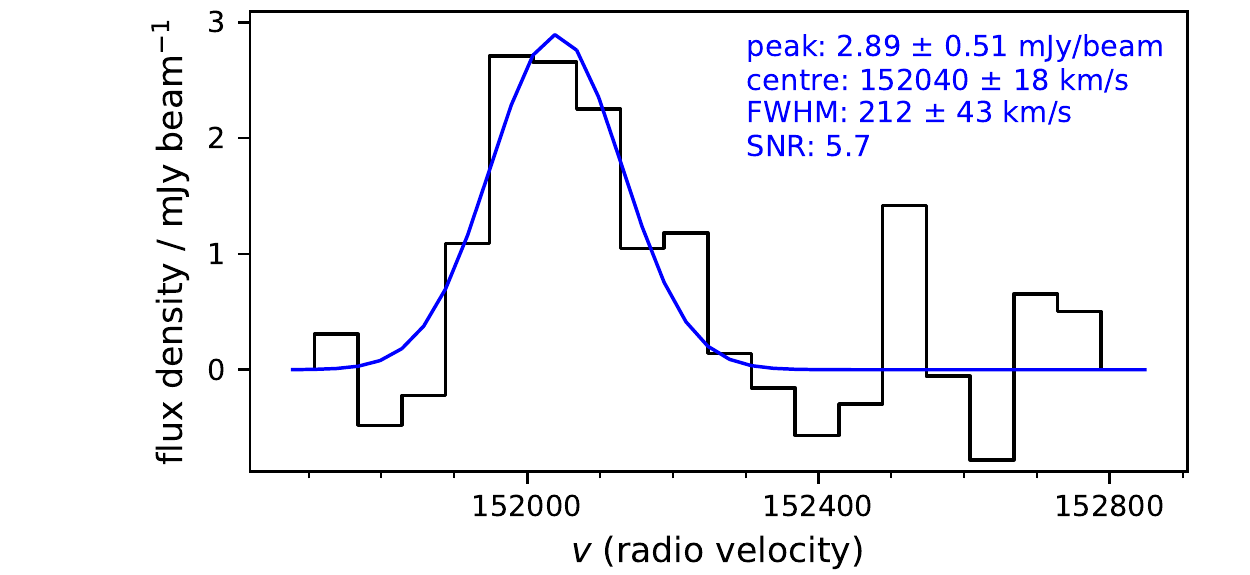}
\includegraphics[width=0.48\textwidth,clip,trim=8mm 0 3mm 0]{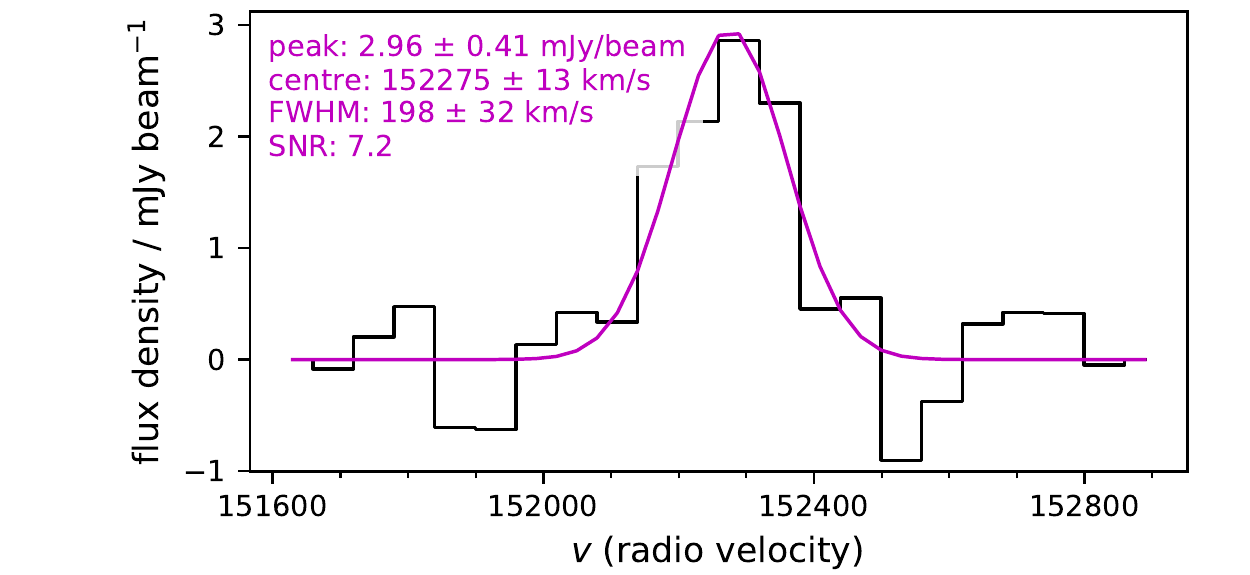}\\
\raisebox{4cm}{2}
\includegraphics[width=0.48\textwidth,clip,trim=8mm 0 3mm 0]{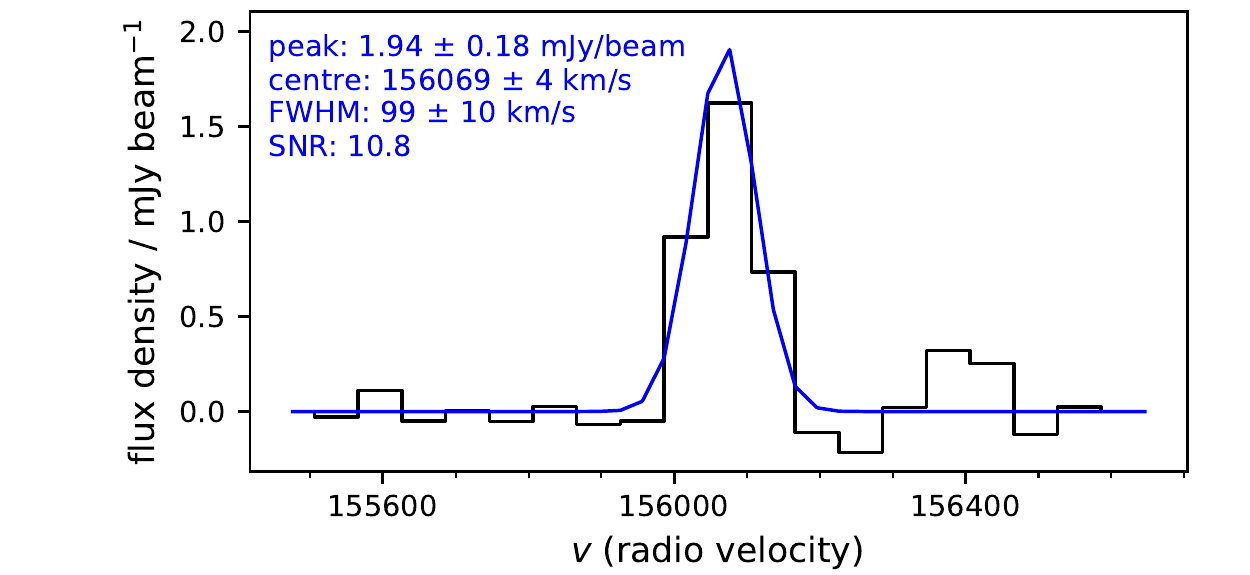}
\includegraphics[width=0.48\textwidth,clip,trim=8mm 0 3mm 0]{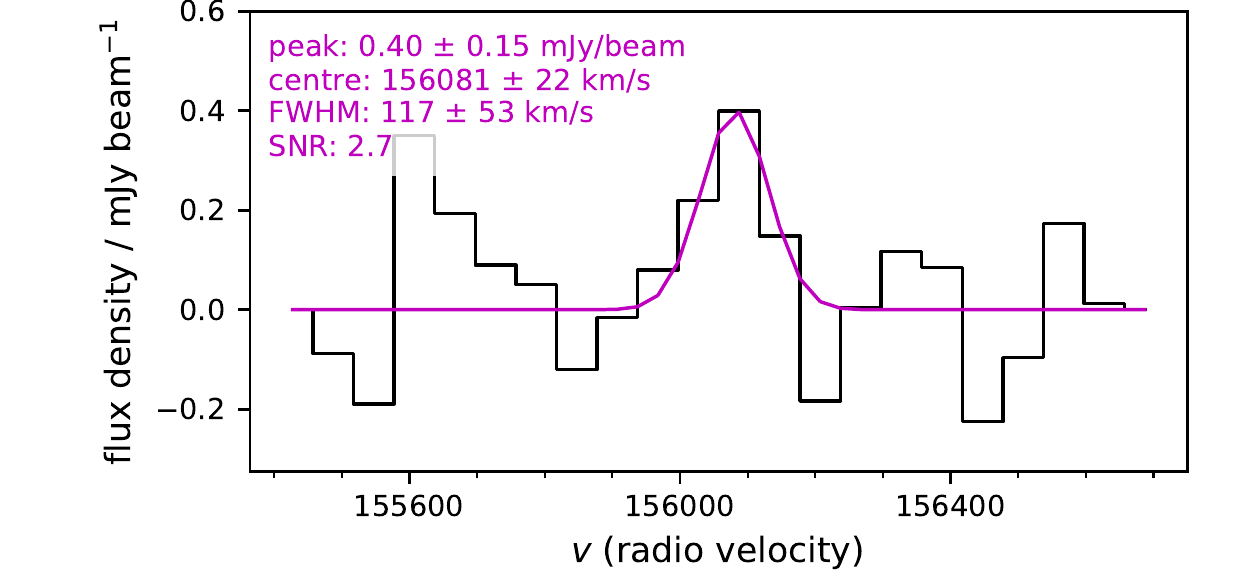}\\
\raisebox{4cm}{3}
\includegraphics[width=0.48\textwidth,clip,trim=8mm 0 3mm 0]{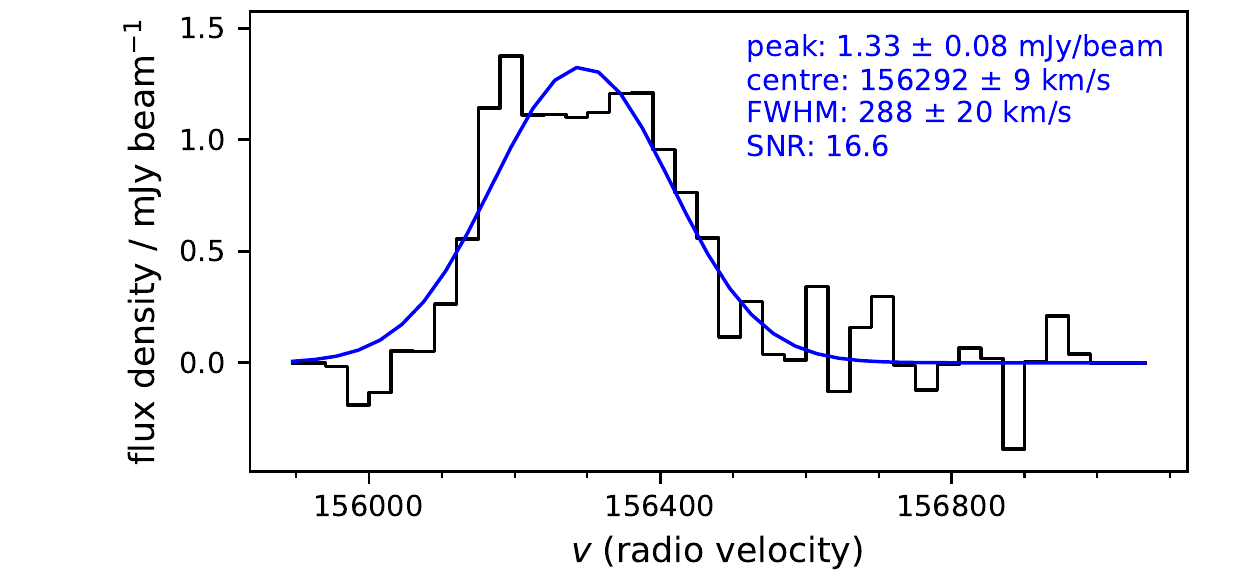}
\includegraphics[width=0.48\textwidth,clip,trim=8mm 0 3mm 0]{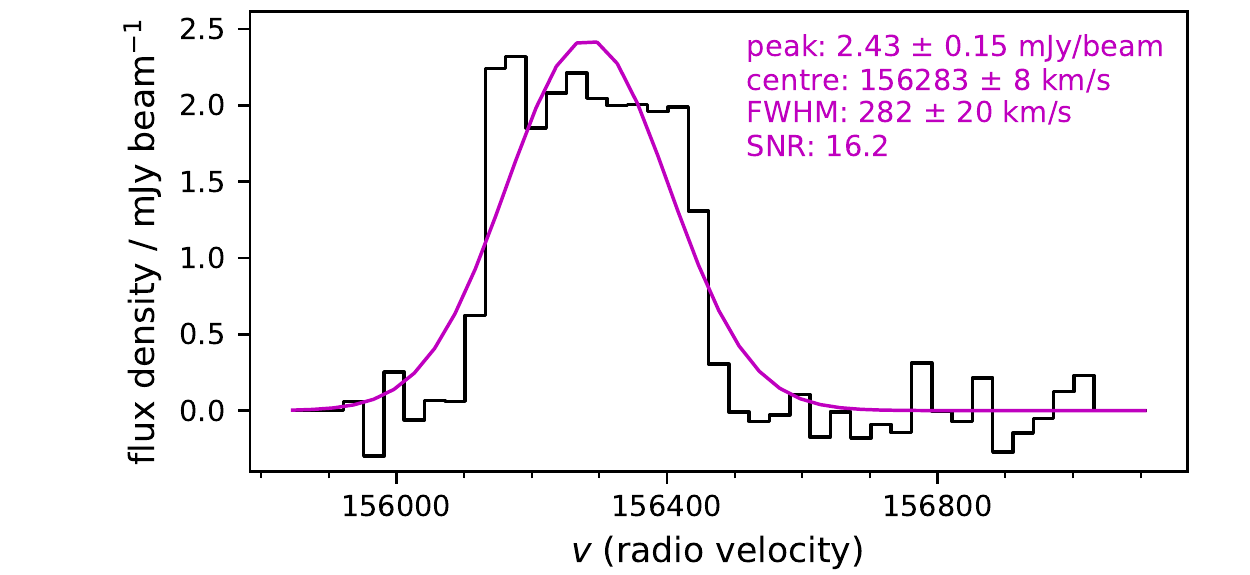}\\
\raisebox{4cm}{4}
\includegraphics[width=0.48\textwidth,clip,trim=8mm 0 3mm 0]{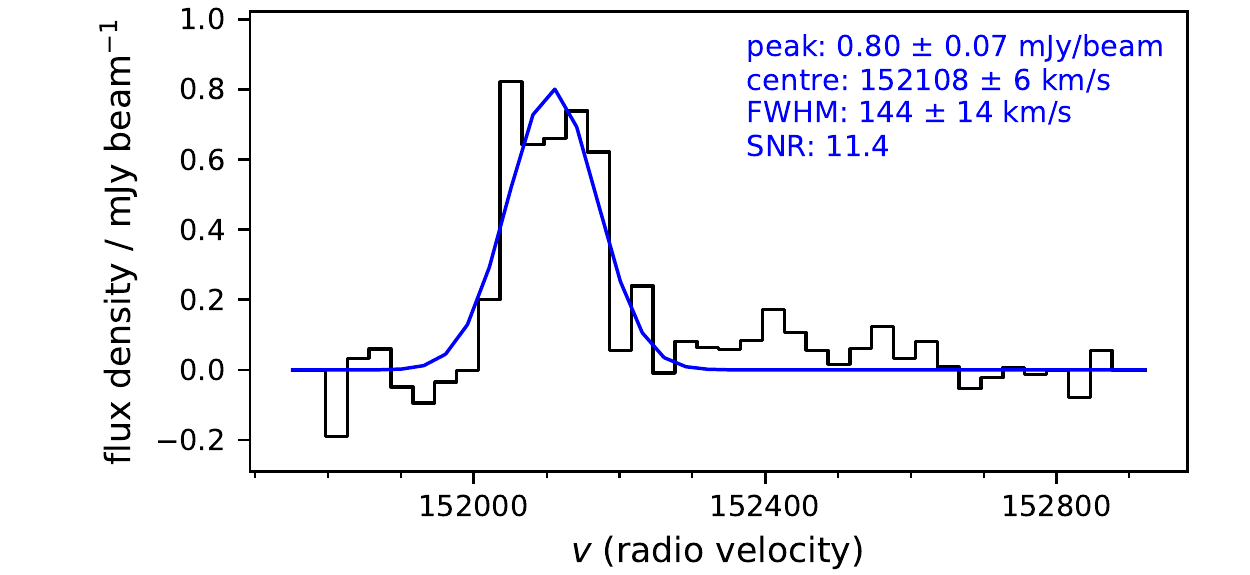}
\includegraphics[width=0.48\textwidth,clip,trim=8mm 0 3mm 0]{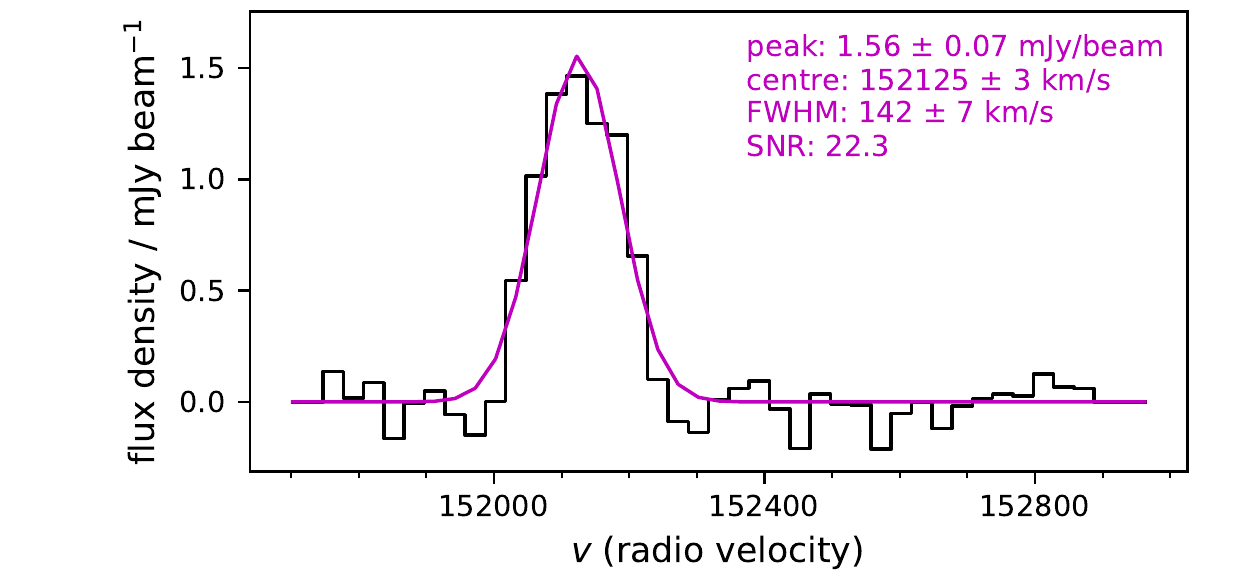}\\
\raisebox{4cm}{5}
\includegraphics[width=0.48\textwidth,clip,trim=8mm 0 3mm 0]{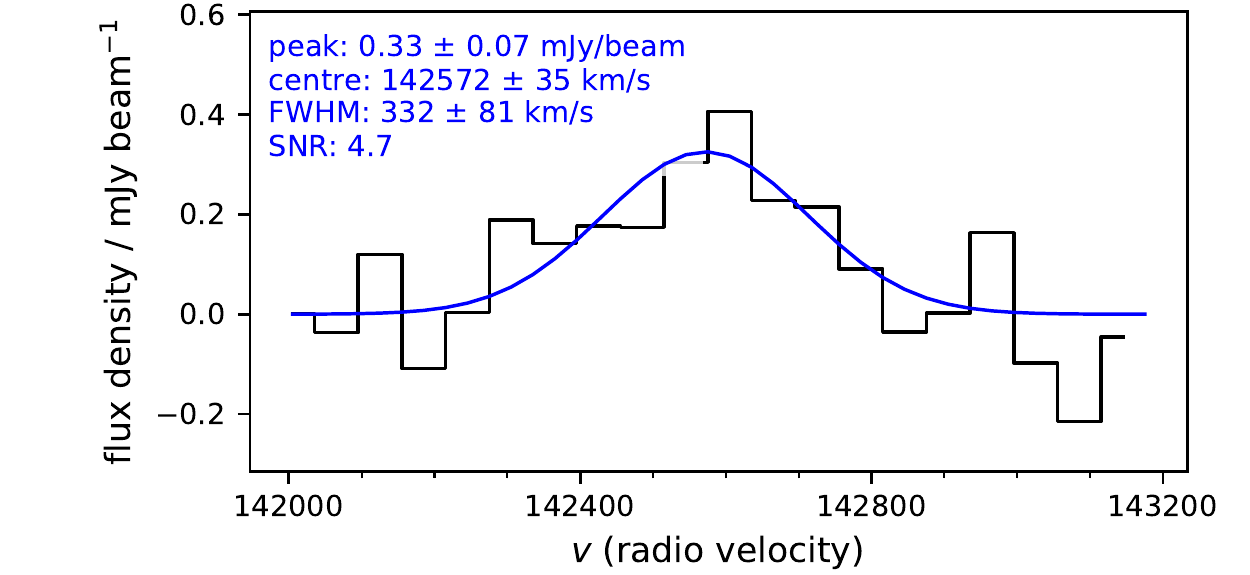}
\includegraphics[width=0.48\textwidth,clip,trim=8mm 0 3mm 0]{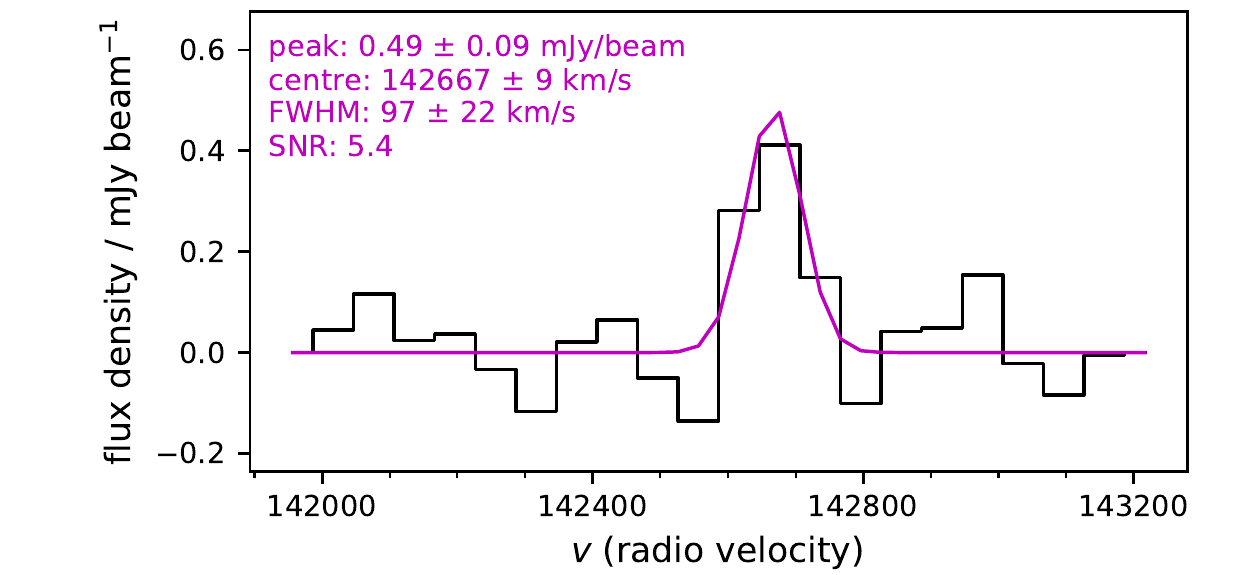}
\caption{Spectra in the peak beam of the \ci\ (left) and \cofour\ (right) spectral windows for each target. Spectra are binned to either 30 or 60 \kms\ resolution depending on SNR and line width. Also shown are the results of fitting Gaussian profiles by least squares.}
\label{fig:spectra}
\end{center}
\end{figure*}

Fig.~\ref{fig:spectra} shows the spectra in the central beam of each source. These were extracted from the peak in the respective moment-0 maps, at the positions indicated by black crosses in Fig.~\ref{fig:mom0}. The \ci\ line is detected with S/N~$>4$ in the spectra of eight of the ten targets, and at S/N~$=3$ in ID 9. \cofour\ is detected at S/N~$>4$ in seven of the nine where it is observable (not including ID 7, where CO falls outside of Band 6), and is additionally detected at S/N~$>2.5$ in ID 2. Source ID 10 was not detected in either line or continuum, and is therefore excluded from the analysis in this paper.

In the nine detected sources, the emission is usually extended on scales larger than the synthesised beam. Fig. \ref{fig:mom0} shows the full extent of the emission in moment-0 maps, which are integrated spectrally across the full line width as shown in the inset spectra. 
The line widths used for integrating the line emission were chosen by inspection of both the peak spectra in Fig. \ref{fig:spectra}, which generally have higher SNR, and the aperture spectra in Fig. \ref{fig:mom0}, which often show a broader line since the aperture encloses the full extent of the galaxy rotation curve. The apertures themselves were chosen to enclose the full extent of the continuum and line emission within the extent of the galaxy as seen in the optical images. Matched apertures were used to extract continuum, \ci\ and \cofour\ fluxes in spatially consistent regions (although see Section~\ref{sec:details} for a discussion of problematic sources). 

The line and continuum fluxes were determined from the integrated emission within these apertures. 
We checked whether a background subtraction was necessary by measuring the median pixel value (in Jy\,beam$^{-1}$) in an annulus enclosing radii 2 to 4 times the semimajor axis of the aperture ellipse, and scaling this by the number of beams in the source aperture to estimate the sky level in the aperture. This was done in the continuum and line moment-0 maps for each source. The absolute sky levels were generally less than 10 per cent of the aperture flux, and within the uncertainties of the aperture flux, so we decided that background subtraction was not necessary, although we note that doing so would make no significant difference to our overall results. 
To estimate uncertainties on aperture fluxes, we took the source aperture and randomised its position 150 times within the field of view, taking the standard deviation of the fluxes measured in the image (using the image without primary-beam correction).

Table~\ref{tab:almaresults} lists the continuuum image RMS noise level, the average RMS per channel in the data cubes, and the integrated aperture measurements for the continuum and spectral lines. We list the centre of the velocity range integrated for each line, rather than the velocity of the line peak, since many of the lines are double-peaked. 
These velocities are relative to the kinematic local standard of rest, and are in the radio convention (the default option in \textsc{casa}), 
\begin{equation}
V_\textrm{radio}=c(\nu_0 -\nu)/\nu_0
\end{equation}
where $\nu_0$ is the rest-frame frequency of the emission line.
To calculate integrated fluxes $S^\prime \Delta V$ (Jy\,\kms), we have integrated the spectrum with respect to the peculiar velocity $V_p$ in the rest frame of the source, 
\begin{equation}
V_p = \dfrac{c (z_\textrm{obs} - z)}{(1+z)}
\end{equation}
\citep[see][]{Peebles1993,Hogg1999}
where $z$ is the redshift of the line centre. This means that $\Delta V_p = \Delta V_\textrm{radio}(1+z)$.
The rest-frame velocity range of the integration limits is listed in the Table ($\Delta V_p$).
Sources which appear on more than one row are discussed in Section~\ref{sec:details}.

\begin{table*}
\caption{Full line and continuum properties from aperture-integrated spectra}
\begin{center}
\begin{tabular}{lcccccccccccr}
\hline
ID & $S_\mathrm{cont}$ & $S^\prime_\mathrm{[CI]}\Delta V$ & $\langle V_r\rangle_\mathrm{[CI]}$ & $\Delta V_\mathrm{[CI]}$ & $z_\mathrm{[CI]}$ & $S^\prime_\mathrm{CO}\Delta V$ & $\langle V_r\rangle_\mathrm{CO}$ & $\Delta V_\mathrm{CO}$ & $z_\mathrm{CO}$ & RMS$_\mathrm{cont}$ & RMS$_\mathrm{cube}$ & $t_\mathrm{int}$\\
 & $\mathrm{mJy}$ & Jy km s$^{-1}$ & km s$^{-1}$ & km s$^{-1}$ &  & Jy km s$^{-1}$ & km s$^{-1}$ & km s$^{-1}$ &  & mJy bm$^{-1}$ & mJy bm$^{-1}$ & m \\
 \hline
1$^S$ & 0.15$\pm$0.17 & 1.43$\pm$0.27 & 152068 & 609 & 1.029 & 0.89$\pm$0.19 & 152259 & 366 & 1.032 & 0.134 & 0.721 & 2 \\
1$^L$ & 1.18$\pm$0.22 & 2.58$\pm$0.29 & 152068 & 609 & 1.029 & 1.22$\pm$0.20 & 152259 & 366 & 1.032 & 0.134 & 0.721 & 2\\
2$^N$ & 1.54$\pm$0.22 & 0.47$\pm$0.30 & 156076 & 250 & 1.086 & 0.61$\pm$0.47 & 156057 & 250 & 1.086 & 0.024 & 0.208 & 16\\
2$^W$ & 1.54$\pm$0.22 & 1.62$\pm$0.61 & 156166 & 626 & 1.087 & 0.61$\pm$0.47 & 155997 & 500 & 1.085 & 0.025 & 0.207 & 16\\
3 & 0.47$\pm$0.11 & 1.06$\pm$0.22 & 156285 & 627 & 1.089 & 2.07$\pm$0.15 & 156266 & 752 & 1.089 & 0.032 & 0.169 & 21\\
4 & 0.29$\pm$0.04 & 0.67$\pm$0.10 & 152126 & 426 & 1.030 & 1.17$\pm$0.07 & 152137 & 426 & 1.030 & 0.016 & 0.099 & 49\\
5 & 0.15$\pm$0.05 & 0.21$\pm$0.09 & 142575 & 686 & 0.907 & 0.09$\pm$0.05 & 142676 & 229 & 0.908 & 0.028 & 0.144 & 52 \\
6 & 0.34$\pm$0.06 & 0.70$\pm$0.15 & 144378 & 694 & 0.929 & 1.59$\pm$0.16 & 144360 & 694 & 0.929 & 0.029 & 0.150 & 25\\
7 & 0.18$\pm$0.03 & 0.64$\pm$0.05 & 164399 & 731 & 1.214 &  &  &  &  & 0.013 & 0.077 & 80\\
8 & 0.19$\pm$0.09 & 0.54$\pm$0.19 & 143950 & 404 & 0.924 & 0.43$\pm$0.15 & 143951 & 404 & 0.924 & 0.032 & 0.206 & 40\\
9 & 0.11$\pm$0.05 & 0.47$\pm$0.15 & 158171 & 508 & 1.117 & 0.49$\pm$0.19 & 158137 & 190 & 1.116 & 0.014 & 0.295 & 175 \\
10 & $<0.10$ & $<0.12$ & -- & 220 & 0.841 & $<0.11$ & -- & 220 & 0.841 & 0.030 & 0.156 & 35\\
 \hline
\end{tabular}
\end{center}
Notes:
$\langle V_r\rangle$ refers to the central velocity (radio convention) in the local standard of rest, and $\Delta V$ ($=\Delta V_p$) refers to the full velocity width of the channels over which the line is integrated, in the rest frame [$=\Delta V_\textrm{radio}(1+z)$]. $S^\prime\Delta V$ is the integrated flux over these channels in the rest frame.
RMS$_\mathrm{cont}$ is the RMS of the continuum image and RMS$_\mathrm{cube}$ is the average RMS per channel in the \ci\ and CO data cubes. 
For ID~1, results from the small and large aperture are indicated by rows 1$^S$ and 1$^L$ respectively. For ID~2, results from the narrow and wide velocity range are indicated by rows 2$^N$ and 2$^W$ respectively (see Section~\ref{sec:details}).
Source ID~10 was not detected so 3-$\sigma$ upper limits were determined in a $2\times1$~arcsec ellipse centred on the optical position, assuming a 220\,\kms\ line width centred on the grism redshift. 
\label{tab:almaresults}
\end{table*}

\renewcommand{\thefigure}{\arabic{figure} (continued)}
\addtocounter{figure}{-1}
\begin{figure*}
\begin{center}
\raisebox{4cm}{6}
\includegraphics[width=0.48\textwidth,clip,trim=8mm 0 3mm 0]{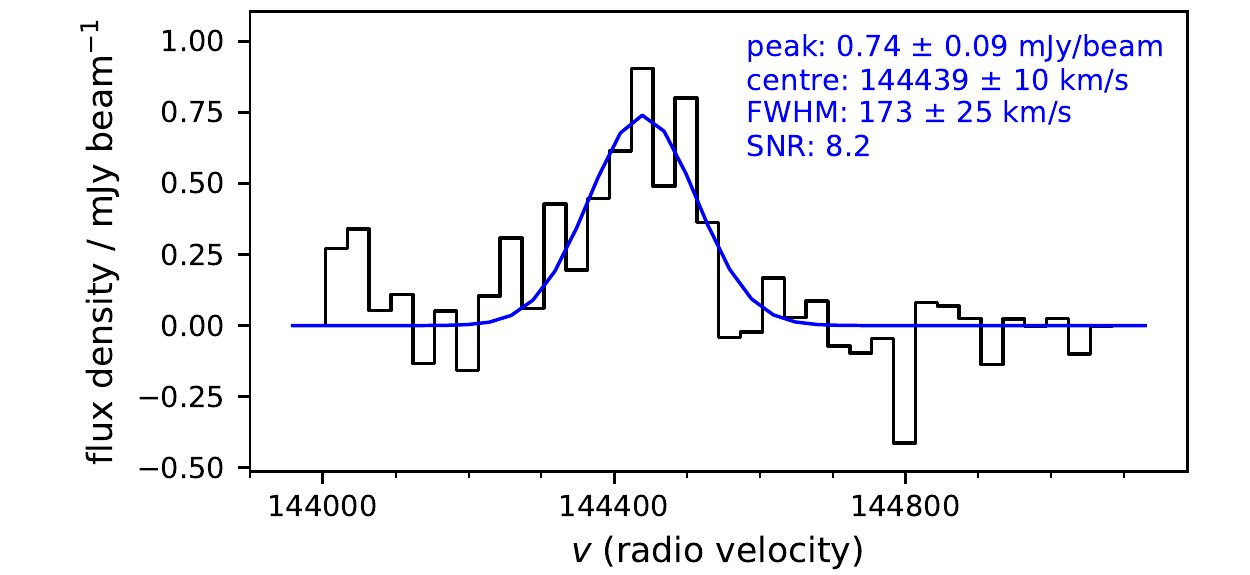}
\includegraphics[width=0.48\textwidth,clip,trim=8mm 0 3mm 0]{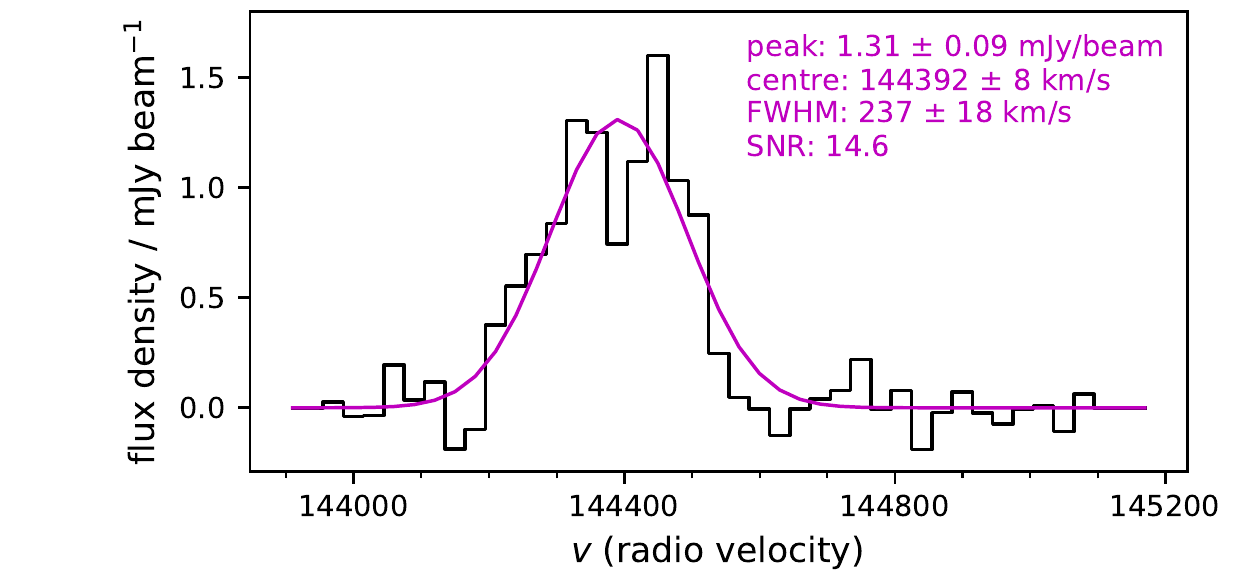}\\
\raisebox{4cm}{7}
\includegraphics[width=0.48\textwidth,clip,trim=8mm 0 3mm 0]{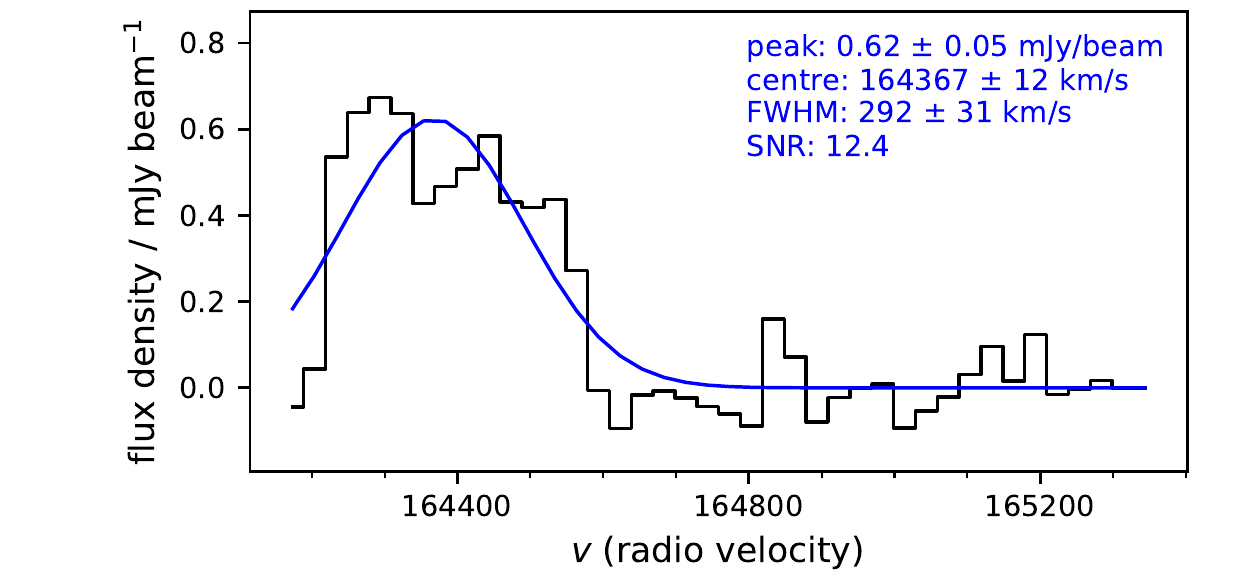} \hspace{0.4715\textwidth}\hphantom{0}\\
\raisebox{4cm}{8}
\includegraphics[width=0.48\textwidth,clip,trim=8mm 0 3mm 0]{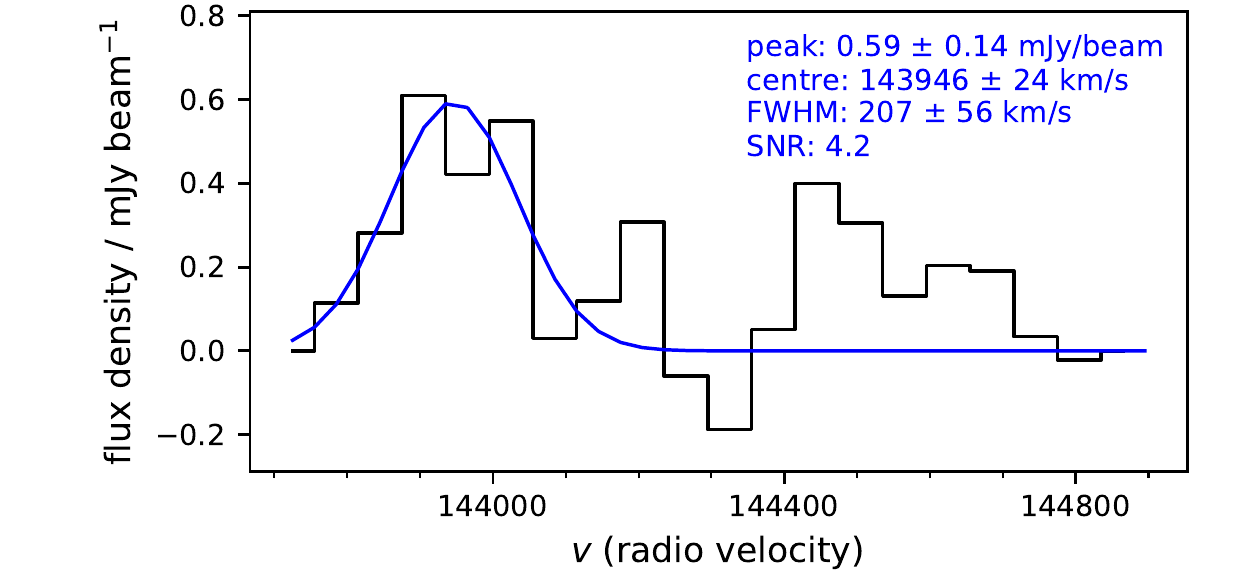}
\includegraphics[width=0.48\textwidth,clip,trim=8mm 0 3mm 0]{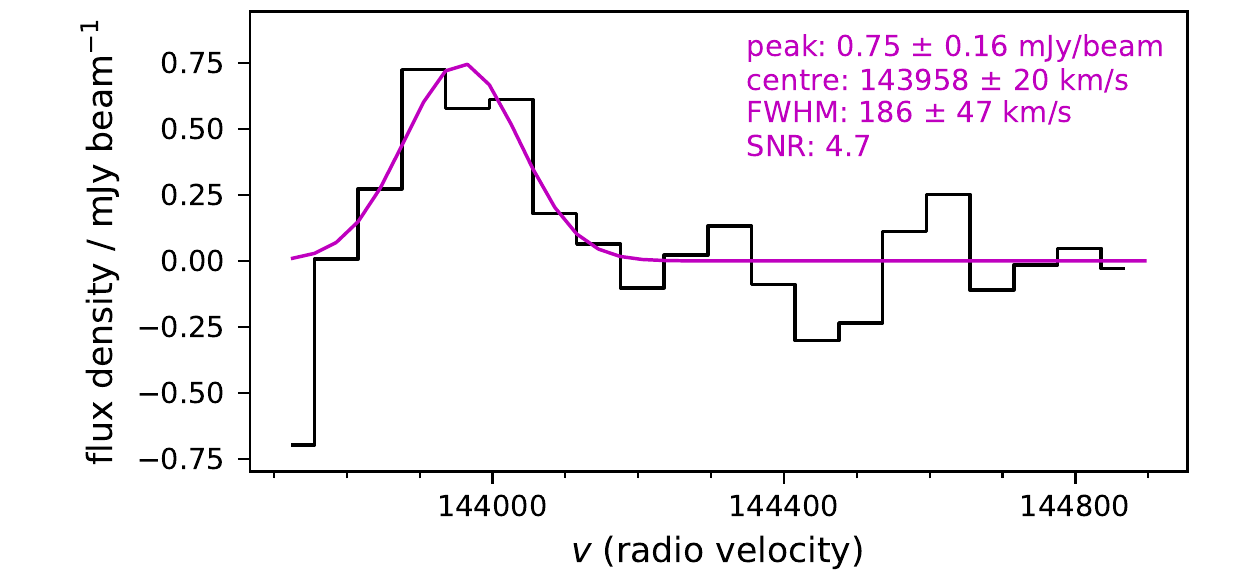}\\
\raisebox{4cm}{9}
\includegraphics[width=0.48\textwidth,clip,trim=8mm 0 3mm 0]{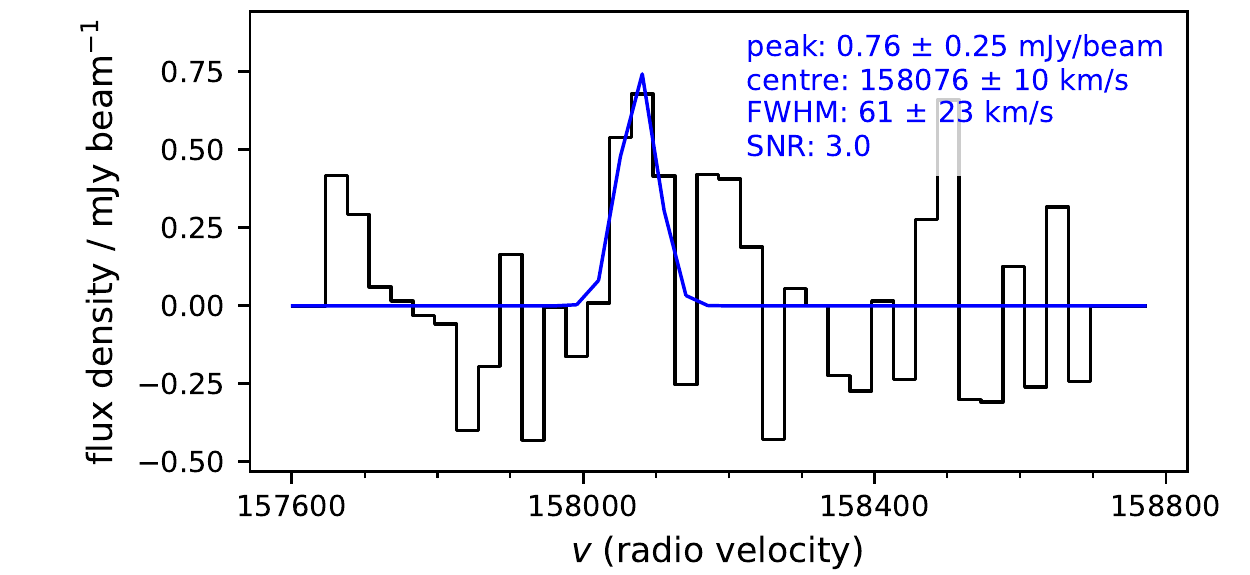}
\includegraphics[width=0.48\textwidth,clip,trim=8mm 0 3mm 0]{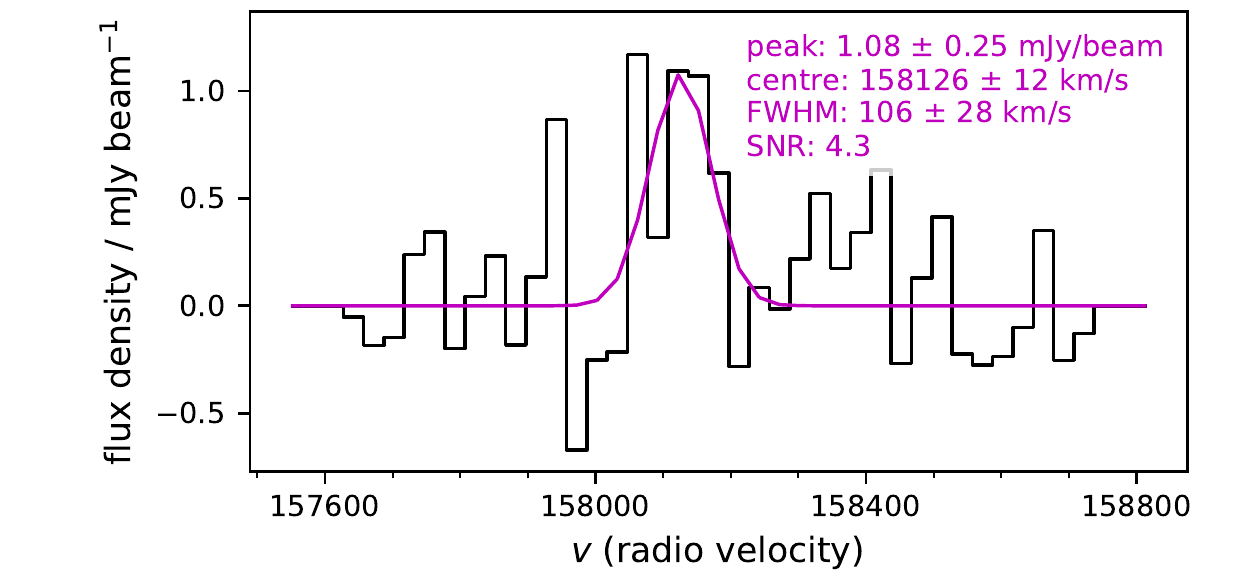}\\
\raisebox{4cm}{10}
\includegraphics[width=0.48\textwidth,clip,trim=8mm 0 3mm 0]{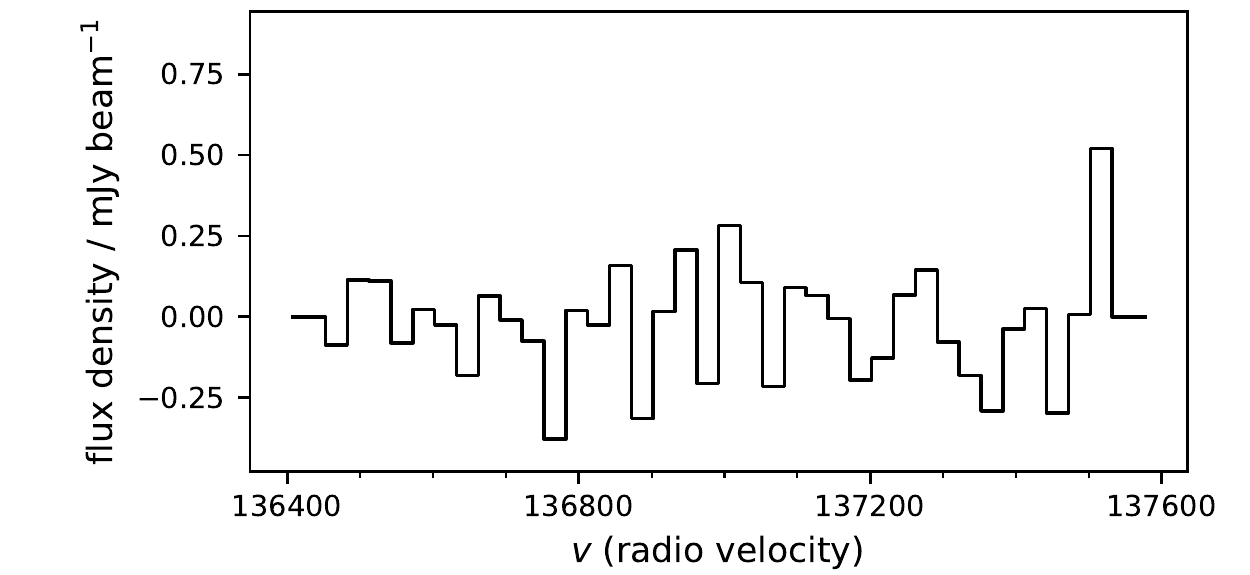}
\includegraphics[width=0.48\textwidth,clip,trim=8mm 0 3mm 0]{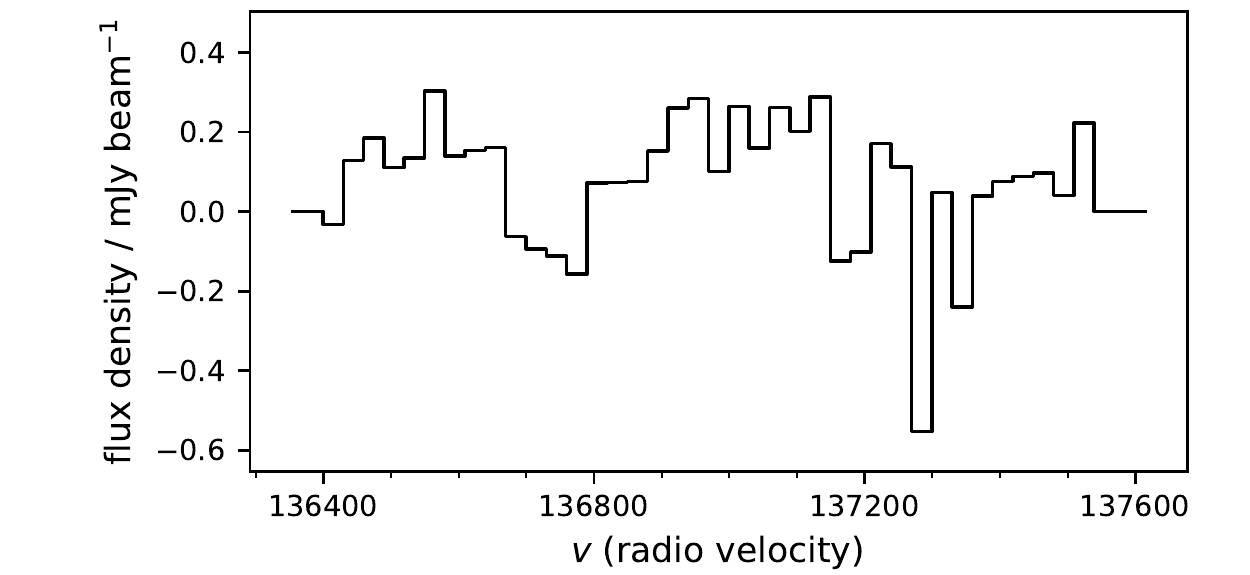}
\caption{Note CO could not be observed in ID 7 due to its redshift.}
\end{center}
\end{figure*}
\renewcommand{\thefigure}{\arabic{figure}}

\subsection{Notes on individual sources}
\label{sec:details}
In this section we discuss the ALMA data for those targets whose flux measurements presented some complications. The reader may wish to skip these details and continue to the results in Section~\ref{sec:discussion}.

\subsubsection{ID 1}
\label{sec:s1}
The first row of images in Fig.~\ref{fig:mom0} contains apertures of varying sizes indicating the varying extent of continuum, \ci\ and CO emission from source ID 1. There is strong \ci\ emission in a broad line extended over an area larger than the optical extent of the galaxy, while the CO and especially the continuum emitting regions are more compact. The central continuum peak is in fact only at the 1-$\sigma$ level, but much stronger continuum emission is apparent in an extended structure to the NW. 
This is unlikely to be a noise peak, since it is extended and has a peak S/N~$=3.2$ and an integrated S/N~$>5.0$. 
Further evidence that it is not spurious is provided by the agreement with SCUBA-2 fluxes as described in Section~\ref{sec:sedfits}.
However, it is unlikely to be a background source, since no object is detected at this location in any of the HST ACS or WFC3/IR images from CANDELS, or in the UDS $K$-band. 
Another possibility is that this emission traces material stripped from the interstellar medium (ISM) of the target galaxy, possibly as a result of a tidal interaction with the neighbouring galaxy/group. The three bright neighbours visible in the HST image (Fig.~\ref{fig:mom0}) are at similar redshifts to ID 1 ($\Delta z_\text{grism}<0.20$). The closest one in projected separation (CID 32863) is 1.2 arcsec away (9.6 kpc), although its redshift is 0.88 compared with 1.032 for ID 1, implying a velocity difference $\Delta V_p=24000$\,\kms.  However, the neighbour to the NE (CID 33216) is at a projected separation of 3.6 arcsec (29 kpc) at $z_\text{grism}=1.04$ ($\Delta V_p=1200$\,\kms) so is probably a satellite of ID 1 (its stellar mass is a factor 8 smaller). If either of these has been involved in an interaction with ID 1 it is possible that some ISM has been stripped in a tidal tail, which would not be visible in the stellar emission seen in the optical.  
Further evidence that this system is not completely relaxed can be seen in the fact that the \ci\ line centre is offset from the CO and from the optical grism redshift by $\Delta V_p \approx 390\kms$ in the rest frame ($\Delta z=0.0026$). In addition, the extended nature of the \ci\ emission in comparison with the optical size of the galaxy may indicate a gaseous outflow.
Multi-phase gaseous outflows can have complex structures, and may not appear cospatial when viewed through different tracers \citep{Pellegrini2013, Morganti2016, Cicone2018}. 

To explore the implications of these possibilities, we have extracted fluxes for ID 1 in two alternative sets of apertures.  The first, referred to in Table~\ref{tab:almaresults} by the label 1$^S$, is a small aperture matched to the continuum emitting region within the optical extent of the target galaxy, and is shown by the small white ellipse over the continuum map in Fig.~\ref{fig:mom0} (the same aperture was used to extract the \ci\ and CO fluxes shown in this row of the table). The second set of apertures, indicated by the label 1$^L$, is shown as the larger set of white ellipses in Fig.~\ref{fig:mom0}, which have been defined separately in the continuum and line emission images. These are based on the assumption that the galaxy has an outflow or tidal tail in which the various ISM phases traced by the dust, \ci\ and \cofour\ are not cospatial. A larger aperture has been placed over the extended continuum peak to the NW, and this flux is added to that in the smaller aperture to account for dust in the galaxy as well as in the outflow/tail. Meanwhile, the \ci\ and CO apertures have each been enlarged to cover the full extent of the significant emission in the respective moment-0 maps.  Note that it is possible that further line emission exists outside of these apertures, although deeper imaging would be required to confirm this.
The results from the two sets of apertures will be discussed in Section~\ref{sec:correlation}.

\subsubsection{ID 2}
\label{sec:s2}

Due to the compactness of this source, the area of the images in Fig.~\ref{fig:mom0} is $6\times6$~arcsec$^2$ for ID 2, compared with $12\times12$~arcsec$^2$ for the other sources. 
The bulk of the \ci\ emission and all of the continuum and CO emission are located in a compact core. A 2-dimensional Gaussian fit to the continuum image has a beam-deconvolved FWHM of $0.37\times0.22$~arcsec, equivalent to $3.0\times1.8$~kpc at $z=1.09$.  
However, we have allowed a larger aperture for this source, since there is weak \ci\ emission extending to the south (comprising around 20 per cent of the total measured flux), consistent with a tail seen in the optical image.

Furthermore, the \ci\ velocity structure is strongly double-peaked, which may be due to two distinct sources separated by $\Delta V_p \approx 400\,\kms$, or possibly a rotating disk-like structure which is either optically thick or \ci-deficient in the centre.
Conversely, the CO emission is very weakly detected, with a central velocity $\approx 156100\,\kms$ close to that of one of the \ci\ peaks, although it is unclear whether there is also weaker CO emission at $\approx 156300\,\kms$ (close to the second \ci\ peak).
The CO deficiency may be related to optical thickness or low gas excitation, and this will be explored in Section~\ref{sec:co_ci}.

The results of integrating the full velocity range of \ci\ (encompassing both peaks) are indicated by ID 2$^W$ in Table~\ref{tab:almaresults}; this is also the line-width that has been assumed in producing the moment-0 map in Fig.~\ref{fig:mom0}.
For comparison, we also list the results of integrating a narrower velocity range encompassing only the peak at $\approx 156100\,\kms$, indicated by ID 2$^N$ in Table~\ref{tab:almaresults}.
We will compare the results from both alternatives in Section~\ref{sec:correlation}.

Note that the optical grism redshift of this galaxy is $z_\textrm{max,grism}=1.078$, compared with the two \ci\ peaks at $z=1.086$ and $1.089$. However, the 68 per cent confidence upper limit is 1.096, which is consistent with our measurement.
Given the weakness of CO, we considered the possibility that the detected line which we identified as \ci\ is in fact \cofour\ at $z=0.954$, or indeed another line at a different redshift, but this is highly unlikely given that both the 3DHST grism redshift and our own photometric redshift fitting using Le Phare \citep{Ilbert2006} indicate $z\approx1.1$ with no secondary solutions.

\subsubsection{ID 5}
Source ID 5 is only weakly detected in \ci\ but we note that there appears to be a mismatch between the \ci, \cofour\ and continuum emitting regions, similar to what was seen in ID 1. This may suggest some kinematic disruption, perhaps related to the proximity of a neighbouring galaxy at a similar redshift (CID 14680), but without deeper imaging we cannot be certain of this. We have therefore simply chosen an aperture similar in size and shape to the optical emission, which encloses the continuum and \ci\ emission.

\subsubsection{ID 7}
We note that this galaxy is very close to a neighbouring galaxy at a similar redshift (CID 15107, most easily seen in Fig.~\ref{fig:rgb} to the left of the central target). This galaxy is much bluer and a factor $\sim300$ less massive than our target according to the 3DHST catalogue, and its disturbed morphology indicates that the pair may be interacting. The aperture we have defined encloses only the target, but the neighbour has no detectable \ci\ flux (either due to its redshift or its low mass) while its continuum flux is roughly 25 per cent that of ID 7.
We also note that ID 7 has no CO data because it is at a redshift where CO is unobservable with ALMA.

\subsubsection{ID 9}
\begin{figure*}
\begin{center}
\raisebox{5.3cm}{1}
\includegraphics[width=0.95\textwidth,clip,trim=5mm 3mm 5mm 3mm ]{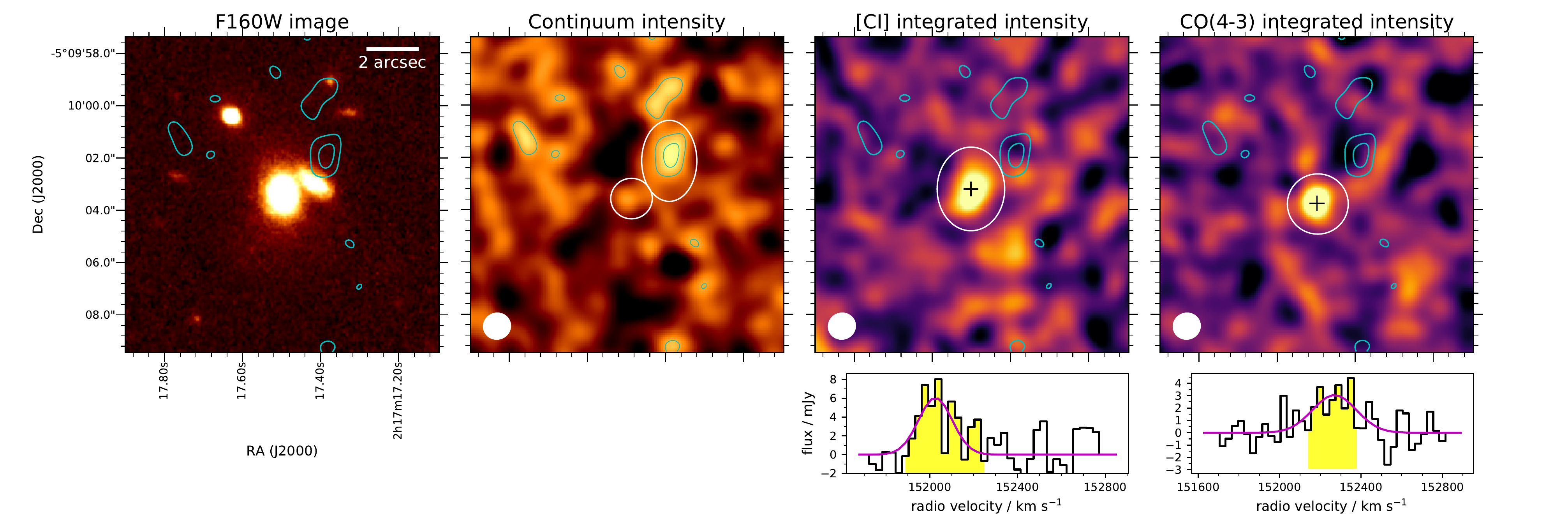}\\
\raisebox{5.3cm}{2}
\includegraphics[width=0.95\textwidth,clip,trim=5mm 3mm 5mm 3mm ]{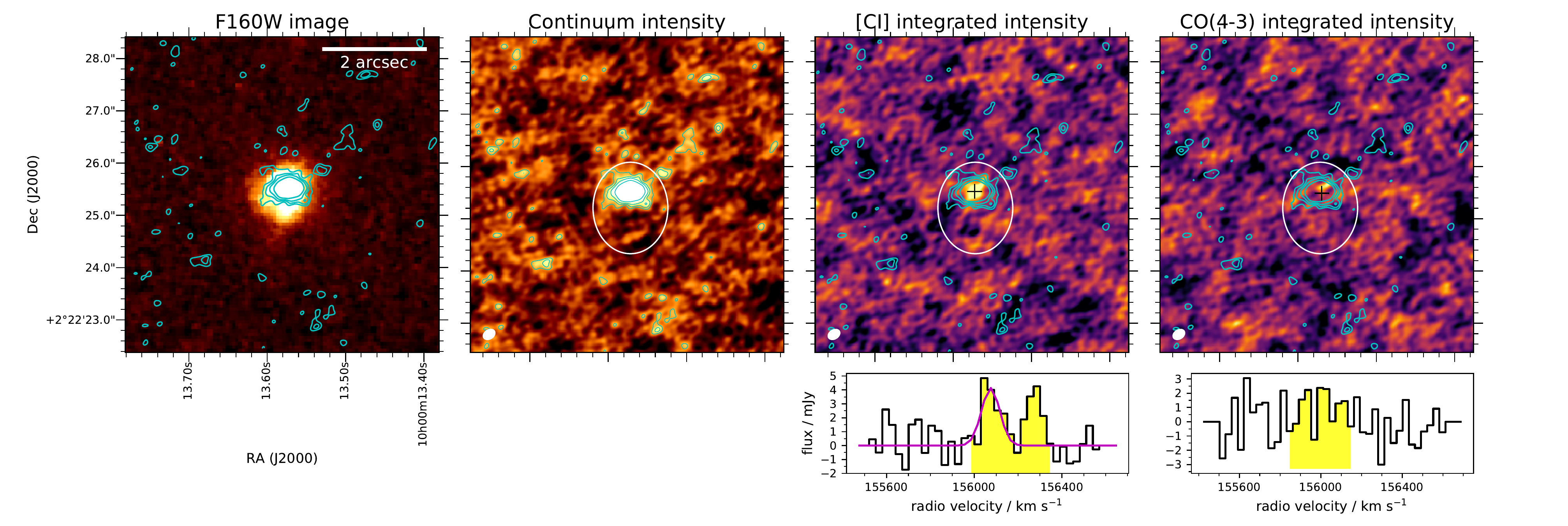}\\
\raisebox{5.3cm}{3}
\includegraphics[width=0.95\textwidth,clip,trim=5mm 3mm 5mm 3mm ]{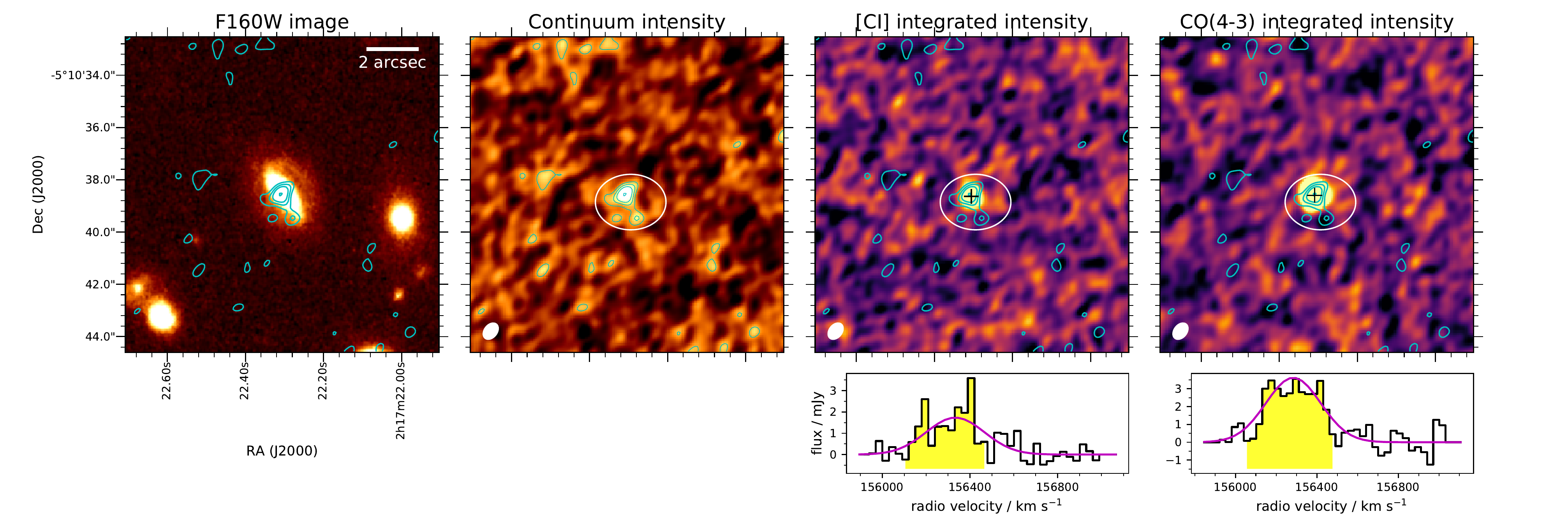}\\
\raisebox{5.3cm}{4}
\includegraphics[width=0.95\textwidth,clip,trim=5mm 3mm 5mm 3mm ]{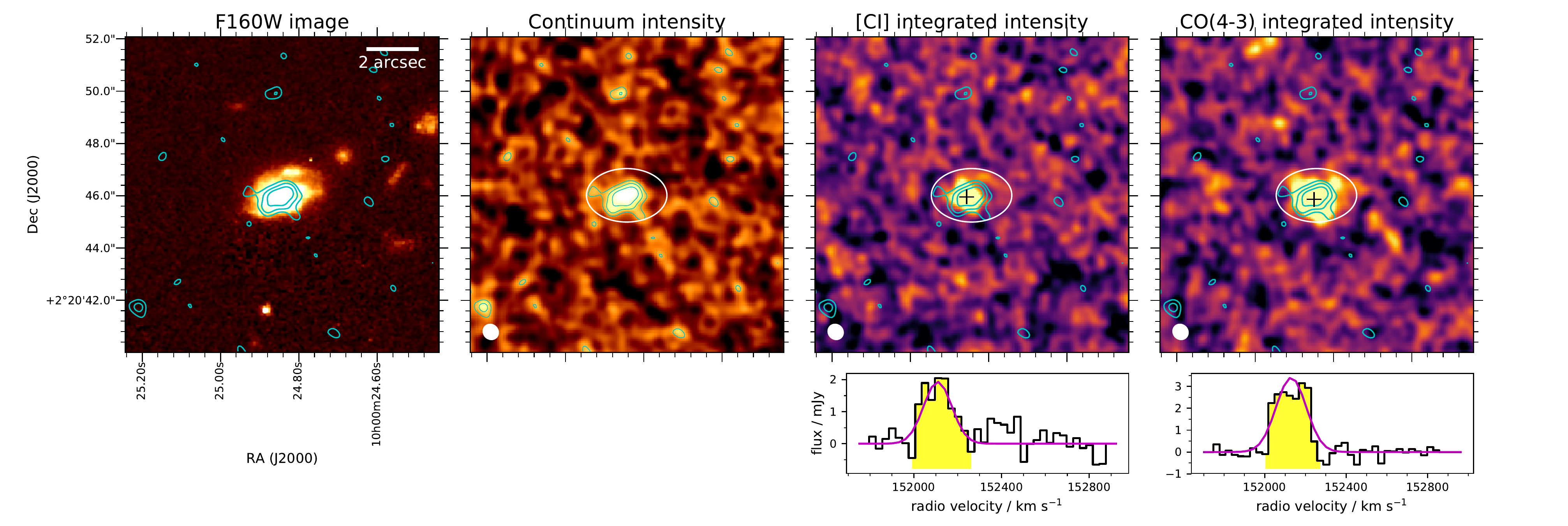}
\vspace{-6pt}
\caption{Optical, ALMA continuum and emission line moment-0 maps, 12 arcsec on a side (6 arcsec for ID 2). Images are produced with natural weighting or (for IDs 1, 5, 8, 9) with a \emph{u-v} taper, as described in Section~\ref{sec:imaging}. 
Continuum contours at 2, 3, 4, 6\,$\sigma$ are shown in cyan on each panel. Black crosses mark the positions at which the peak spectra in Fig.~\ref{fig:spectra} were extracted. White ellipses show the apertures used to measure the total flux, and the spectra in these apertures are shown beneath. The filled regions under these spectra indicate the velocity ranges integrated in the moment-0 maps. The multiple apertures shown for ID 1 are discussed in Section~\ref{sec:s1}.
}
\label{fig:mom0}
\end{center}
\end{figure*}

As described in Section~\ref{sec:imaging}, the continuum image for ID 9 is the result of combining data taken in Cycles 4 and 5, in order to maximise signal-to-noise, although we note that consistent fluxes were measured in each of the individual data sets.
We have imaged the continuum and emission lines with a beam size of $1.1\times1.1$ arcsec due to the extended nature of the source, and we have used an aperture matched to the optical size of the galaxy. The full \ci\ line in the aperture spectrum in Fig.~\ref{fig:mom0} has a broad and skewed profile, with a peak at around 158220\,\kms, and weaker but significant emission extending to 158050\,\kms\ (hence the central velocity is 158140\,\kms). The spectrum in the central beam peaks at 158080\,\kms, as shown in Fig.~\ref{fig:spectra}. 
The profile of the CO aperture spectrum is more closely approximated by a narrow Gaussian centred at 158140\,\kms.
The broad line-width and extended spatial configuration of \ci\ leads to a low signal-to-noise image, but the line is nevertheless detected at S/N~$\approx2.5$ as shown in Table~\ref{tab:almaresults}.

\renewcommand{\thefigure}{\arabic{figure} (continued)}
\addtocounter{figure}{-1}
\begin{figure*}
\begin{center}
\raisebox{5.3cm}{5}
\includegraphics[width=0.95\textwidth,clip,trim=5mm 3mm 5mm 3mm ]{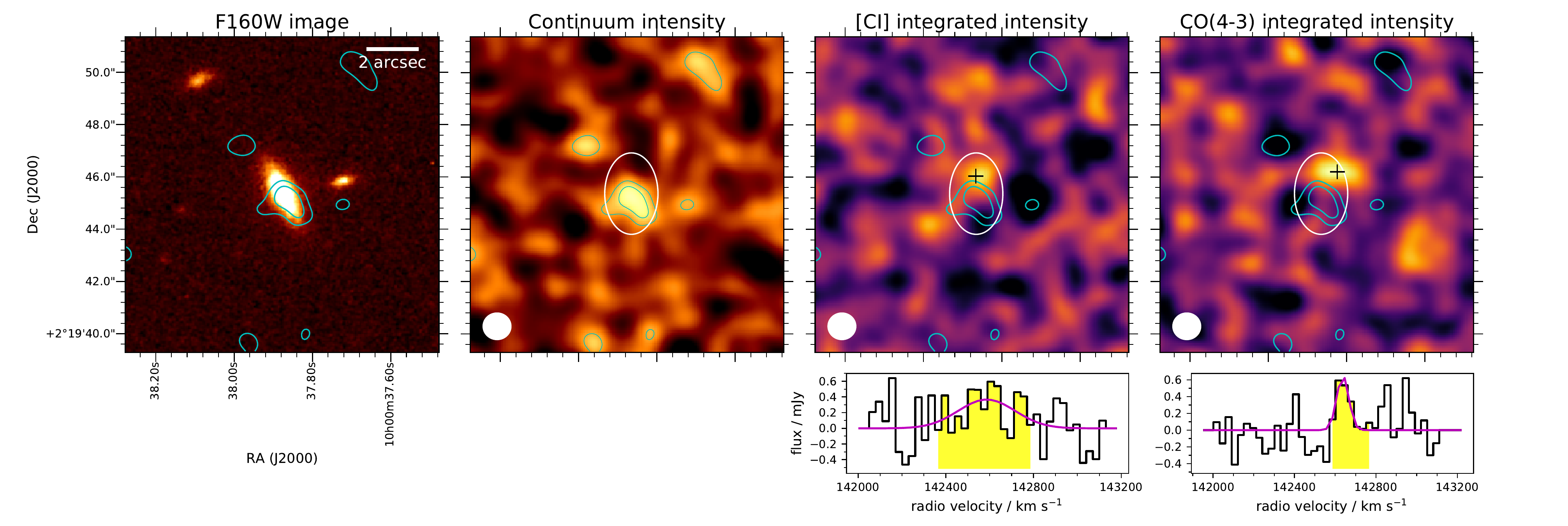}\\
\raisebox{5.3cm}{6}
\includegraphics[width=0.95\textwidth,clip,trim=5mm 3mm 5mm 3mm ]{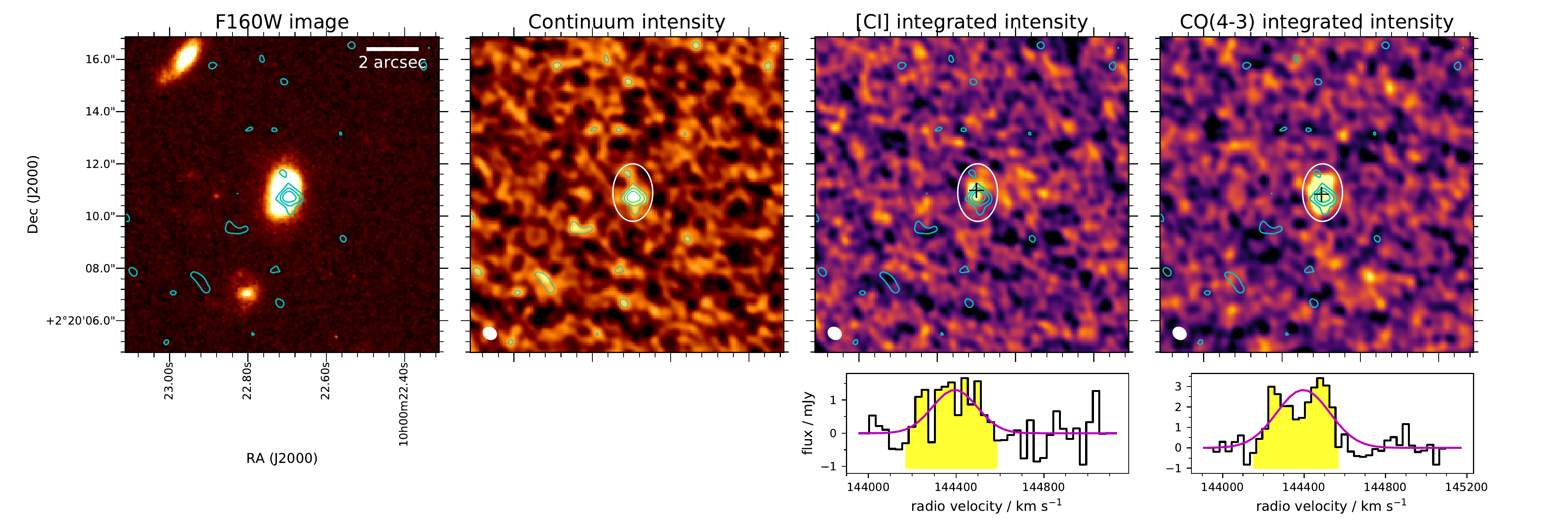}\\
\raisebox{5.3cm}{7}
\includegraphics[width=0.95\textwidth,clip,trim=5mm 3mm 5mm 3mm ]{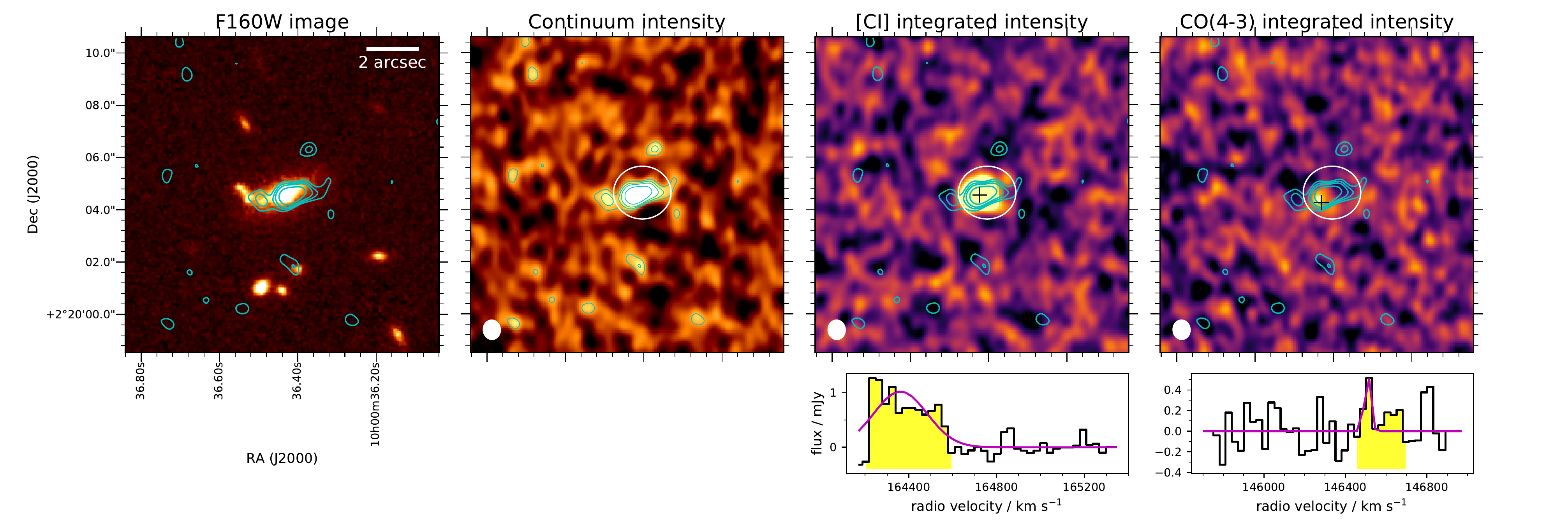}\\
\raisebox{5.3cm}{8}
\includegraphics[width=0.95\textwidth,clip,trim=5mm 3mm 5mm 3mm ]{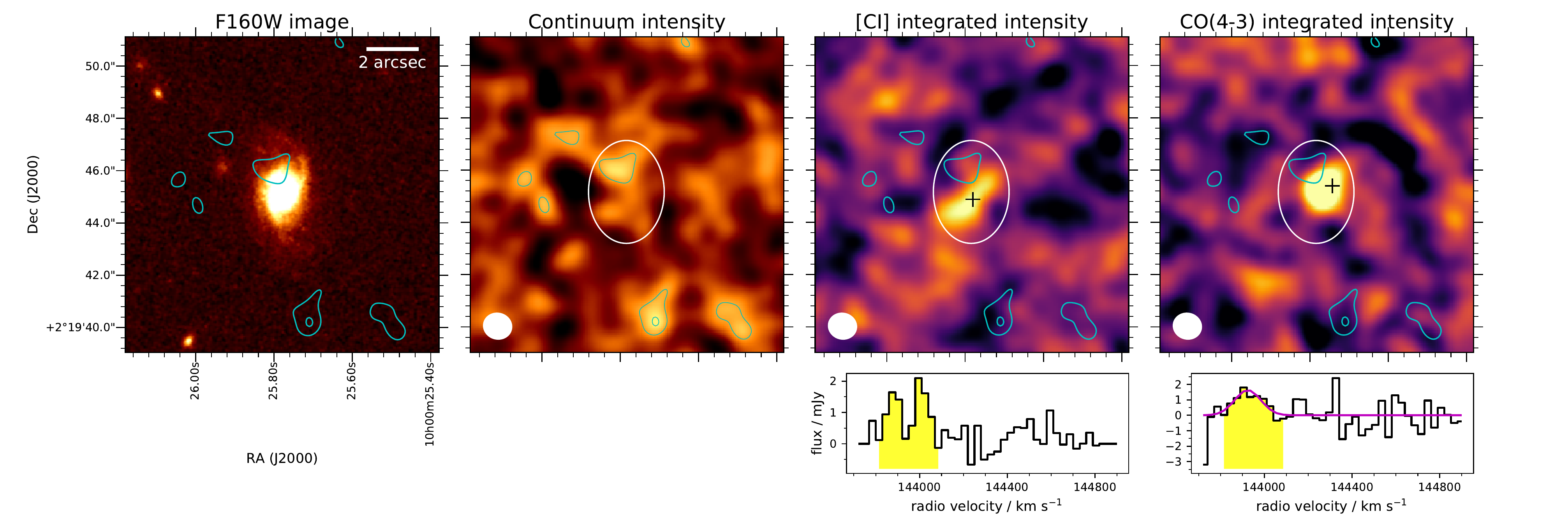}\\
\caption{}
\end{center}
\end{figure*}

\addtocounter{figure}{-1}
\begin{figure*}
\begin{center}
\raisebox{5.3cm}{9}
\includegraphics[width=0.95\textwidth,clip,trim=5mm 3mm 5mm 3mm ]{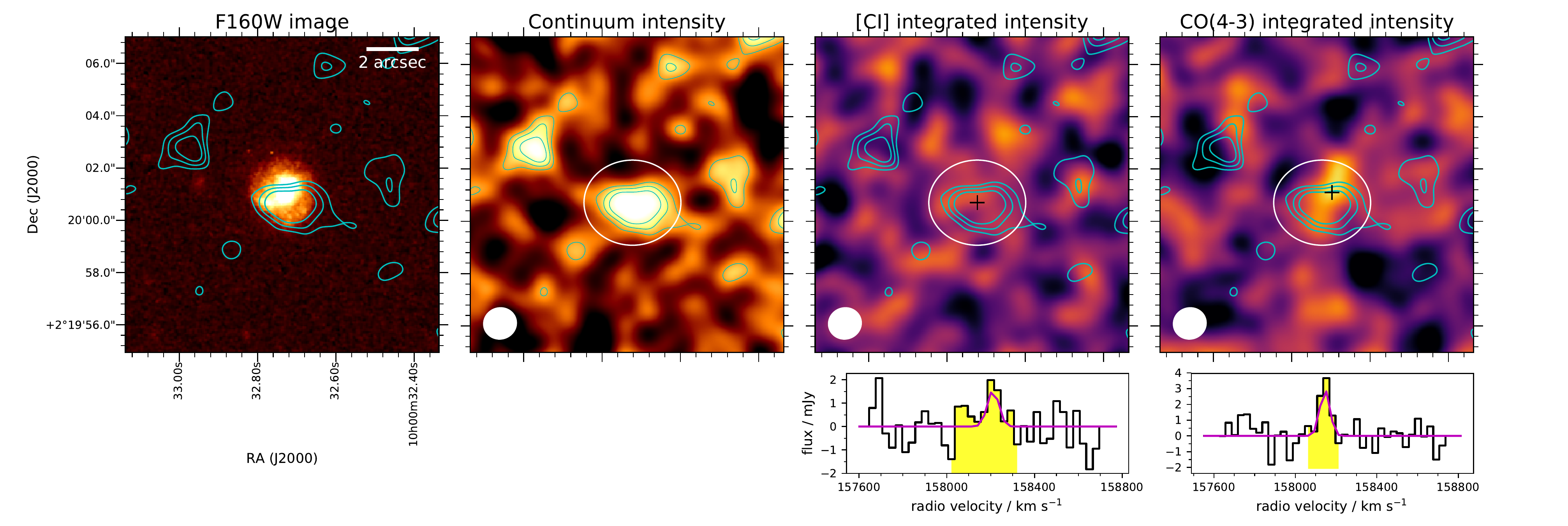}\\
\raisebox{5.3cm}{10}
\includegraphics[width=0.95\textwidth,clip,trim=5mm 3mm 5mm 3mm ]{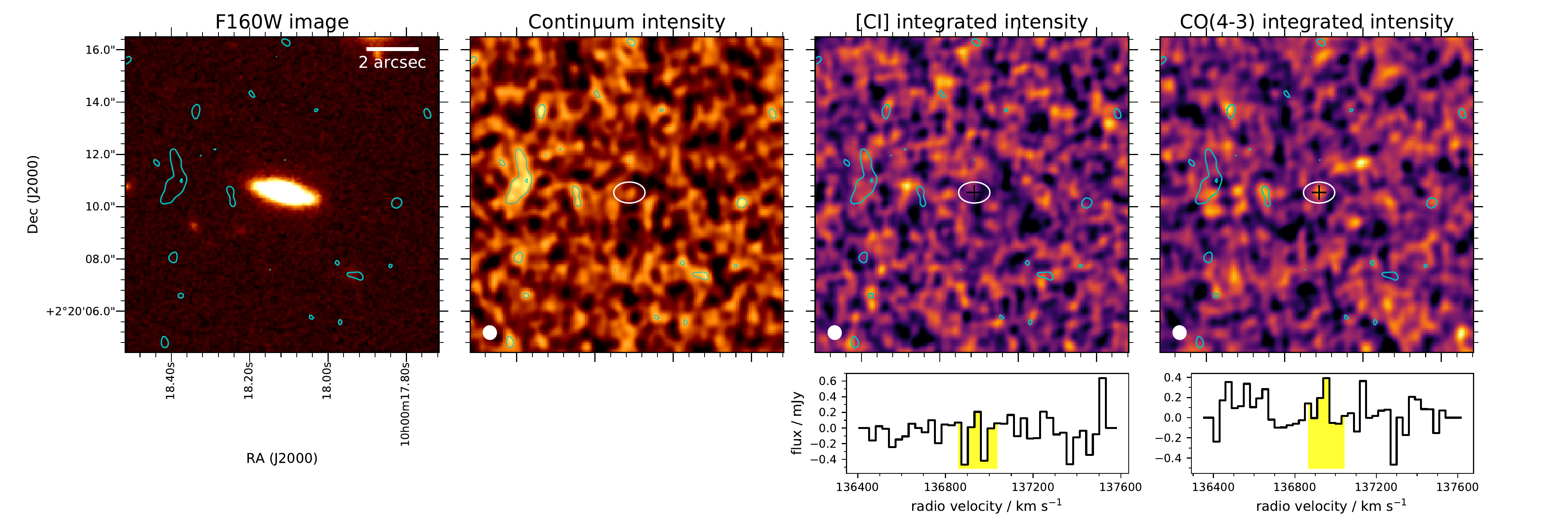}
\caption{}
\end{center}
\end{figure*}
\renewcommand{\thefigure}{\arabic{figure}}

\section{Results and discussion}
\label{sec:discussion}

\begin{table*}
\caption{Summary of correlation analyses between selected parameters. For each pair we list the Spearman rank correlation coefficient $r_s$ and its associated $p$-value; the inverse-variance weighted average ratio; and (for parameter pairs with significant correlations $p<0.01$) the parameters of total-least-squares fits to the relations, together with the residual variance $\sigma_r^2$ of the fits.}
\renewcommand{\arraystretch}{1.1}
\begin{tabular}{lclcccccc}
\hline
Correlation $y-x$  				&   \multicolumn{2}{c}{Spearman rank correlation} &  Average ratio              &  \multicolumn{2}{c}{Linear fit $y = a_0 x$}  & \multicolumn{3}{c}{Power-law fit $y = a_1 x^{a_2}$} \\
  (Figure no.)        							 &  $r_s$           	& $p$-value  &  $\langle y/x \rangle$   &       $a_0$ & $\sigma^2_r$                              &       $a_1$ & $a_2$ & $\sigma^2_r$       \\
\hline
$S^\prime_\textrm{[CI]}-S_\textrm{cont}$  (7)      	&    0.95 $\pm$ 0.12 & 0.000088 &      2.01 $\pm$ 0.20 	&   2.39 $\pm$ 0.21          &      1.8 &    1.90 $\pm$ 0.22 &    0.746 $\pm$ 0.096 &    1.2 \\
\dots excl. ID 1,2  							&    0.93 $\pm$ 0.23 &    0.0025 &      2.34 $\pm$ 0.27 	&   2.67 $\pm$ 0.28          &      1.0 &    1.81 $\pm$ 0.66 &    0.71 $\pm$ 0.26 &    1.1 \\
$L^\prime_\textrm{[CI]}-M_d^{\textrm{cont}}$  (10a)  &    0.92 $\pm$ 0.12 &  0.00051 &      4.65 $\pm$ 0.47 	&   5.52 $\pm$  0.49   	&      1.8 &    0.071 $\pm$ 1.6 &    0.79 $\pm$ 0.10 &    1.6 \\
$L^\prime_\textrm{[CI]}-M_d^\textrm{SED}$  (10b) 	&    0.80 $\pm$ 0.24 &    0.0096 &      5.51 $\pm$ 0.67 	&   9.82 $\pm$  0.84 		&      6.2  &   14.5 $\pm$ 1.6 &    0.646 $\pm$ 0.077 &   4.7 \\
$L^\prime_\textrm{[CI]}-10^{-3}\Lir$    (11)   		&    0.88 $\pm$ 0.24 &    0.0016 &     3.57 $\pm$ 0.36 	&  4.06 $\pm$ 0.36 		 &      1.9 &    1.49$\pm$0.95 &    1.60 $\pm$ 0.38 &    1.4 \\
$L^\prime_\textrm{[CI]}-10^{16}L_\textrm{cont}$ (12) 						&    0.92 $\pm$ 0.10 &  0.00051 &     22.6 $\pm$ 2.3 &   26.8 $\pm$ 2.4 & 1.8 &   28.0 $\pm$ 2.2 & 0.79 $\pm$ 0.10 &      1.6 \\
\hline
$S^\prime_\textrm{CO}-S^\prime_\textrm{[CI]}$   (9)  &    0.71 $\pm$ 0.17 &       0.047 &      0.556 $\pm$ 0.056     &     -- &    -- &    -- &   -- \\
\dots excl. ID 1,2       						&    0.94 $\pm$ 0.16 &     0.0048 &      0.528 $\pm$ 0.057 &  0.549 $\pm$ 0.057      &      1.4 &    0.638 $\pm$ 0.063 &    0.46 $\pm$ 0.15 &   0.33 \\
\hline
$S^\prime_\textrm{CO}-S_\textrm{cont}$     (8)     	&    0.62 $\pm$ 0.18 &        0.10 &      1.24 $\pm$ 0.17 &   --     &     -- &    -- &    -- &   -- \\
\dots excl. ID 1   							&    0.68 $\pm$ 0.23 &      0.094 &      1.37 $\pm$ 0.21 &       --     &     -- &    -- &    -- &   -- \\
$L^\prime_\textrm{CO}-M_d^{\textrm{cont}}$   		&    0.62 $\pm$ 0.19 &        0.10 &      3.25 $\pm$ 0.43 &  --     &     -- &    -- &    -- &   -- \\
$L^\prime_\textrm{CO}-M_d^\textrm{SED}$ 		&    0.62 $\pm$ 0.23 &        0.10 &      3.26 $\pm$ 0.61 &  --     &     -- &    -- &    -- &   -- \\
$L^\prime_\textrm{CO}-10^{-3}\Lir$      			&    0.76 $\pm$ 0.18 &      0.028 &      3.54 $\pm$ 0.37 &  --     &     -- &    -- &    -- &   -- \\
\hline
\end{tabular}
\label{tab:correlations}
\end{table*}

In this section we explore correlations between various tracers and derived physical parameters. For easy reference, we summarise all the correlation results discussed below in Table~\ref{tab:correlations}.
We quantify the strength of correlations between $x$ and $y$ data using the Spearman rank correlation coefficient $r_s$ and its associated $p$-value. 
To give an estimate of the sensitivity of $r_s$ to the measurement errors, we estimated the variance of $r_s$ in a Monte-Carlo simulation. We created 100 realisations, and in each realisation we randomly resampled every value of $x$ and $y$ from a normal distribution with mean and standard deviation given by its real measurement and error bar. The uncertainty on the final value of $r_s$ is given by the standard deviation of the 100 $r_s$ values obtained from the simulation, and is typically $\approx0.2$.

In cases where the measured correlation is significant ($p<0.01$), we also attempt to fit the relationship using total-least-squares minimisation (which takes into account the large error bars in both $x$ and $y$ data). For this purpose we use the orthogonal distance regression method implemented in \textsc{scipy.odr}. The quality of these fits is parameterised by the residual variance $\sigma^2_r$, quantifying deviations between the data and the best-fitting model. These results are also listed in Table~\ref{tab:correlations}.

\subsection{The [CI]--continuum correlation}
\label{sec:correlation}
\begin{figure*}
\begin{center}
\includegraphics[width=0.7\textwidth,clip,trim=0 3mm 0 0]{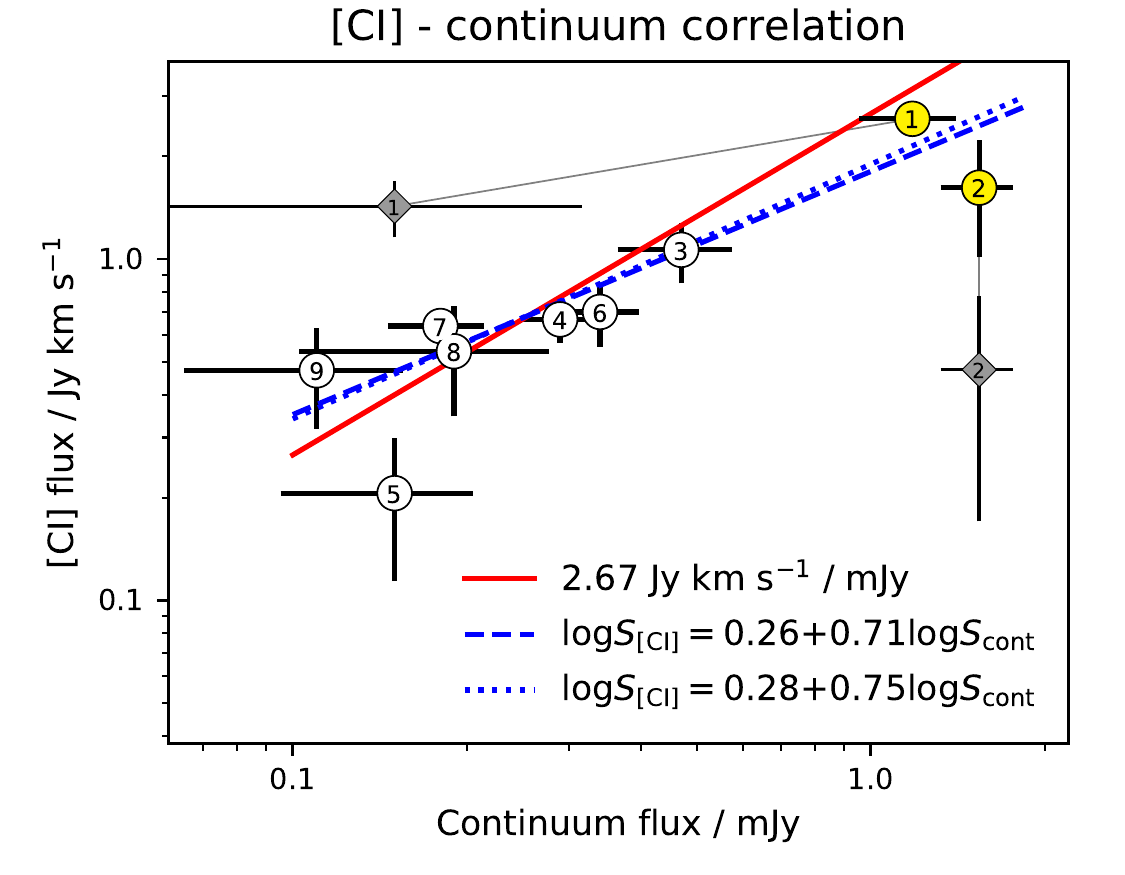}
\caption{The \ci--continuum correlation for the nine detected sources. Two alternative measurements are plotted for ID 1: the grey diamond and yellow circle respectively indicate the smaller and larger apertures described in Section~\ref{sec:s1}. Similarly for ID 2, the grey diamond and yellow circle indicate the narrow and wide velocity ranges discussed in Section~\ref{sec:s2} (see also Table~\ref{tab:almaresults}). The model fits shown by coloured lines are described in Section~\ref{sec:correlation}: the red line is a single-parameter linear model assuming a fixed flux ratio, and is fitted only to the white symbols with unambiguous fluxes; the blue dashed line is a power-law fit to the white symbols, and the blue dotted line is a power-law fit to the white and yellow symbols.}
\label{fig:correlation}
\end{center}
\end{figure*}

The correlation between \ci\ and continuum fluxes measured in apertures is shown in Fig.~\ref{fig:correlation}. 
The plot shows multiple data points for two of the sources as grey and yellow symbols. Source ID 1 has two alternative sets of apertures defined as discussed in Section~\ref{sec:s1}. The grey diamond indicates a conservative aperture matched to the size of the optical and continuum source, while the yellow circle is the result of enlarging the \ci\ aperture to enclose the full size of the emitting region, and also accounting for bright continuum emission that is spatially offset from the source in what appears to be an outflow or tidal tail (potentially resulting from an interaction with one of the neighbouring galaxies in the field of view).
For ID 2, we consider the \ci\ emission in a conservatively narrow velocity range consistent with the weakly detected CO emission (shown by the grey diamond in Fig.~\ref{fig:correlation}) as well as that in a wider velocity range encompassing both peaks in the \ci\ spectrum (yellow circle).  As discussed in Section~\ref{sec:s2}, the double-peaked profile is somewhat unusual, but given that the peaks are offset by only 400\,\kms, and are not separated in angular location, the most likely explanation is that both result from emission within the same galaxy, and therefore the larger value of \ci\ flux should be considered as the total flux of this source (yellow symbol in Fig.~\ref{fig:correlation}).

Several model fits are plotted alongside the data in Fig.~\ref{fig:correlation}. The red line shows the best-fitting linear model of the form $S^\prime_\mathrm{[CI]}\Delta V = a_0\,S_\mathrm{cont}$, which yields $a_0=2.67\pm0.28$~Jy\,\kms\,mJy$^{-1}$ as the best-fitting single flux ratio ($\sigma^2_r=1.0$; $d.o.f.=6$). This model has been fitted only to the white data points, to avoid confusion/uncertainty resulting from the alternative flux measurements of IDs 1 and 2.
The blue lines show best-fitting power-law models of the form $S^\prime_\mathrm{[CI]}\Delta V = a_{1}\,S_\mathrm{cont}^{a_2}$. 
The blue dashed line is fitted to the white data points, with best-fitting parameters $a_1=1.81\pm0.66$ and $a_2=0.71\pm0.26$ ($\sigma^2_r=1.1$; $d.o.f.=5$). 
The blue dotted line is fitted to the white and yellow data points, including ID 1 (accounting for the extended emission) and ID 2 (accounting for both peaks in the \ci\ line profile). This model is very similar to the previous one which excluded those objects, and has best-fitting parameters $a_1=1.90\pm0.22$ and $a_2=0.746\pm0.096$ ($\sigma^2_r=1.2$; $d.o.f.=7$).

Overall, these results indicate a strong correlation between continuum flux and \ci\ flux.  The Spearman rank coefficient and significance level respectively are $0.93$, $2.5\times10^{-3}$ (for the seven white data points) or $0.95$, $8.8\times10^{-5}$ (for the nine white and yellow data points). A single \ci/continuum flux ratio of $a_0=2.67$~Jy\,\kms\,mJy$^{-1}$ provides a good fit to the data, although a shallower power-law model with index $a_2\approx0.7$ provides a similarly good fit to all objects except for ID 5.
IDs 1 and 2 would be outliers to this correlation if the conservatively small aperture for ID 1 and the narrow velocity range for ID 2 are considered, but if the full fluxes measured for each are considered to be genuine then they fall close to the same relationship as the other seven detected sources. 

\subsection{The \cofour--continuum correlation}
\begin{figure*}
\begin{center}
\includegraphics[width=0.7\textwidth,clip,trim=0 3mm 0 0]{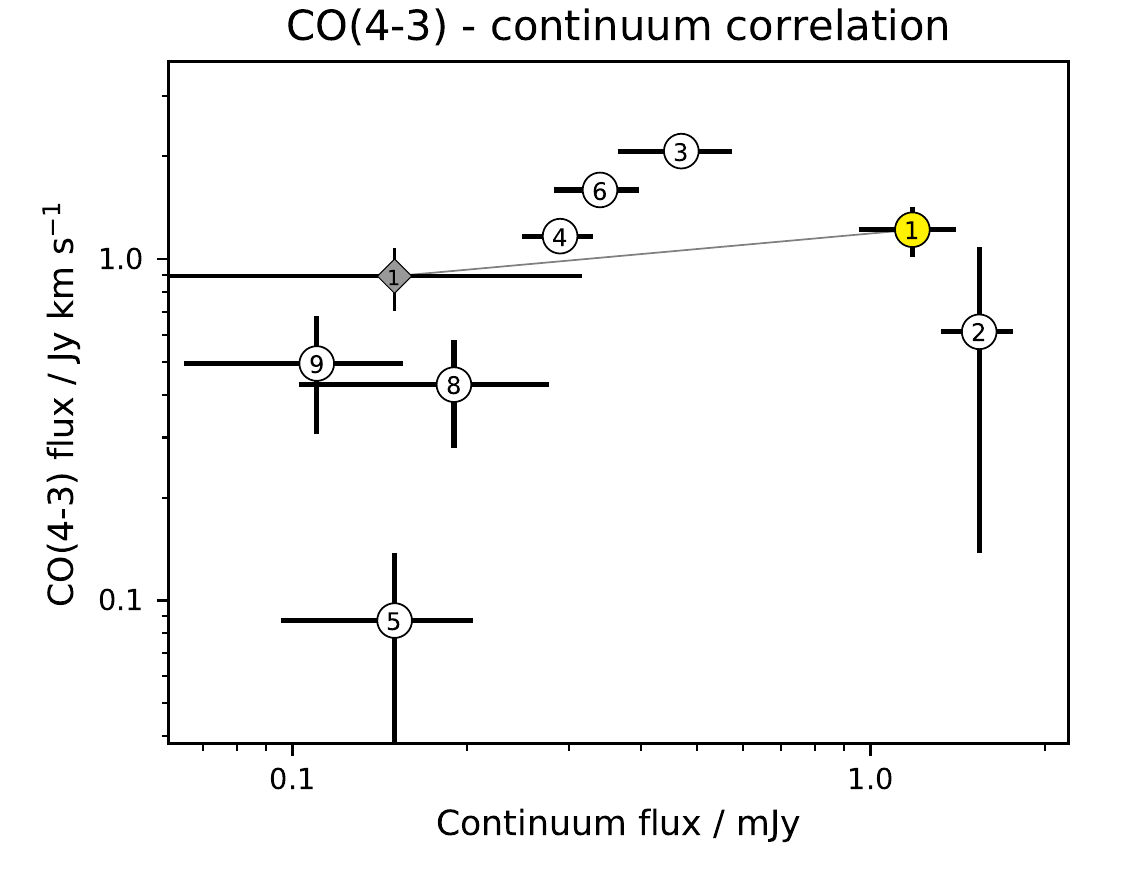}
\caption{The \cofour--continuum correlation for the eight sources detected in CO (compare Fig.~\ref{fig:correlation}). 
For ID 1, the grey diamond and yellow circle respectively indicate the smaller and larger apertures described in Section~\ref{sec:s1}. 
}
\label{fig:cocorrelation}
\end{center}
\end{figure*}

Since we hypothesise that the dust and atomic carbon trace the same phase of the ISM, a basic test we can conduct is to compare the correlation between these tracers with the correlation between dust and \cofour, since \cofour\ is expected to trace a denser and warmer phase of the ISM. This correlation is shown in Fig.~\ref{fig:cocorrelation}.
In this sample, the \cofour--continuum correlation is significantly weaker than the \ci--continuum correlation: the Spearman rank coefficient and significance level respectively are 0.68, 0.094 (for the seven white data points) or 0.62, 0.10 (for the eight white and yellow data points). 
The dust emission is more closely correlated with the low-density, low-excitation gas traced by \ci, and thus more closely traces the total molecular gas content than the warm, dense gas traced by \cofour.


\subsection{Discussion of scatter and outliers: clues from line ratios and dust temperature}
\label{sec:co_ci}
\begin{figure}
\begin{center}
\includegraphics[width=0.49\textwidth]{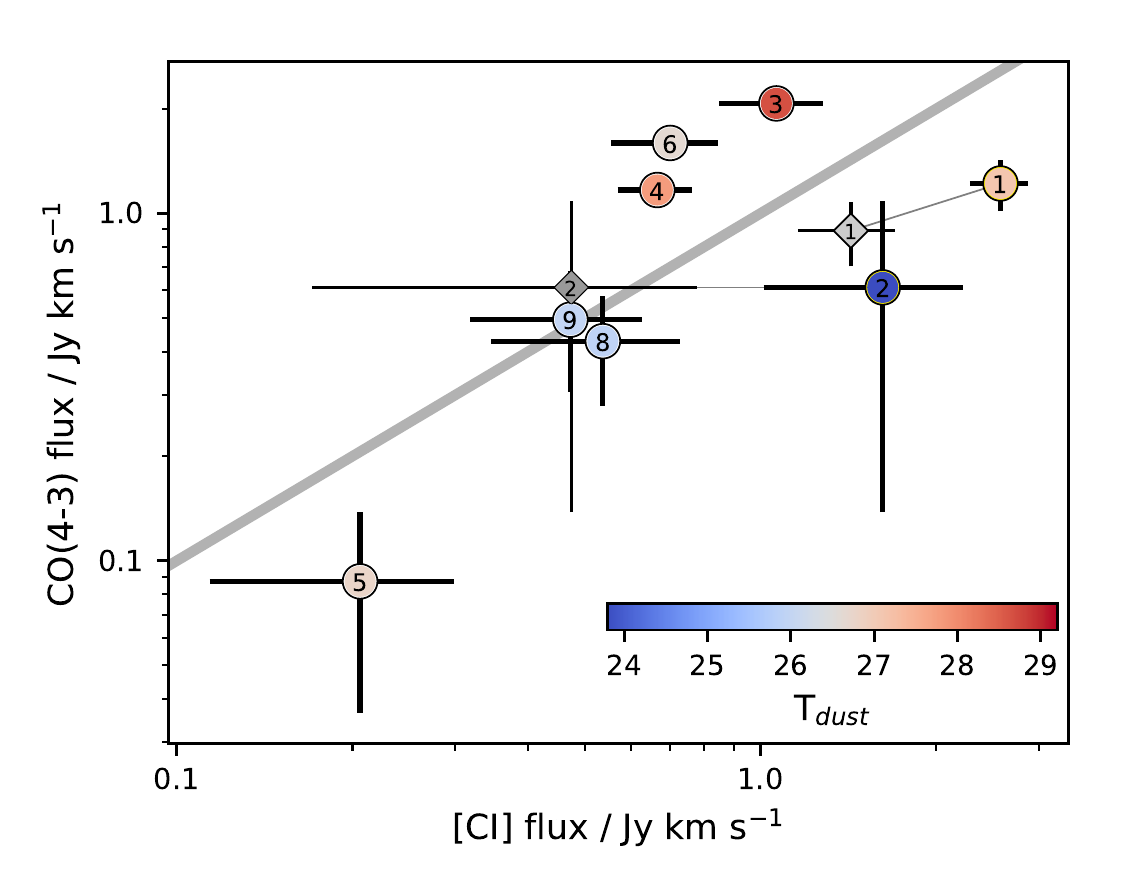}
\caption{The relationship between \ci\ and \cofour\ fluxes. Circle and diamond symbols are as in Fig.~\ref{fig:correlation}.
The colour scale is dust temperature from the SED fits described in Section~\ref{sec:sedfits}. The light grey diagonal line is a constant flux ratio of 1, shown for reference. Scatter perpendicular to this line is driven by a combination of gas excitation, density and optical depth. 
}
\label{fig:ci_co}
\end{center}
\end{figure}

Our results so far suggest a good correlation between \ci\ and continuum fluxes, and a weaker relationship between \cofour\ and continuum, although two sources (IDs 2 and 5) are notably strong in continuum relative to both emission lines, and ID 1 is particularly strong in \ci\ relative to continuum.
To complete the comparison between these tracers, Fig.~\ref{fig:ci_co} shows the relationship between \ci\ and \cofour.
Symbols are coloured according to the median dust temperature from the MCMC posteriors from SED fitting; uncertainties on these are typically $\pm$1--3\,K ($\pm$9\,K for ID 1) as shown in Table~\ref{tab:physparams}. Most sources have median dust temperatures around 26--29\,K, so the scatter between these is similar to the size of their uncertainties.
The physical relationship between \ci\ and \cofour\ is complicated, and can be influenced by a number of factors including optical depth and cloud filling factor (since CO is generally optically thick), as well as gas density and temperature. Gas temperature is not necessarily linked to dust temperature since collisions are generally rare and the dust temperature is governed by radiative heating by the ISRF.

Source IDs 1, 2 and 5 have the lowest \cofour/\ci\ ratios, and it is interesting that these all show unusual morphological distributions of gas and dust (see Section~\ref{sec:details}). IDs 1 and 5 both show evidence for interactions, with ID 1 appearing to have a tidal tail or outflow of dust, and an extended distribution of \ci\ that may be outflowing, while ID 5 has \ci\ and CO emission offset from the continuum and from the optical light. Relatively weak \cofour\ in IDs 1 and 5 may therefore be explained by the presence of \ci\ and continuum emission in regions where \cofour\ is not seen; such regions would therefore be characterised by diffuse/low-excitation ISM. ID 2 has a very weak CO detection, but a double-peaked \ci\ line profile in which one of the peaks has no associated CO emission. This galaxy also has the highest SSFR in the sample (see Fig.~\ref{fig:msfr}), is very compact and has a much lower dust temperature ($18\pm1$\,K) and higher dust mass ($2.3^{+0.3}_{-0.2}\times10^9\Msun$) than the rest of the sample (see Table~\ref{tab:physparams}). It appears therefore to have a very high surface density of dust and gas, and so it is likely that the CO is more optically thick than in the other galaxies in the sample, leading to a suppressed \cofour/\ci\ ratio.

The \ci--continuum correlation in Fig.~\ref{fig:correlation} is very strong, although the sample is small and ambiguities in the total fluxes of IDs 1 and 2 add some complication to the interpretation. The objects which appear to deviate most from the overall relationship are ID 5 and possibly 1 and/or 2. As previously mentioned, these three all show unusual properties in terms of their spatial distribution of gas and dust, and relatively low \cofour/\ci\ ratios. ID 2 has an especially low dust temperature, but the scatter in dust temperatures for the rest of the sample is small and comparable to the errors, so there is no evidence that dust temperature plays any role in the scatter in the \ci--continuum correlation.

\begin{figure}
\begin{center}
\includegraphics[width=0.49\textwidth]{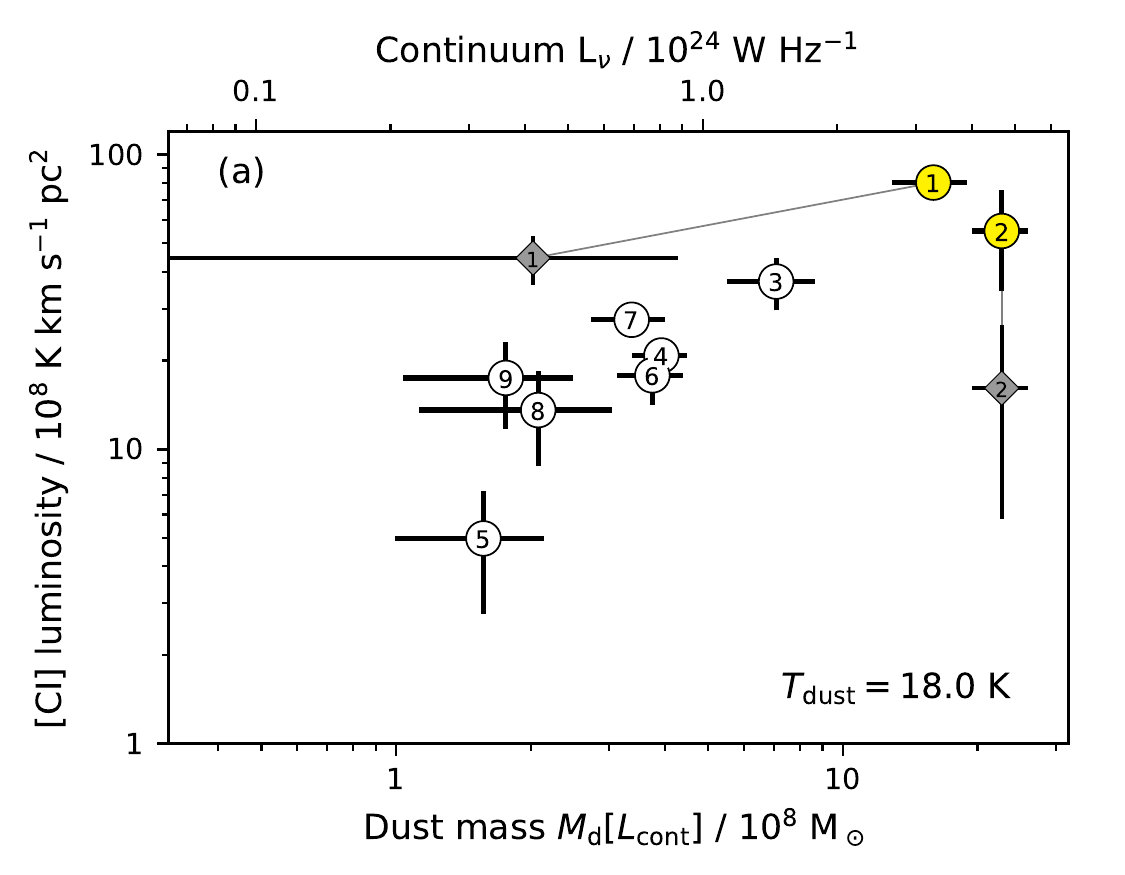}
\includegraphics[width=0.49\textwidth,clip,trim=0 0 0 3mm]{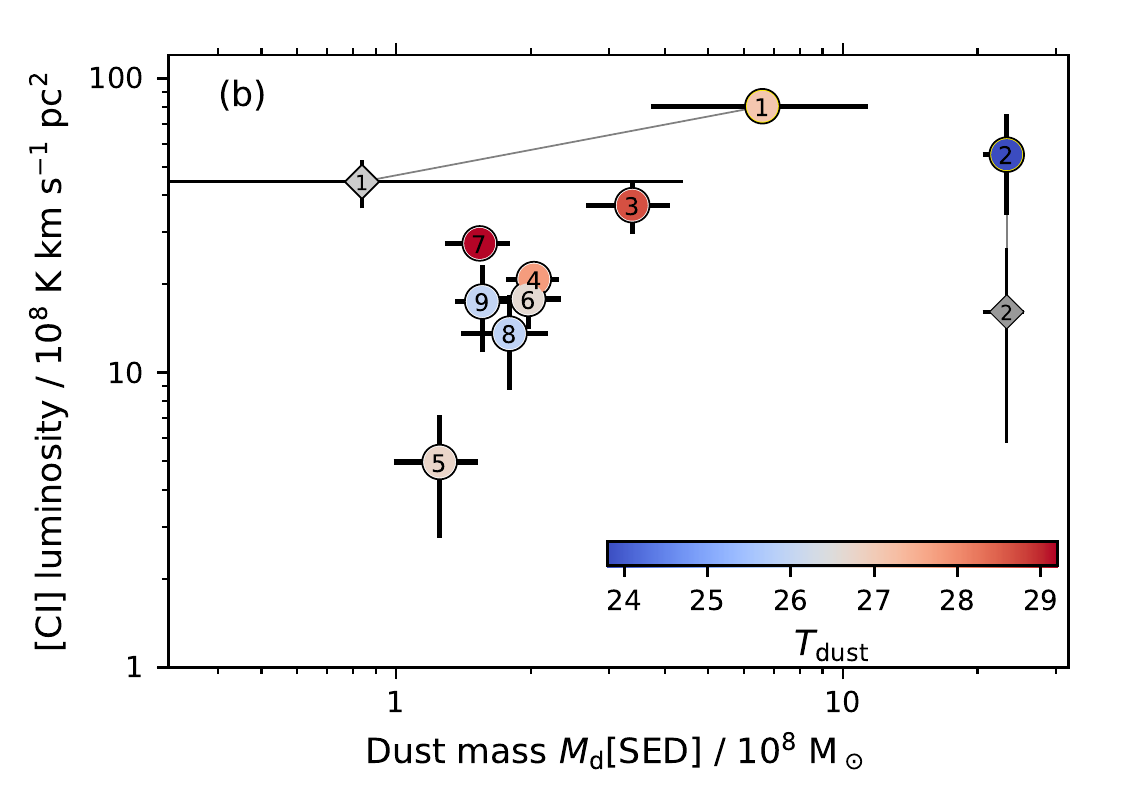}
\caption{The relationship between \ci\ luminosity and continuum luminosity (upper axis, top panel), which can be converted into dust mass assuming a fixed temperature (lower axis, top panel) or assuming the best-fitting dust temperatures from the SED fits in Table~\ref{tab:physparams} (bottom panel). Symbols are as in Fig~\ref{fig:ci_co}.
}
\label{fig:ci_mdust}
\end{center}
\end{figure}

\subsection{The correlation between \ci\ and dust mass}
\label{sec:ci_mdust}
In Fig.~\ref{fig:ci_mdust}a (top axis) we compare \ci\ luminosity against the continuum monochromatic luminosity at rest-frame 609\,\micron. These two tracers should be well correlated if the dust/\Htwo\ and \ci/\Htwo\ ratios are constant and the dust mass/light ratio is constant. This mass/light ratio is primarily dependent on the mass-weighted temperature of the dust. Assuming a fixed temperature $T_d=18$\,K,  $\beta=1.8$ and optically-thin dust emission, we can convert $L_\nu$ directly to dust mass as shown on the lower axis of Fig.~\ref{fig:ci_mdust}a.
For this conversion, we have chosen to use the lowest temperature from the best-fitting values in our sample ($T_d=18\pm1$\,K; Table~\ref{tab:physparams}). This provides a reasonable upper limit for the dust mass, since the SED fitting is unlikely to under-estimate the mass-weighted temperature, but can easily over-estimate it in the presence of a warmer, less-massive dust component that will dominate the luminosity at $\lambda\lesssim250\,\micron$. 

For comparison, in Fig.~\ref{fig:ci_mdust}b, we plot the dust masses derived from the best-fitting SEDs (with the parameters listed in Table~\ref{tab:physparams}). While the dust masses shown in Fig.~\ref{fig:ci_mdust}a provide a reasonable upper limit on the true dust mass, those in Fig.~\ref{fig:ci_mdust}b provide a reasonable lower limit, since they are derived under the assumption that a single-temperature SED describes all the photometry at observed wavelengths $100<\lambda<850\,\micron$, and the temperature derived is therefore luminosity-weighted rather than mass-weighted. 

The correlation between dust mass and \ci\ luminosity is noticeably weaker when assuming the variable dust temperatures given by the full SED fits (Fig.~\ref{fig:ci_mdust}b: $r_s=0.80, p=0.0096$) compared with assuming a fixed dust temperature (Fig.~\ref{fig:ci_mdust}a: $r_s=0.92, p=0.00051$). 
This indicates that the assumption of the luminosity-weighted temperature (from the SED fitting) introduces additional scatter, and is a less reliable measurement of the true dust mass than is obtained from assuming a fixed temperature.  The true value of the mass-weighted temperature is lower than or equal to the luminosity-weighted temperature, but is difficult to estimate accurately.  Assuming a fixed temperature of 25\,K (instead of 18\,K) would result in lower dust masses by a factor 0.72.

The correlation in Fig.~\ref{fig:ci_mdust}a has a Spearman rank coefficient and significance level of $0.86$, $0.014$ (including the seven white data points) or $0.92$, $5.1\times10^{-4}$ (including the nine white and yellow data points). This is slightly weaker than the correlation between \ci\ and continuum fluxes, but accounting for the effects of measurement errors the difference is not significant (see Table~\ref{tab:correlations}). 
These results are limited by the modest dynamic range of luminosities in the current sample, but in Section~\ref{sec:interp} we explore a wider range using samples from the literature.

\subsection{The correlations between \ci, \cofour\ \&\ \Lir}

\begin{figure}
\begin{center}
\includegraphics[width=0.49\textwidth,clip,trim=0 2mm 0 6mm]{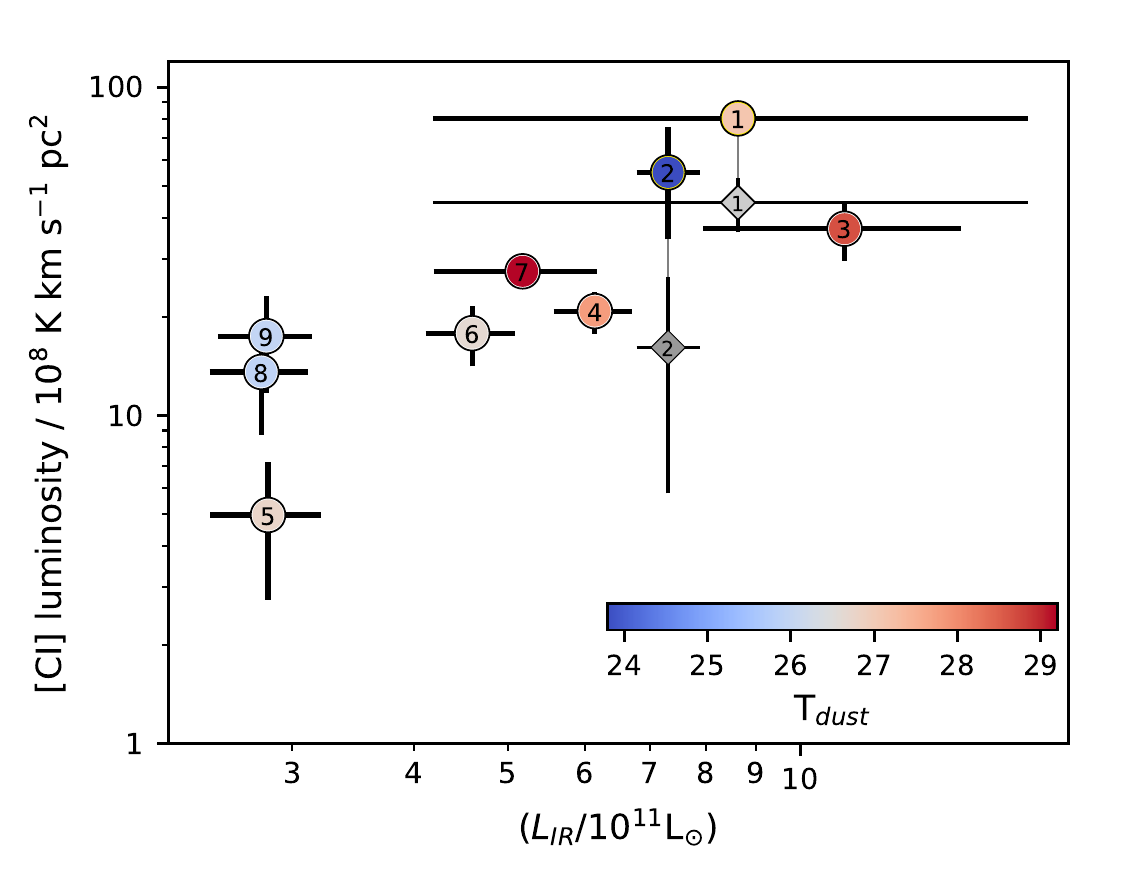}
\caption{The relationship between \ci\ luminosity and total IR luminosity. Symbols are as in Fig~\ref{fig:ci_co}.}
\label{fig:ci_ltir}
\end{center}
\end{figure}

In Fig.~\ref{fig:ci_ltir} we plot \ci\ luminosity against the total IR luminosity from the SED fitting ($\Lir\propto$ SFR).  This relationship contains more scatter than the \ci--continuum or \ci--\Mdust\ correlations, but there remains a significant correlation ($r_s=0.88, p=0.0016$). The relationship is weaker because \ci\ traces molecular gas while \Lir\ traces SFR; these two parameters are strongly correlated but are not as closely linked as the dust and gas masses.
As shown in Table~\ref{tab:correlations}, the correlation between \cofour\ (a tracer of dense gas) and \Lir\ is also weak in this sample ($r_s=0.76, p=0.028$), but this is stronger than the correlation between \cofour\ and any other parameter.

\section{Dust and CI as tracers of gas mass}
\label{sec:discussion2}
\subsection{Gas mass calibrations}
\label{sec:m_H2}
Dynamical PDR modelling strongly suggests that \ci\ emission is a robust and direct tracer of molecular gas mass. 
Following \citet{Papadopoulos2004}, we calibrate the molecular ($\Htwo$ + He) gas mass as a function of $S^\prime_\mathrm{[CI]}\Delta V$ (assuming optically thin emission) as
\begin{equation}
\dfrac{M_\mathrm{mol}^\mathrm{[CI]}}{\Msun} = \dfrac{4\pi\,\mu\, m_{\Htwo}}{hc\, A_{10}\, \Xci \, Q_{10}} \left( \dfrac{d_L^2}{1+z} \right) \dfrac{S^\prime_{\mathrm{[CI]}} \Delta V}{\textrm{Jy\,km\,s}^{-1}}
\label{eqn:MmolCI}
\end{equation}
where $m_{\Htwo}$ is the molecular mass of \Htwo\ (in \Msun), $\mu=1.36$ accounts for the mass of helium and \Xci\ is the C\textsc{i}/\Htwo\ abundance ratio.
For the \mbox{\ci(1--0)} transition, 
$A_{10}=7.93\times10^{-8}$s$^{-1}$ is the Einstein coefficient, while $Q_{10}$ is the excitation factor, which is a function of gas density and temperature. 
We cannot measure $Q_{10}$ without measurements of both \ci(1--0) and \ci(2--1), but in typical conditions of kinetic temperature $20<T_\mathrm{kin}<40$\,K and density $300<n< 10^{4}$\,cm$^{-3}$, $Q_{10}$ remains in the narrow range of approximately $0.25-0.45$ \citep{Jiao2017}. 
We adopt a moderate value of $Q_{10}=0.35$, following \citet{Papadopoulos2012a}.
The other unknown is \Xci;
observational constraints on this range between $\sim1-5\times10^{-5}$ \citep{Papadopoulos2004a}.
For example, $\Xci=2.2\times10^{-5}$ in the Ophiuchus molecular cloud complex \citep{Frerking1989};
$3\times10^{-5}$ in the M82 starburst nucleus \citep{Weiss2003};
$5\times10^{-5}$ in the Cloverleaf quasar at $z=2.5$ \citep{Weiss2005};
while the average value in a sample of high-redshift SMGs and quasars is $\Xci=3.9\times10^{-5}$ \citep{Alaghband-Zadeh2013}.
We adopt an intermediate value of $3\times10^{-5}$.

There is also strong empirical evidence that the long-wavelength dust-continuum luminosity is tightly correlated with CO-derived molecular gas masses \citep{Scoville2016}, in quantitative agreement with results from some cosmological zoom simulations \citep{Liang2018, Privon2018}. 
\citet{Scoville2016} provide the following calibration for molecular gas mass as a function of observed 850-\micron\ flux density:
\begin{equation}
\dfrac{M_\mathrm{mol}^\mathrm{cont}}{10^{10}\Msun} = 1.78 \dfrac{S_{\nu_{obs}}}{\mathrm{mJy}} (1+z)^{-4.8} \left(\dfrac{\nu_{850\mu\mathrm{m}}}{\nu_{obs}}\right)^{3.8} \left(\dfrac{d_L}{\mathrm{Gpc}}\right)^2 \dfrac{6.7\times10^{19}}{\alpha_{850}} \dfrac{\Gamma_0}{\Gamma_\mathrm{RJ}}
\label{eqn:Mmol850}
\end{equation}
assuming $T_d=25$\,K and $\beta=1.8$ \citep{Scoville2016err}. Here, $\Gamma_\mathrm{RJ}$ is the deviation of the modified blackbody SED from the Rayleigh Jeans law, which is given by 
\begin{equation}
\Gamma_\mathrm{RJ}(T_d,\nu_{obs},z) = \dfrac{h \nu_{obs}(1+z)/kT_d}{\exp[h \nu_{obs}(1+z)/kT_d]-1},
\end{equation}
and $\Gamma_0=\Gamma_\mathrm{RJ}(T_d,\nu_{850\mu\mathrm{m}},0)$.
\citet{Scoville2016} show that normal star-forming galaxies, local ULIRGs and high-$z$ SMGs are broadly consistent with a single conversion factor $\alpha_{850}=6.7\times10^{19}$~erg\,s$^{-1}$\,Hz$^{-1}$\,\Msun$^{-1}$, or alternatively 
\begin{equation}
\alpha_{850}=6.2\times10^{19}\left(\dfrac{L_{\nu 850}}{10^{31}~\textrm{erg~s}^{-1}~\textrm{Hz}^{-1}}\right)^{0.07}~\textrm{erg~s}^{-1}~\textrm{Hz}^{-1}~\Msun^{-1}.
\label{eqn:alpha850}
\end{equation}
Either calibration of $\alpha_{850}$ is subject to the assumption of a single Galactic CO/\Htwo\ conversion factor, $X_\textrm{CO}=3\times10^{20}$ N(H$_2$)\,cm$^{-2}$\,(K\,\kms)$^{-1}$, for all galaxies in their sample.
This method calibrates the molecular gas mass from a single long-wavelength photometric data point rather than from the dust mass from SED fitting (see discussion in Section~\ref{sec:ci_mdust}). 
We therefore use our ALMA continuum measurements at rest-frame $609\,\micron$ to calculate $M_\mathrm{mol}^\mathrm{cont}$ using this method.

\subsection{Comparing independent gas mass estimates}
\label{sec:interp}

\begin{table}
\caption{Results of gas and dust mass calculations: $M_d$ calculated from SED fitting and from $L_\textrm{cont}$; and $M_\textrm{mol}$ calculated from continuum and from \ci\ (assuming $Q_{10}=0.35$, $\Xci=3\times10^{-5}$).}
\renewcommand{\arraystretch}{1.1}
\begin{tabular}{cccccc}
\hline
ID  &  $M_d$[SED]  & $M_d$[$L_\textrm{cont}$] & $M_\mathrm{mol}^\mathrm{[CI]}$ & $M_\mathrm{mol}^\mathrm{cont}$ & \multirow{2}{*}{$\dfrac{M_\mathrm{mol}^\mathrm{cont}}{M_\mathrm{mol}^\mathrm{[CI]}}$}  \\ 
     &  $10^7\Msun$ & $10^7\Msun$  & $10^9\Msun$                    & $10^9\Msun$                   & \\ 
\hline
1 &  66 $\pm$ 38 & 159 $\pm$ 30 &  99 $\pm$ 11 & 184 $\pm$ 35 &  1.9 \\ 
2 & 232 $\pm$ 25 & 227 $\pm$ 32 &  68 $\pm$ 25 & 255 $\pm$ 36 &  3.8 \\ 
3 &  34 $\pm$  7 &  71 $\pm$ 16 &  46 $\pm$  9 &  87 $\pm$ 19 &  1.9 \\ 
4 &  20 $\pm$  3 &  39 $\pm$  6 &  26 $\pm$  4 &  50 $\pm$  7 &  1.9 \\ 
5 &  13 $\pm$  3 &  16 $\pm$  6 &   6 $\pm$  3 &  21 $\pm$  8 &  3.5 \\ 
6 &  20 $\pm$  4 &  37 $\pm$  6 &  22 $\pm$  5 &  48 $\pm$  8 &  2.2 \\ 
7 &  15 $\pm$  3 &  34 $\pm$  6 &  34 $\pm$  3 &  43 $\pm$  8 &  1.3 \\ 
8 &  18 $\pm$  4 &  21 $\pm$ 10 &  17 $\pm$  6 &  28 $\pm$ 13 &  1.7 \\ 
9 &  16 $\pm$  2 &  18 $\pm$  7 &  22 $\pm$  7 &  24 $\pm$ 10 &  1.1 \\ 
\hline
\end{tabular}
\label{tab:gasmass}
\end{table}

The results of calculating molecular gas masses from \ci, using \eqnref{eqn:MmolCI}, and from the continuum, using equations (\ref{eqn:Mmol850}--\ref{eqn:alpha850}), are listed in Table~\ref{tab:gasmass}.
The continuum-derived gas masses are larger by a factor ranging from 1 to 4 across the sample.
For comparison, Table~\ref{tab:gasmass} also lists the dust masses estimated from full IR SED-fitting to each galaxy's photometry (Section~\ref{sec:sedfits}) and from the ALMA continuum flux assuming $T_d=18$\,K (the lowest temperature in the sample according to our SED fitting; see also Section~\ref{sec:ci_mdust}). These two estimates of dust mass can be seen as upper and lower bounds on the true value, since the bulk of the dust is unlikely to be colder than 18\,K, and cannot be hotter than the luminosity-weighted temperature of the full IR SED.
The molecular gas mass derived from the continuum following \citet{Scoville2016}
implicitly assumes a constant dust/gas ratio of approximately 120. Using our \ci-derived gas masses, we measure the mean dust/gas ratio in our sample to be between 70 and 130 (the range given by the two alternative dust mass estimates in Table~\ref{tab:gasmass}).

\begin{figure*}
\begin{center}
\includegraphics[width=0.8\textwidth]{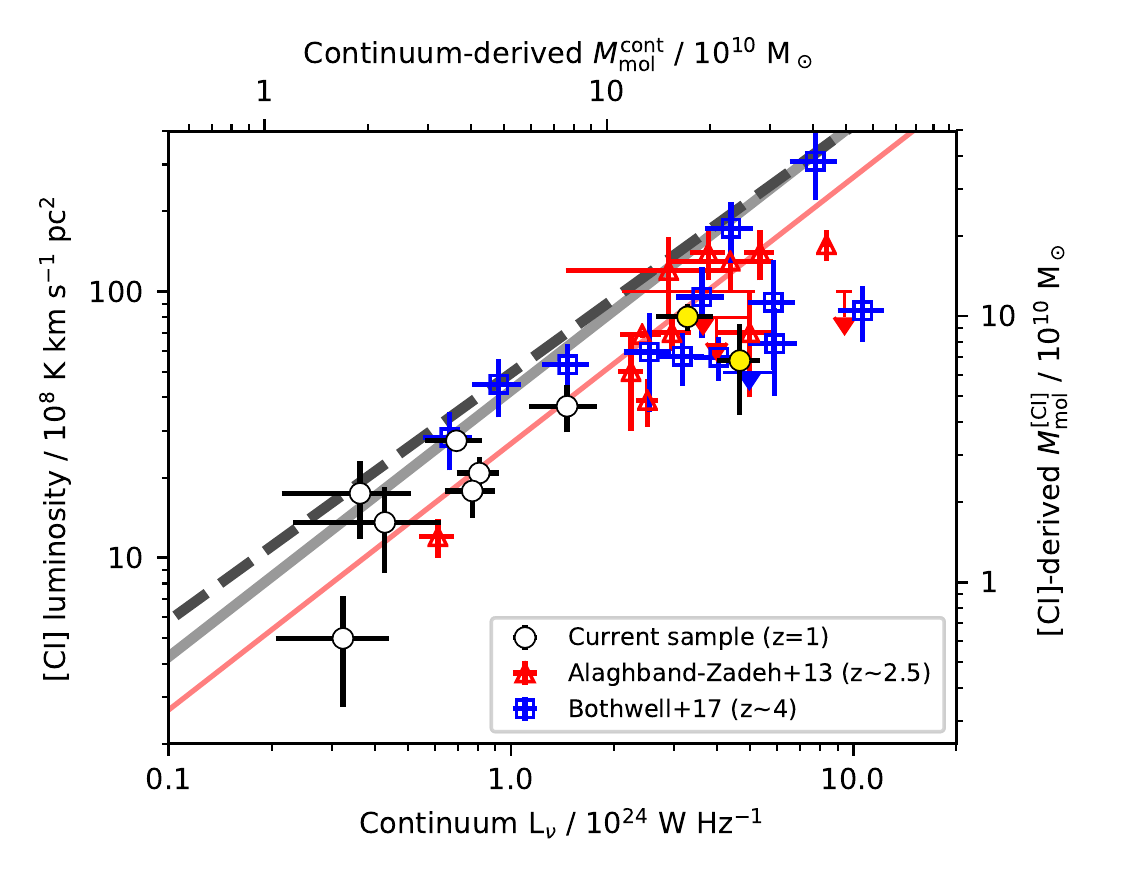}
\caption{Direct comparison of molecular gas tracers provided by continuum and \ci\ in our sample of $z=1$ galaxies in comparison to $z>2$ galaxy samples from the literature. Our sample is shown by the white and yellow circles (as in Fig.~\ref{fig:ci_mdust}a), while red triangles show the $z\sim2.5$ SMG sample from \citet{Alaghband-Zadeh2013} and blue squares show the sample of 1.4-mm-selected, lensed, dusty, star-forming galaxies at $z\sim4$ from \citet{Bothwell2017}.  Upper limits are plotted at $3\sigma$ where \ci\ was undetected. Continuum luminosities at rest-frame 609\,\micron\ are measured from the photometry provided by \citet{Alaghband-Zadeh2013} and references therein, and by \citet{Weiss2013}. Where necessary, we have extrapolated these from the nearest available wavelength assuming an SED with $T_d=35$\,K and $\beta=1.8$.
All luminosities are corrected for lens magnification. 
The top and right-hand axes indicate the molecular gas mass scales under the assumptions described in Section~\ref{sec:m_H2}, in particular $\alpha_{850}=6.7\times10^{19}$ and $\Xci=3\times10^{-5}$.
The solid grey line indicates where the data should lie if \ci-derived gas masses agree with continuum-derived gas masses under these same assumptions. The dashed line is the equivalent assuming $\alpha_{850}=6.2\times10^{19}(L_{\nu 850}/10^{31})^{0.07}$. The thin pink line is the linear fit to our data, given by $L^\prime_\textrm{[CI]}/L_\textrm{cont}=2.68\times10^{-15}$~K\,\kms\,pc$^{2}$\,(W\,Hz$^{-1}$)$^{-1}$, which implies $\Xci=1.9\times10^{-5}$.
}
\label{fig:lit}
\end{center}
\end{figure*}

Fig.~\ref{fig:lit} shows a direct comparison of the results of calibrating molecular gas mass from the continuum and \ci\ luminosities. In addition to our sample of $z=1$ star-forming galaxies close to the main sequence, we also show two samples of high-redshift dusty star-forming galaxies from the literature. These are the $z\sim2.5$ SMG sample from \citet{Alaghband-Zadeh2013}, which also includes SMGs from \citet{Cox2011}, \citet{Danielson2011} and \citet{Walter2011b}; and the 1.4-mm-selected sample of lensed, $z\sim4$ galaxies from \citet{Bothwell2017} and \citet{Weiss2013}. For the continuum fluxes of these samples, we have used the values at the wavelength of \ci\ (609\,\micron) where available, otherwise we have used the continuum flux at the nearest available wavelength to estimate the 609-\micron\ flux assuming an SED with $T_d=35$\,K and $\beta=1.8$. Continuum and line luminosities of lensed sources in these papers have been corrected for magnification using the published values, but we have not accounted for uncertainties in the magnification because we are interested the line/continuum relationship, which is unaffected as long as the \textsc{Ci} and dust share a common spatial distribution, i.e. assuming no differential magnification.

The striking result in Fig.~\ref{fig:lit} is that the \ci\ and continuum luminosities are strongly correlated over the wide dynamic range encompassed by these three samples, and that there appears to be a common relationship over all luminosities and redshifts sampled. This indicates that the dust and \ci\ are both tracing the same phase of the ISM, and are doing so in similar ways at all redshifts.

The top and right-hand axes of Fig.~\ref{fig:lit} indicate the results of calibrating molecular gas mass from continuum and \ci\ luminosities respectively in all of these samples, using the assumptions described in Section~\ref{sec:m_H2}.
The solid, light-grey line indicates the relation that would be expected if the \ci\ and continuum calibrations yielded the same gas mass ($M_\textrm{mol}^\textrm{[CI]}=M_\textrm{mol}^\textrm{cont}$), using the continuum calibration with constant $\alpha_{850}=6.7\times10^{19}$~erg\,s$^{-1}$\,Hz$^{-1}$\,\Msun$^{-1}$. The dashed, dark-grey line is the equivalent assuming the luminosity-dependent $\alpha_{850}$ in \eqnref{eqn:alpha850} (which is almost identical).

We find that all of the galaxies considered here, from all three samples, are on or below the relation that would be expected given our assumptions in Section~\ref{sec:m_H2}. The average offset is $-0.25$~dex (i.e. $\langle M_\textrm{mol}^\textrm{[CI]}/M_\textrm{mol}^\textrm{cont} \rangle=0.56$), although the data span a range of offsets between $-0.6$ and 0~dex (scatter $1\sigma=0.21$\,dex).
There are several possible explanations for the offset and scatter in the relation:
\begin{enumerate}
\item Firstly, the gas mass traced by dust may be overestimated, since the \citet{Scoville2016} calibration is based on an assumed constant dust/gas ratio and $X_\textrm{CO}$. However, the types of galaxies in the samples discussed here (e.g. $L_{\nu850}\sim10^{30}$\,erg\,s$^{-1}$\,Hz$^{-1}$ in our sample) are well-represented in the sample that \citet{Scoville2016} used to calibrate their relation, so this is unlikely to lead to a systematic offset, although it likely contributes to the scatter. 
\item Variations in $\Xci$ would introduce scatter, and if $\Xci<3\times10^{-5}$ (systematically) then our \ci\ luminosities would be underestimated. The value has been observed to vary by a factor $\sim2$ above or below the value assumed here (e.g. \citealp{Papadopoulos2004a}), but there are still only a small number of measurements in the literature. 
\item Varying excitation conditions would lead to a range of $Q_{10}$. $L^{\prime}_\textrm{[CI]}$ would be underestimated for galaxies with $Q_{10}<0.35$ (the value we assumed).  This will lead to scatter in the relation, although it is unlikely to explain the offset since luminous galaxies are more likely to have higher excitation (greater $Q_{10}$). 
\item We have assumed that \ci\ is optically thin, but if this is not the case then $M_\textrm{mol}^\textrm{[CI]}$ may be under-estimated. For example, if the emission emerges from radiatively decoupled cells (i.e. well-separated optically thick clouds) then \eqnref{eqn:MmolCI} underestimates gas mass by a factor $\beta=(1-e^{-\tau})/\tau$; e.g. $0.6<\beta<1$ for $\tau<1$ \citep{Papadopoulos2004}.  This may explain why many high-$z$ SMGs (and our source ID 2) have lower \ci-derived gas masses, due to their high gas densities. 
However, if this is the explanation for the systematic offset in our sample, then it would imply high optical depths throughout the sample. This is inconsistent with the fact that most of the sample have dust obscuration factors consistent with the average for their stellar mass (Fig.~\ref{fig:msfr}). It is also inconsistent with the extended nature of both \ci\ and continuum emission in most galaxies (Fig.~\ref{fig:mom0}).
\end{enumerate}
Overall, any of these effects is likely to contribute towards scatter in the relation, but the systematic offset is most likely explained by the unknown value of $\Xci$, or (in some cases) by optical depth effects.

\subsection{A prediction for the \ci/H$_2$ conversion}
In Section \ref{sec:interp} we compared the results of calibrating the molecular gas mass from \ci\ and continuum fluxes independently, making assumptions about \Xci, $Q_{10}$ and optical depth on the one hand, and $\alpha_{850}$ (encoding the dust/gas ratio and the dust mass/light ratio) on the other.
An alternative approach would be to assume that \ci\ and continuum both trace exactly the same gas mass (i.e. $M_\mathrm{mol}^\mathrm{[CI]} = M_\mathrm{mol}^\mathrm{cont}$), and make a prediction for the unknown value of \Xci, holding all other parameters constant. Our conclusion in Section \ref{sec:interp} was that \Xci\ was the parameter most likely to deviate systematically from the assumed value.
The best-fitting linear relation between $L_\textrm{cont}$ and $L^\prime_\textrm{[CI]}$ is indicated by the thin pink line in Fig.~\ref{fig:lit} (see also Table~\ref{tab:correlations}). Under the assumption that $M_\mathrm{mol}^\mathrm{[CI]} = M_\mathrm{mol}^\mathrm{cont}$, this implies $\Xci=(1.9\pm0.1)\times10^{-5}$. This is subject to the additional systematic uncertainties in $Q_{10}$, optical depth and $\alpha_{850}$. These may be investigated by future studies of the \ci\ (2--1)/(1--0) ratio, resolved observations of \ci, and improved understanding of the relationship between continuum and CO(1--0) emission, and of $\alpha_\mathrm{CO}$ (on which $\alpha_{850}$ is calibrated), at intermediate and high redshifts.

\section{Conclusions}
\label{sec:conclusions}
We have used ALMA to measure \ci(1--0), \cofour\ and dust continuum emission in a sample of star-forming galaxies distributed across the main sequence at $z=1$. 
We uncover a strong correlation between the total \ci\ and continuum fluxes of these galaxies. This translates into a similarly strong correlation between \ci\ luminosity and dust mass when a single dust temperatue is assumed to calibrate dust mass from the Rayleigh-Jeans continuum.

Considering global fluxes, we find that the \cofour--continuum and \cofour--\ci\ correlations are much weaker than the \ci--continuum correlation. \cofour\ emission traces a warmer, denser phase of the ISM, while \ci\ and dust continuum emission both appear to trace the same phase. This indicates that dust continuum is well correlated with the total molecular gas mass, since results from observations and dynamical PDR modelling indicate that \ci\ exists throughout the molecular ISM.
A deeper understanding of the relationships between ISM phases traced by CO, \ci\ and continuum can be achieved by studying their spatial extents and kinematics, which we leave for future work.  

By combining our sample of $z=1$ normal star-forming galaxies with high-redshift SMGs and lensed, mm-selected galaxies from the literature \citep{Alaghband-Zadeh2013, Bothwell2017}, we find that the correlation between continuum and \ci\ luminosities remains strong over the full range of luminosities and redshifts up to $z\sim4$. All three samples are consistent with a single relationship with a scatter of $1\sigma=0.2$\,dex. 

We use the \ci\ data to derive molecular gas masses for the three samples, and compare these to independent results from the dust continuum assuming the correlation from \citet{Scoville2016}, which is calibrated on CO. 
If we assume a typical value for the C\textsc{i}/\Htwo\ abundance ratio, $\Xci=3\times10^{-5}$, we find that \ci-derived gas masses in our sample are 1--4 times smaller than continuum-derived gas masses, and imply dust/gas ratios between 70--130.
Alternatively, if we assume that \ci\ and continuum luminosities trace the same gas mass, with fiducial values for the dust continuum--\MHtwo\ conversion ($\alpha_{850}$) and \ci\ excitation ($Q_{10}$), and assuming that \ci\ is optically thin, we predict $\Xci=(1.9\pm0.1)\times10^{-5}$. 

This study is the first analysis of \ci\ in a sample of ordinary star-forming galaxies outside the local Universe. The strength of the \ci--continuum correlation that we measure clearly motivates further analysis in extended samples of high-redshift star-forming galaxies, to explore a wider dynamic range in luminosity and determine the tightness and universality of the correlation. Achieving a better understanding of the relationship between dust emission and independent, low-excitation tracers of molecular gas including \ci(1--0) and \coone\ will enable accurate calibration of molecular gas mass from submm photometry, which is available for very large samples of high-redshift galaxies.


\section*{Acknowledgements}
The authors wish to thank Padelis Papadopoulos for useful discussions, and the staff at ALMA for their technical and scientific support. We also thank the anonymous referee for comments which improved the paper.
NB acknowledges the support of the UK Science and Technology Facilities Council and of the European Research Council in the form of the Advanced Investigator Program, \textsc{cosmicism} (ERC-2012-ADG\_20120216, PI: R. J. Ivison).
This paper makes use of the following ALMA data: ADS/JAO.ALMA\#2016.1.01184.S and ADS/JAO.ALMA\#2017.A.00013.S. ALMA is a partnership of ESO (representing its member states), NSF (USA) and NINS (Japan), together with NRC (Canada), NSC and ASIAA (Taiwan), and KASI (Republic of Korea), in cooperation with the Republic of Chile. The Joint ALMA Observatory is operated by ESO, AUI/NRAO and NAOJ.
This research made use of \textsc{astropy}, a community-developed core Python package for Astronomy \citep{Astropy2018}, and \textsc{aplpy}, an open-source plotting package for Python \citep{Robitaille2012}.

\bibliographystyle{mn2e}
\bibliography{MyLibrary20180423}



\bsp	
\label{lastpage}
\end{document}